\documentclass[s5paper_withbleed,11pt]{su_thesis}	
\usepackage[usenames,dvipsnames,svgnames]{xcolor}
\usepackage{pdfpages}
\usepackage{ifthen}
\usepackage{amssymb}
\usepackage{graphicx}
\usepackage[bf, sf, FIGTOPCAP, normal, tight]{subfigure}
\usepackage{changepage}
\usepackage{calc}
\usepackage{cancel}
\usepackage{multicol}
\usepackage{multirow}
\usepackage{bigstrut}
\usepackage{rotating}
\usepackage{booktabs}
\usepackage{hhline}
\usepackage{braket}
\usepackage{epigraph}
\usepackage{subfigure}
\usepackage{psfrag}
\usepackage{times}
\usepackage{amsfonts}
\usepackage{amsmath}

\usepackage{caption}
\usepackage{wrapfig}

\usepackage[utf8]{inputenc} 

\usepackage{array}
\usepackage{mathrsfs}
\usepackage{tikz}

\usepackage{natbib}
\bibpunct[, ]{(}{)}{;}{a}{}{,}
\renewcommand{\bibname}{References}

\usepackage[T2A,OT2,T1]{fontenc}

\newcommand\crussian[1]{{\fontencoding{T2A}\fontfamily{cmr}\selectfont #1}}

\def\aj{AJ}%
\def\araa{ARA\&A}%
\def\apj{ApJ}%
\def\apjl{ApJ}%
\def\apjs{ApJS}%
\def\ao{Appl.~Opt.}%
\def\apss{Ap\&SS}%
\def\aap{A\&A}%
\def\aapr{A\&A~Rev.}%
\def\aaps{A\&AS}%
\def\azh{AZh}%
\def\baas{BAAS}%
\def\jcap{JCAP}%
\def\jrasc{JRASC}%
\def\mnras{MNRAS}%
\def\memras{MmRAS}%
\def\pra{Phys.~Rev.~A}%
\def\prb{Phys.~Rev.~B}%
\def\prc{Phys.~Rev.~C}%
\def\prd{Phys.~Rev.~D}%
\def\prl{Phys.~Rev.~Lett.}%
\def\pasp{PASP}%
\def\pasj{PASJ}%
\def\qjras{QJRAS}%
\def\skytel{S\&T}%
\def\solphys{Sol.~Phys.}%
\def\sovast{Soviet~Ast.}%
\def\ssr{Space~Sci.~Rev.}%
\def\zap{ZAp}%
\def\nat{Nature}%
\newcounter{papercount}

\newlength{\paperboxheight}
\newlength{\paperboxwidth}
\newlength{\verticalpaperboxoffset}

\newcommand{\paperboxbackgroundcolor}{black}
\newcommand{\paperboxtextcolor}{Goldenrod}
\newcommand{\papersummary}[8]{

  \begin{description}
    \item[\bfseries\sffamily #8] {#1}. \emph{{#2}}, {#3}~\textbf{{#4}}, {#5} ({#6}); {#7}.
  \end{description}

}

\newcommand{\papersummaryNP}[5]{

  \begin{description}
    \item[\bfseries\sffamily #5] {#1}. \emph{{#2}}, {#3}; {#4}.
  \end{description}

}

\newcommand{\papersummaryNA}[4]{

  \begin{description}
    \item[\bfseries\sffamily #4] {#1}. \emph{{#2}}, {#3}.
  \end{description}

}

\newcommand{\paper}[9]{

  \ifodd \value{page}
  \else
    \newpage
    \null
    \newpage
  \fi
  
  \addtocounter{papercount}{1}
  \vspace*{\verticalpaperboxoffset}

  \begin{adjustwidth*}{}{-41mm}
    \begin{flushright}
      \huge\sffamily\bfseries \color{\paperboxtextcolor} \colorbox{\paperboxbackgroundcolor}
        {\parbox[c][\paperboxheight]{\paperboxwidth}{\hspace{1cm}Paper \Roman{papercount}}}
    \end{flushright}
  \end{adjustwidth*}

  \vfill
  
  \begin{flushleft}
{\textcolor{black}{
    {#1}\\
    \emph{{#2}}\\
    {#3}~\textbf{{#4}}, {#5} ({#6}) {#7}.
}}
  \end{flushleft}

  \phantomsection
  \addcontentsline{toc}{chapter}{Paper \Roman{papercount}: #2}
  \label{#9}

  \newpage
  \null
  \newpage
  \if1\includepapers
\nopagecolor
  \includepdf[pages=-]{#8}
  \fi
  \addtolength{\verticalpaperboxoffset}{1.2\paperboxheight} 
  
}

\newcommand{\paperNP}[6]{

  \ifodd \value{page}
  \else
    \newpage
    \null
    \newpage
  \fi
  
  \addtocounter{papercount}{1}
  \vspace*{\verticalpaperboxoffset}

  \begin{adjustwidth*}{}{-41mm}
    \begin{flushright}
      \huge\sffamily\bfseries \color{\paperboxtextcolor} \colorbox{\paperboxbackgroundcolor}
        {\parbox[c][\paperboxheight]{\paperboxwidth}{\hspace{1cm}Paper \Roman{papercount}}}
    \end{flushright}
  \end{adjustwidth*}

  \vfill
  
  \begin{flushleft}
{\textcolor{black}{
    {#1}\\
    \emph{{#2}}\\
    \emph{{#3}} {#4}.
}}
  \end{flushleft}

  \phantomsection
  \addcontentsline{toc}{chapter}{Paper \Roman{papercount}: #2}
  \label{#6}

  \newpage
  \null
  \newpage
  \if1\includepapers
\nopagecolor
  \includepdf[pages=-]{#5}
  \fi
  \addtolength{\verticalpaperboxoffset}{1.2\paperboxheight} 
  
}

\def\includepapers{0}

\newcommand\newc{\newcommand}

\newc{\beqn}{\begin{eqnarray}}
\newc{\eeqn}{\end{eqnarray}}
\newc{\be}{\begin{equation}}
\newc{\ee}{\end{equation}}
\newc{\nn}{\nonumber}
\newc{\dd}{\mathrm{d}}
\newc{\md}{\mathrm{d}}
\newc{\gmn}{g_{\mu\nu}}
\newc{\bgmn}{\bar g_{\mu\nu}}
\newc{\emn}{\eta_{\mu\nu}}
\newc{\fmn}{f_{\mu\nu}}
\newc{\bfmn}{\bar f_{\mu\nu}}
\newc{\Gmn}{G_{\mu\nu}}
\newc{\Mmn}{M_{\mu\nu}}
\newc{\hmn}{h_{\mu\nu}}
\newc{\meff}{M_\mathrm{eff}}
\newc{\mA}{\mathbb{A}}
\newc{\mB}{\mathbb{B}}
\newc{\dx}{\mathrm{d}^4x}
\newc{\dg}{\delta{g}}
\newc{\df}{\delta{f}}
\newc{\dm}{\delta{M}}
\newc{\dS}{\delta{S}}
\newc{\dMG}{\delta M^G}
\newc{\dMg}{\delta M}
\newc{\bphi}{\bar{\Phi}}
\newc{\bpsi}{\bar{\Psi}}
\newc{\dphi}{\delta{\Phi}}
\newc{\dpsi}{\delta{\Psi}}
\newc{\dgg}{\delta{G}}
\newc{\bgg}{\bar{G}}
\newc{\bg}{\bar{g}}
\newc{\Tr}{\mathrm{Tr}}
\newc{\tr}{\mathrm{Tr}}
\newc{\Lag}{\mathcal{L}}
\newc{\E}{\bar{\mathcal{E}}}
\newc{\EE}{\tilde{\mathcal{E}}}
\newc{\mfp}{m_{\mathrm{FP}}}
\newc{\Mc}{M^{\mathrm{c}}}
\newc{\te}{\tilde\epsilon}
\newc{\cN}{{\cal N}}
\newc{\bbeta}{\bar{\beta}}
\newc{\mpl}{M_{\mathrm{Pl}}}

\newcommand{\lbrac}{\left\{}
\newcommand{\rbrac}{\right\}}
\newcommand{\qu}[1]{``#1''}

\newcommand{\mbTh}{\mathbf{\Theta}}

\title{\LARGE The supernova cosmology cookbook:\\ Bayesian numerical recipes}
\author{Natallia V. Karpenka}
\date{September 2014}
\shortdate{2014}
\type{Doctoral Thesis in Theoretical Physics}

\division{Oskar Klein Centre for\\ Cosmoparticle Physics\\
  and  Cosmology,\\ 
  Particle Astrophysics \\and String Theory\\\vskip 0.5em}
\department{Department of Physics}
\address{SE-106 91 Stockholm}
\city{Stockholm}
\country{Sweden}
\cplogo{\includegraphics[height=4cm, trim = 0 0 250 200, clip=true]{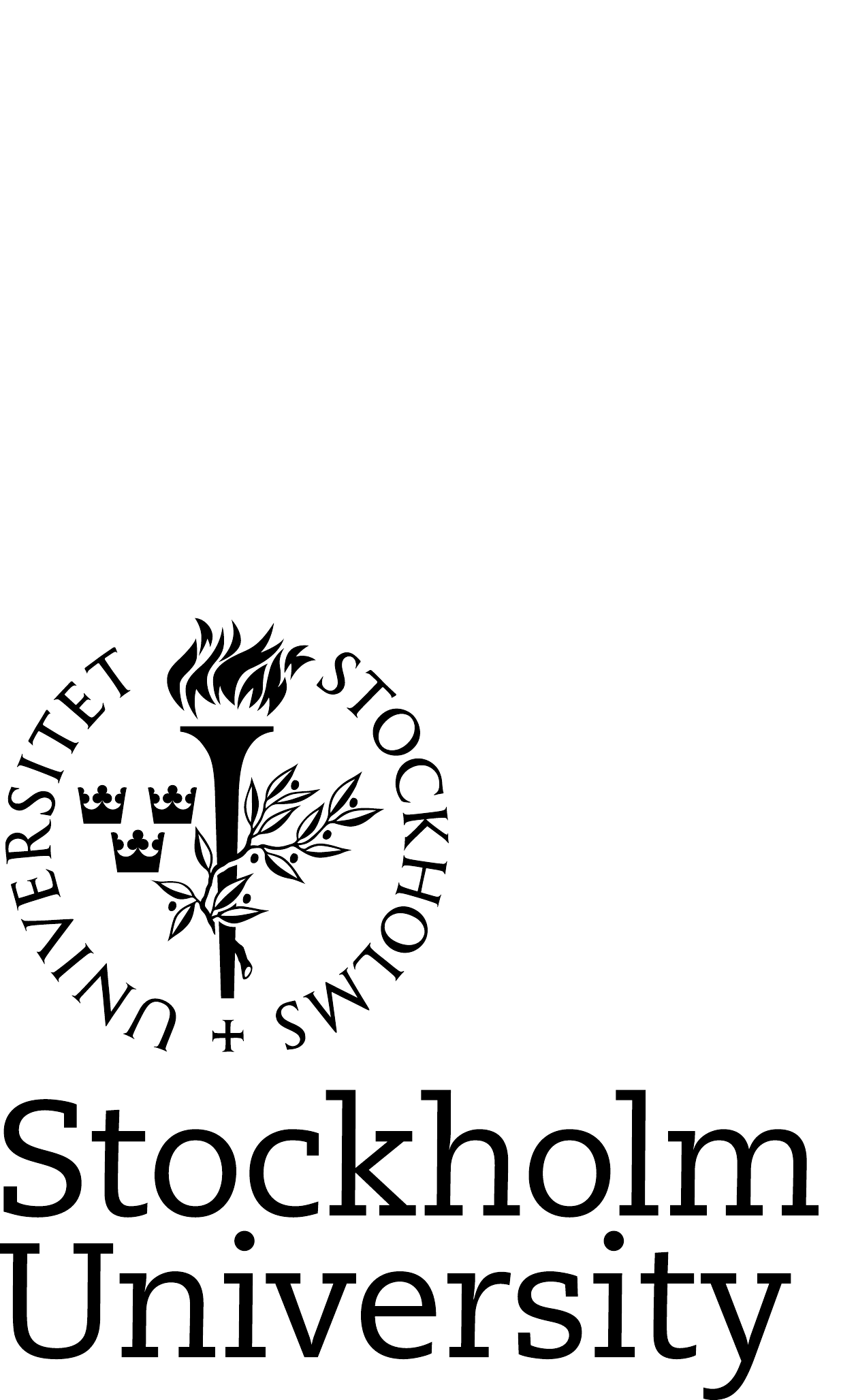}}

\publisher{\vspace{2mm}\\Printed by Universitetsservice US-AB, Stockholm, Sweden, 2014.\vspace{5mm}
\vspace{2mm}\\Figures 1.2, 1.3, 1.5, 2.1, 2.2, 2.3, 3.1, 3.3 and 4.4 used with permission.
\vspace{5mm}\\Typeset in pdf\LaTeX}
\copyrightline{pp.\ i--xxii, 1--98 \copyright\ Natallia V. Karpenka, 2014\vspace{2mm}} 
\trita{xx}
\issn{aa}
\isrn{bb}
\isbn{978-91-7447-953-9 (pp.\ i--xxii, 1--98)}
\comment{\emph{Cover image:} Cosmology princess cake. See Appendix A. Credit: Dr.~Oscar Larsson}
\innerlogo{\vspace{3.5cm}\includegraphics[height=3cm]{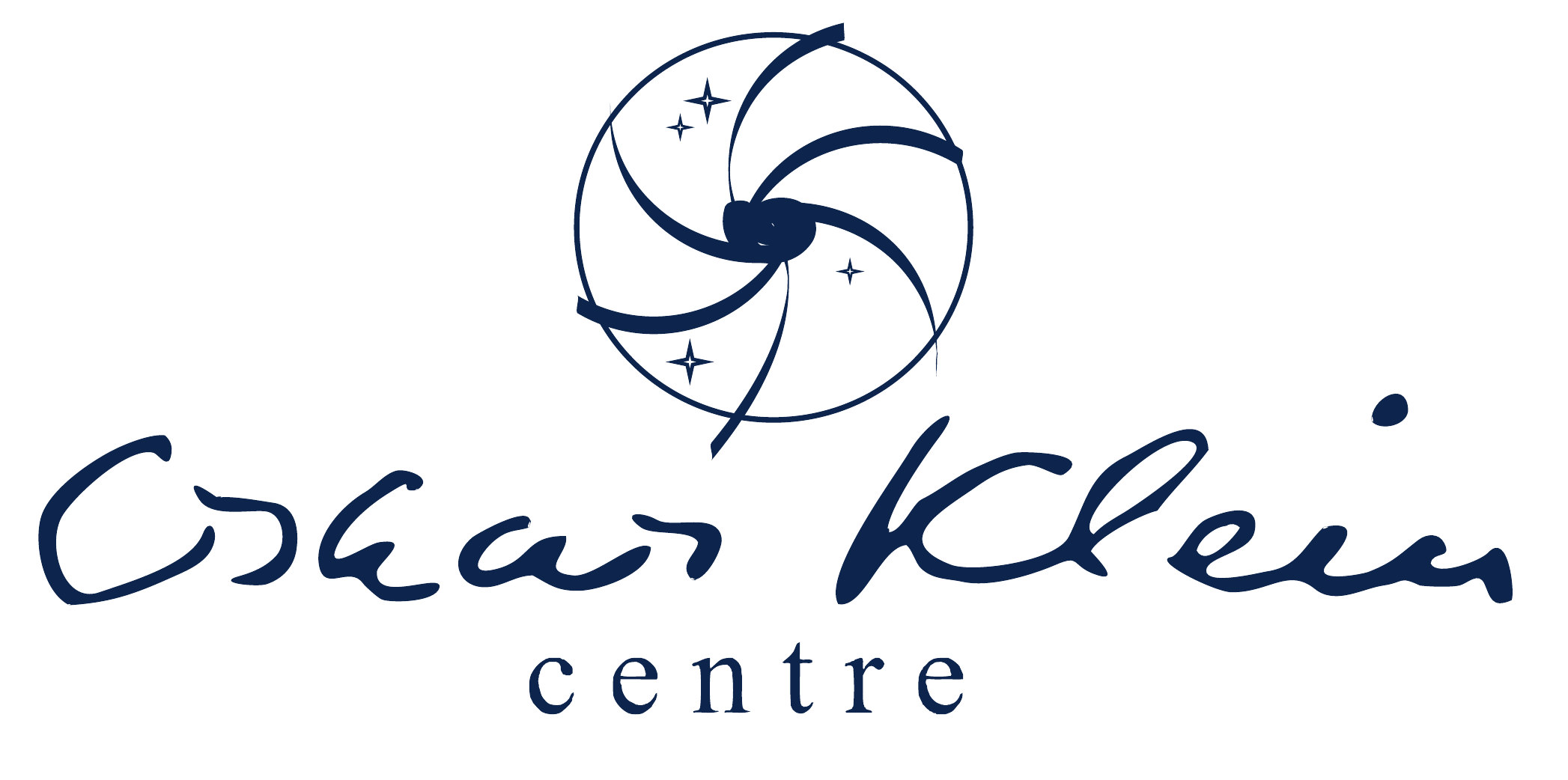}}
\extrainnerlogo{\includegraphics[height=4cm, trim = 0 0 250 400, clip=true]{Figures/sulogo}}

\newc{\papi}{{\protect{\hyperref[papI]{Paper I}}}}
\newc{\papii}{{\protect{\hyperref[papII]{Paper II}}}}
\newc{\papiii}{{\protect{\hyperref[papIII]{Paper III}}}}
\newc{\papiv}{{\protect{\hyperref[papIV]{Paper IV}}}}
\newc{\papv}{{\protect{\hyperref[papV]{Paper V}}}}
\newc{\papvi}{{\protect{\hyperref[papVI]{Paper VI}}}}
\newc{\papvii}{{\protect{\hyperref[papVII]{Paper VII}}}}
\newc{\papviii}{{\protect{\hyperref[papVIII]{Paper VIII}}}}

\newc\AIPCP[3] {{\em AIP Conf. Proc.} {\bf #1} (#2) #3}
\newc\AJ[3] {{\em Astrophys. J.} {\bf #1} (#2) #3}
\newc\AMS[3] {{\em Ann. Math. Statist.} {\bf #1} (#2) #3}                
\newc\AP[3] {{\em Ann. Phys.} {\bf #1} (#2) #3}
\newc\APJ[3] {{\em Astropart. J.} {\bf #1} (#2) #3}
\newc\APP[3] {{\em Astropart. Phys.} {\bf #1} (#2) #3}
\newc\APS[3] {{\em Astrophys. J. Suppl.} {\bf #1} (#2) #3}
\newc\ARNPS[3] {{\em Ann. Rev. Nucl. Part. Sci.} {\bf C#1} (#2) #3}
\newc\BA[3] {{\em Class. Quant. Grav.} {\bf C#1} (#2) #3}              
\newc\CPC[3] {{\em Comput. Phys. Commun.} {\bf C#1} (#2) #3}
\newc\CP[3] {{\em Contemp. Phys.} {\bf #1} (#2) #3}                     
\newc\EPJ[3] {{\em Euro. Phys. Journ.} {\bf C#1} (#2) #3}
\newc\JCAP[3] {{\em JCAP} {\bf #1} (#2) #3}
\newc\JTAP[3] {{\em Class. Quant. Grav.} {\bf #1} (#2) #3}
\newc\JHEP[3] {{\em JHEP} {\bf #1} (#2) #3}
\newc\JPG[3] {{\em J. Phys.} {\bf G #1} (#2) #3}
\newc\IJMP[3] {{\em Int. J. Mod. Phys.} {\bf A #1} (#2) #3}
\newc\MNRAS[3] {{\em Mon. Not. Roy. Astron. Soc.} {\bf #1} (#2) #3}
\newc\MPL[3] {{\em Mod. Phys. Lett.} {\bf A #1} (#2) #3}
\newc\NAR[3] {{\em New Astron. Rev.} {\bf #1} (#2) #3}                  
\newc\NCA[3] {{\em Nuovo Cimento} {\bf #1} (#2) #3}
\newc\NIM[3] {{\em Nucl. Instrum. Methods} {\bf #1} (#2) #3}
\newc\NIMA[3] {{\em Nucl. Instrum. Methods} {\bf A #1} (#2) #3}
\newc\NAT[3] {{\em Nature} {\bf #1} (#2) #3}
\newc\NPB[3] {{\em Nucl. Phys.} {\bf B #1} (#2) #3}
\newc\NPA[3] {{\em Nucl. Phys.} {\bf A #1} (#2) #3}
\newc\NPPS[3] {{\em Nucl. Phys. Proc. Suppl.} {\bf #1} (#2) #3}                
\newc\PLB[3] {{\em Phys. Lett.} {\bf B #1} (#2) #3}
\newc\PR[3] {{\em Phys. Rep.} {\bf #1} (#2) #3}
\newc\PRL[3] {{\em Phys. Rev. Lett.} {\bf #1} (#2) #3}
\newc\PRD[3] {{\em Phys. Rev.} {\bf D #1} (#2) #3}
\newc\PRC[3] {{\em Phys. Rev.} {\bf C #1} (#2) #3}
\newc\PTP[3] {{\em Prog. Theor. Phys.} {\bf #1} (#2) #3}
\newc\RMP[3] {{\em Rev. Mod. Phys.} {\bf #1} (#2) #3 }
\newc\RPP[3] {{\em Rept. Prog. Phys.} {\bf #1} (#2) #3 }
\newc\SC[3] {{\em Science} {\bf #1} (#2) #3 }
\newc\ZPC[3] {{\em Z. Phys.} {\bf C #1} (#2) #3}
\newc\Err[3] {{\em Erratum-ibid.} {\bf #1} (#2) #3 }

\definecolor{purple}{RGB}{85,26,139}
\usepackage[colorlinks=true,linkcolor=blue]{hyperref}

\begin{document}

\begin{poem}

\vspace*{35mm}

\noindent



\vspace*{60mm}
\raggedleft
\bf \crussian{Посвящается\\ дедушке Михаилу Тимофеевичу Назарчуку \\и бабушке Эмме Владимировне Карпенко.}

\end{poem}

\begin{abstract}

Theoretical and observational cosmology have enjoyed a number of
significant successes over the last two decades. Cosmic microwave
background measurements from the Wilkinson Microwave Anisotropy
Probe and Planck, together with large-scale structure and supernova (SN) searches, have put very tight constraints on cosmological
parameters. Type Ia supernovae (SNIa) played a central role in the
discovery of the accelerated expansion of the Universe, recognised by
the Nobel Prize in Physics in 2011.

 The last decade has seen an
enormous increase in the amount of high quality SN observations, with
SN catalogues now containing hundreds of objects. This number is
expected to increase to thousands in the next few years, as data from
next-generation missions, such as the Dark Energy Survey and
Large Synoptic Survey Telescope become available. In order to
exploit the vast amount of forthcoming high quality data, it is
extremely important to develop robust and efficient statistical
analysis methods to answer cosmological questions, most notably
determining the nature of dark energy. 

To address these
problems my work is based on nested-sampling approaches to parameter
estimation and model selection and neural
networks for machine-learning. Using advanced Bayesian techniques, I constrain the
properties of dark-matter haloes along the SN lines-of-sight via
their weak gravitational lensing effects, develop methods for
classifying SNe photometrically from their lightcurves, and present results on
more general issues associated with constraining cosmological
parameters and testing the consistency of different SN
compilations.

\end{abstract}

\begin{svensksammanfattning}

Teoretisk och observationell kosmologi har åtnjutit många viktiga framgångar de senaste årtiondena. Mätningar av den kosmiska mikrovågsbakgrunden från Wilkinson Microwave Anisotropy Probe and Planck, tillsammans med undersökningar av Universums storskaliga struktur och supernovor, har satt stränga begränsningar på de kosmologiska parametrarna. Supernovor av Typ Ia spelade en central roll i upptäckten av Universums accelererade expansion, en upptäckt som belönades med Nobelpriset 2011.

 Det senaste årtiondet har fört med sig en enorm ökning av mängden högkvalitativa observationer av supernovor, och kataloger innehåller nu hundratals objekt. Detta antal förväntas öka till tusentals inom de närmsta åren i och med att data från nästa generations observationer som Dark Energy Survey och Large Synoptic Survey Telescope blir tillgängliga. För att kunna utnyttja den stora mängden kommande data är det extremt viktigt att utveckla robusta och effektiva tekniker för statistisk analys för att kunna svara på de kosmologiska frågeställningarna, framför allt gällande den mörka energins beskaffenhet. 
 
 För att angripa dessa problem är mitt arbete baserat på parameteruppskattning och modellval via nested sampling, samt neurala nätverk för maskininlärning. Med hjälp av avancerade Bayesianska metoder har jag satt gränser på egenskaperna hos halor av mörk materia längs med supernovors siktlinjer via deras svaga gravitationella effekt, utvecklat metoder för fotometrisk klassificering av supernovor från deras ljuskurvor, samt arbetat med mera allmänna frågor associerade med bestämningen av de kosmologiska parametrarna samt undesökt förenligheten av olika sammanställningar av supernovor.

\end{svensksammanfattning}

\chapter*{List of accompanying papers}
\pagestyle{plain}
\addcontentsline{toc}{chapter}{List of accompanying papers}

\papersummary{M.C. March, N.V. Karpenka, F. Feroz and
  M.P. Hobson}{\\Comparison of cosmological parameter inference methods
  applied to supernovae lightcurves fitted with
  SALT2}{\mnras}{437}{4}{p. 3298-3311}{\\\href{http://arxiv.org/abs/1207.3705}{arXiv:1207.3705}}{\protect{\hyperref[papI]{Paper
      I}}}

\papersummary{N.V. Karpenka, M.C. March, F. Feroz and
  M.P. Hobson}{\\Bayesian constraints on dark matter halo properties
  using gravitationally-lensed
  supernovae}{\mnras}{433}{4}{p. 2693-2705}{\href{http://arxiv.org/abs/1207.3708}{arXiv:1207.3708}}{\protect{\hyperref[papII]{Paper
      II}}}

\papersummary{N.V. Karpenka, F. Feroz and M.P. Hobson}{\\A simple and
  robust method for automated photometric classification of supernovae
  using neural networks}{\mnras}{429}{2}{p. 1278-1285}{\\\href{http://arxiv.org/abs/arXiv:1208.1264}{arXiv:1208.1264}}{\protect{\hyperref[papIII]{Paper
      III}}}

\papersummaryNP{N.V. Karpenka, F. Feroz and M.P. Hobson}{\\Testing
  the mutual consistency of different supernovae surveys}{submitted to
  \mnras}{\href{http://arxiv.org/abs/arXiv:1407.5496}{arXiv:1407.5496}}{\protect{\hyperref[papIV]{Paper
      IV}}}

\papersummaryNA{N.V. Karpenka, F. Feroz and M.P. Hobson}{\\Photometric classification of supernovae using hierarchical neural network method}{to be
  submitted to \mnras }{\protect{\hyperref[papV]{Paper V}}}

\begin{acknowledgments}

I thank my supervisor, Prof. Joakim Edsj\"o, for his advice and
support, and for allowing me the academic freedom to pursue a diverse
range of projects.

I thank all the people who have been a part of the fantastic OKC
family during my PhD years; it is impossible to name you all. Thank
you for the lunch chats, Tuesday fikas, coffee breaks, poker nights
and midsummer, New Years and the other celebrations we shared. And a
little cheeky thanks to Chris Savage for all the time you wasted on
teaching me basic programming during my first year.

Another amazing academic family whom I thank is the Cavendish
Astrophysics group. Your tea-times in the old tea room and the new tea
corridor will stay with me forever. Thanks for making me feel so at
home during all my long visits. A special thank you goes to Karen, who
made my visits so easy by taking care of all the arrangements. I
cannot thank Michael Hobson and Farhan Feroz enough for everything I
have learned from them over the last few years; thank you for making
me part of the Bayesian club. Farhan, thank you for introducing me to
the ``giant puri''; my life was incomplete without it!  Michael, I
could not think of a better mentor and friend than you are to me. The
Collison family is also an inseparable part of my Cambridge
experience; thanks for your incredible hospitality and all the wine I
drank at your house.

I thank my friends from home for visiting me in Stockholm so often and
for bringing part of home with them to me, when I didn't have the
chance to go there myself, especially Tanya, Chris, \crussian{географy-строителю} Zhenya and \crussian{``всем теоретикам''}. Talking of home, I give huge thanks to two
women, the first of whom gave a solid base to my physics knowledge and
the second of whom taught me English and in so doing opened up a whole
new world for me. Natallia Dmitrievna Emelyanchenko and Svetlana
Vasil'evna Luzgina, I would never have had the chance to complete this
work without you! 

My dearest thanks go to Johannes. It was a great pleasure to have you
in my life for all these years; we lived together, we learned
together, and we even raised a fantastic Weakly Interacting Massive
Katt Type A (Wimka) together. It would have been so different without
you! I also thank Johannes' family: pappa Pontus, Gisela, mormor Hilde,
Charles, Majsan, Elsa, Sabine and everyone else --- tack s\aa ~mycket.

Last, but not least, my thanks go to my family for always being there
for me. Babushka Nina, dedushka Misha, and all my aunts, cousins and nephews,
thank you for following my successes and for never really asking what
it is that I do at work.  Mama and Papa, you have always been the
greatest example of academics, couple and family for me. I still can't
believe my luck that I am your daughter.

\end{acknowledgments}

\begin{preface}

\subsubsection*{Thesis plan}

This doctoral thesis consists of two major parts: (i) a short summary
of SN observations, cosmological constraints, the statistical
methods used for these analyses and my results; and (ii) the corresponding 5
articles published, submitted or ready to be submitted for publication.

The first part of the thesis is further divided into seven chapters, the
contents of which are briefly summarised below.
\begin{itemize}

\item[$\ast$] Chapter~\ref{intro} motivates this work from
  a cosmological point of view. It gives a short introduction to the
  history of cosmology, highlights its major discoveries and outlines
  remaining significant challenges.

\item[$\ast$] Chapter~\ref{ch:ST} gives an overview of the numerical
  methods used in my work: Bayesian methods and their use in parameter
  estimation and model selection; scanning algorithms, concentrating
  on \textsc{MultiNest}; and neural
networks (NNs) and their applications.

\item[$\ast$] Chapter~\ref{ch:SN_Clas} discusses SN discovery,
  classification and the study of progenitor models. It presents a
  historical overview of SN surveys and an outlook on the current
  state and problems that future surveys will bring. It also
  summarises the techniques for
  the standardisation of SNIa, and their associated shortcomings.

\item[$\ast$] Chapter~\ref{ch:Cosmology} discusses different
  techniques, including the $\chi^2$-method and Bayesian Hierarchical Method (BHM),
  for cosmological parameter inference, and an assessment of their
  advantages and drawbacks.

\item[$\ast$] Chapter~\ref{ch:lensing} reviews gravitational lensing
  effects in astronomy; it gives a short derivation of the lens
  equation and shows how one can view SNe through gravitational telescopes.

\item[$\ast$] Chapter~\ref{ch:Summ} summarises the main results in my
  papers and unpublished studies.
  
 \item[$\ast$] Chapter~\ref{ch:Outlook} gives a brief outlook on
   future challenges in SN cosmology and outlines, in this context,
   how one can develop further the methods presented in this work to
   address these issues.

\end{itemize}

\newpage
\subsubsection*{My contribution to the papers}

\papi \\ I made the original discovery of the previously unknown
difference in the best-fit cosmological parameters obtained from the
3-year SuperNova Legacy Survey (SNLS) SN catalogue using the standard
$\chi^2$-method and the BHM. I
initiated the project to explore this difference systematically, using
SN simulations as well as real SN data. I worked closely with Marisa
March in interpreting the results and led numerous discussions
regarding the direction of the project and the content of the paper. I
also wrote and revised large sections of the manuscript.\\

\vspace*{-2mm}

\noindent\papii \\ I proposed the project to perform a Bayesian
analysis of gravitationally-lensed SNe to constrain the
properties of dark matter haloes along their lines-of-sight. I wrote
the code to perform the simulations and the analysis of real data. I
ran the code and performed the numerical calculations to produce the
results, made the plots for the paper and wrote most of the
manuscript.
\newline

\vspace*{-2mm}

\noindent\papiii \\ I suggested the original idea of creating a
homogeneous training set by fitting SN lightcurves with a
parameterised function. I wrote the code to perform this fitting and
ran it to produce the training sets and plots of the fits. I
interpreted the results of the NN classifier and led discussions
regarding the direction of the research and the content of the
paper. I wrote most of the manuscript.
\newline

\vspace*{-2mm}

\noindent\papiv \\ I created the realistic simulations. I wrote the
code used for the analysis of the simulations and real data. I also
interpreted and summarised all the results. I created all the plots
used in the paper. I wrote most of the manuscript.
\newline

\vspace*{-2mm}

\noindent\papv \\I created the data-sets for the hierarchical NN (HNN) training. 
I interpreted the results of the NN classifier.  I created all the plots
of the paper. I wrote most of the manuscript.
I simulated realistic Dark Energy Survey (DES) data-sets to investigate this method further.
\newline

\hspace{0cm}\\
\raggedleft
\noindent Natallia V. Karpenka\\
Stockholm, August 2014

\end{preface}

\begin{abr}

\begin{tabular}{ll}

BHM & Bayesian Hierarchical Method\\
CCD & charge-coupled device\\
CDM & cold dark matter\\
CMB & cosmic microwave background\\
CSP & Carnegie Supernova Project \\
DES & Dark Energy Survey \\
HNN & hierarchical neural network \\
HST & Hubble Space Telescope\\
HZT & High-Z team \\
MCMC & Markov chain Monte Carlo \\
MLCS & Multi Light Curve Shape\\
NFW &  Navarro--Frenk--White \\
NN & neural network\\
iPTF & intermediate Palomar Transient Factory\\
SALT & Spectral Adaptive Lightcurve Template\\
SCP & Supernova Cosmology Project \\
SDSS & Sloan Digital Sky Survey\\
(t)SIS & (truncated) singular isothermal sphere \\
SLSN & superluminous supernova \\
SN & supernova\\
SNIa & supernova type Ia\\
SNII-P & Type II Plateau SN\\
SNLS & SuperNova Legacy Survey\\ 
WMAP & Wilkinson Microwave Anisotropy Probe\\
ZTF & Zwicky Transient Factory 
\end{tabular}
\end{abr}

\tableofcontents

\clearpage

\mainmatter

\part[Background material and summary of my results]{Background material \\and\\ summary of my results}



\chapter{Introduction}
\label{intro}

\epigraph{`You never know when you'll luck out. Take it from one who knows.'}{\textit{Max Frei}}

In this chapter, I present a broad introduction to the standard model
of cosmology, on which the research presented in this PhD
thesis is based. Rather than presenting a detailed analytical
description, which can be found in numerous textbooks, I present a
more qualitative, chronological account of its development, which
hopefully makes this material more accessible and places into context
how we have arrived at our current understanding of the Universe.  The
original research in this thesis is focussed on observations of
SNe and the application of novel statistical methods to their
analysis, and I will describe these topics in subsequent chapters.

\section{Relativistic gravitation, cosmology and the expanding universe}

Nearly a century has passed since Einstein published his completed
theory of general relativity in November 1915. Within a month,
Einstein discovered that his new theory could account precisely for a
well-known ``anomaly'' in the orbit of Mercury. Moreover, in 1919, his
prediction for the deflection of light from distant stars by the Sun
was experimentally verified by Arthur
Eddington, and Einstein became internationally famous.

As early as 1917, Einstein realised that he had the necessary tools
with which to derive the first fully self-consistent model of the
Universe as a whole. He immediately faced a problem, however, in that
his equations predicted the Universe to expand or contract, which ran
contrary to the prevailing belief at the time that the Universe was
static. In order to construct a static model for the Universe,
Einstein added to his equations an extra term that contained a new
constant of nature called the ``cosmological constant''. 
By carefully fine-tuning the value of the
cosmological constant, Einstein constructed a static model of the
Universe. Through the mid 1920s, however, Friedmann, Lema\^itre and
Robertson all independently obtained the general solution to Einstein's
equations of general relativity for an isotropic universe, each
finding that, without fine-tuning the value of the cosmological
constant, the generic predicted behaviour was for the universe to
expand or contract.

\begin{figure}
\centerline{\includegraphics[width=9cm]{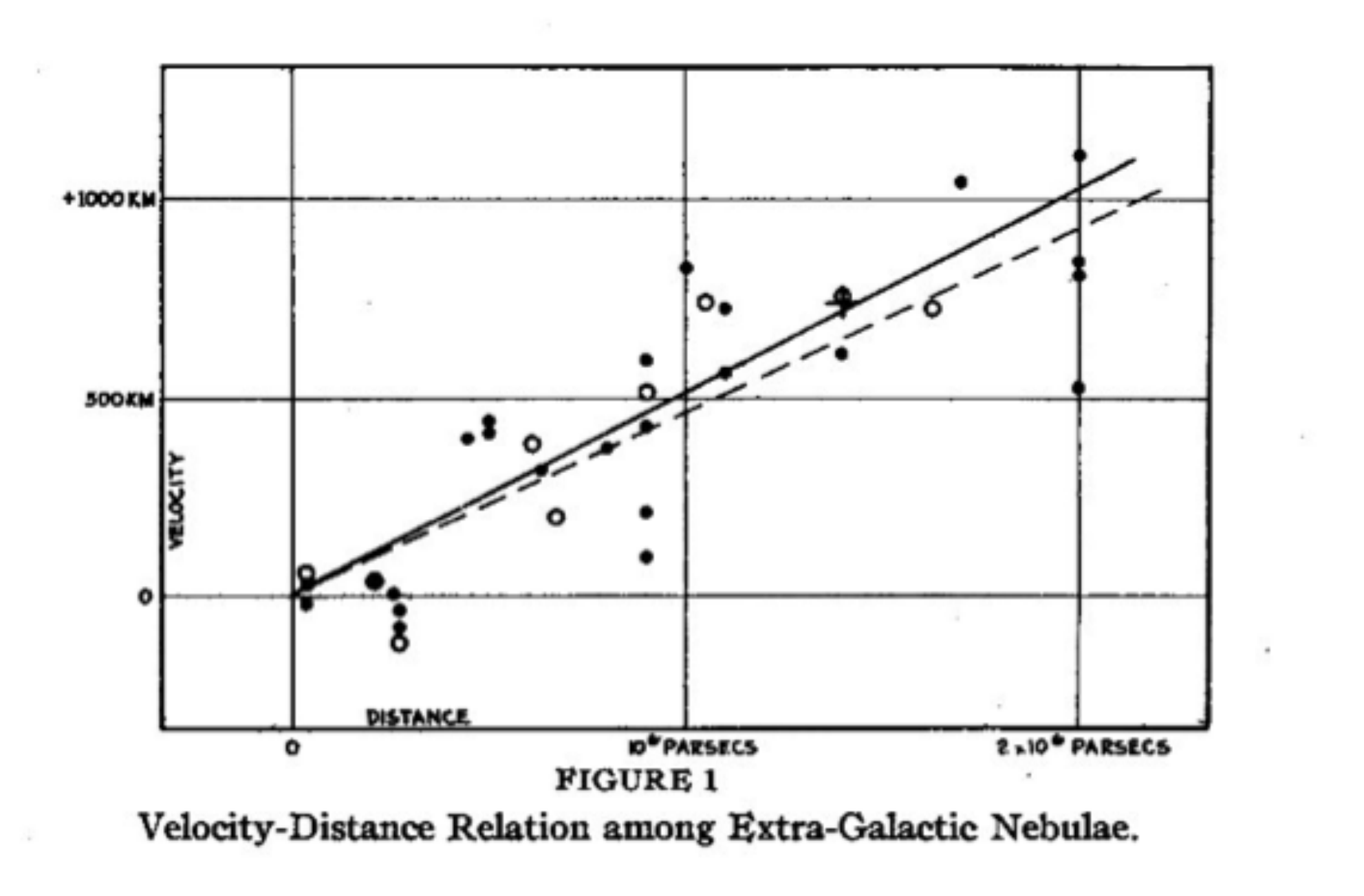}}
\caption{Plot from Hubble (1929), which shows that the redshift of a
  galaxy, interpreted by Hubble as a speed of recession, is
  proportional to its distance.}
\label{f:hubble}
\end{figure}

Theory and observation came together in 1929, when Edwin Hubble
combined his distance estimates to a selection of spiral galaxies with
exquisite spectroscopic studies of the galaxies made nearly 20 years
earlier by Vesto Slipher. Such spectra may be used as a cosmic
``bar-code'' to identify particular atoms from the pattern of narrow
lines in the spectrum, and also as a ``radar-gun'' to determine the
velocity of the emitting material along the line-of-sight by measuring
the Doppler shift in the wavelength of the spectral lines as compared
with laboratory measurements on Earth.  Slipher found that the ``spiral
nebulae'' were made from normal matter, but also discovered that their
observed spectral lines were all shifted significantly to longer
wavelengths (towards the red end of the visible spectrum of light).
From these so-called ``redshifts'' in the spectral lines\footnote{The
  redshift $z$ is defined by $1+z =
  \lambda_\text{obs}/\lambda_\text{em}$, where $\lambda_\text{obs}$ is
  the observed wavelength of the spectral line and $\lambda_\text{em}$
  is its emitted wavelength, i.e. that measured in a laboratory
  experiment.}, he thus deduced that the galaxies were all moving away
from us at considerable speeds.  When Hubble compared the speeds of
recession of these galaxies with the distances that he had measured to
them, obtained by observing the periods of the Cepheid variable stars
that they contained, he made the astonishing discovery that the speed
of recession of an object is proportional to its distance, as shown in
the Figure~\ref{f:hubble}. This was interpreted as resulting from the
Universe expanding uniformly in all directions.  When Einstein learned
of Hubble's results, he is said to have described his inclusion of the
additional cosmological constant term in the equations of general
relativity as ``the biggest blunder'' of his life. As we will see
later, however, posterity may judge otherwise.

\begin{figure}
\centerline{\includegraphics[width=9cm]{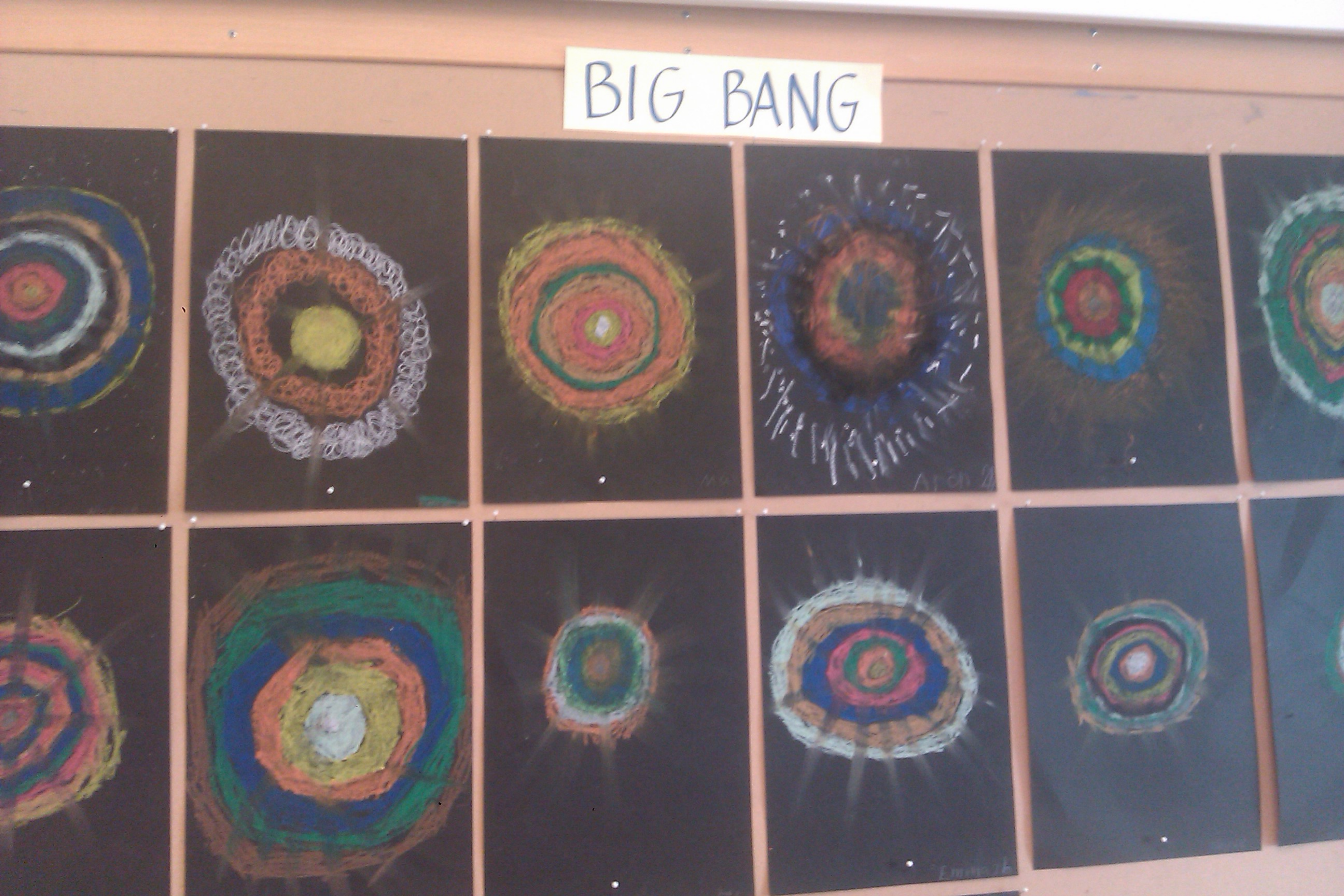}}
\caption{Eight-year-old artists' impressions of the Big
  Bang. Photograph courtesy of Pontus Bergstr\"om.}\label{f:bigbang}
\end{figure}

The expansion of the Universe, when combined
with Einstein's theory of general relativity (with the simplifying
assumptions of the large-scale homogeneity and isotropy of space) laid
the foundations for the development of the standard Big Bang theory of
cosmology (Figure~\ref{f:bigbang}), which remains to the present day
our best description of the Universe. The idea was first proposed in
1932 by Georges Lema\^itre, who suggested that
the observed expansion of the Universe implied that, moving backwards
in time, it must contract and would continue to do so until all the
matter in Universe was contained in a single point, a ``primeval
atom'', which marked the origin of the spacetime fabric itself.  Thus,
running the expansion of the Universe backwards according to the laws
of general relativity, and extrapolating, implies that the matter had
an infinite density and temperature at a {\em finite} time in the
past. Indeed, the presence of such a {\em singularity} indicates the
limit of applicability of the theory of general relativity. Based on a
range of cosmological observations, it is currently estimated that this
occurred around 13.7 billion years ago. It is worth noting, however,
that the Big Bang theory cannot and does not provide any explanation
for the initial singularity, but instead describes and explains the
general evolution of the Universe since that instant, as it expanded
from an extremely hot and dense state at very early times to its
cool and diluted state today.

The Big Bang theory was advocated and developed further in the late
1940s and early 1950s by George Gamow, who introduced the idea that
the nuclei of the light elements, such as helium, deuterium and
lithium, could be formed from nuclear processes occurring in the
rapidly expanding and cooling first minutes of the Universe, following
the Big Bang.  His colleagues, Ralph Alpher and Robert Herman, also
determined the thermal history of the Universe in this model and
predicted the existence of the cosmic microwave
background (CMB) radiation, a near-uniform bath of thermal radiation, sometimes called ``the
afterglow of creation'', that pervades the Universe. Alpher and Hermann
calculated that, just 300,000 years after the Big Bang, the Universe
would cool sufficiently for the ionised gas of mostly free protons and
electrons (plus other light nuclei) to combine to form neutral atoms
(predominantly hydrogen), marking a sharp transition between an opaque
charged-particle plasma to a transparent neutral gas through which
photons can travel unhindered, stretching as the Universe expands,
until they are observed today as the CMB, a thermal blackbody
radiation characterised by a temperature of just a few Kelvin.

By the early 1960s, observations also revealed that the percentage by
mass of Helium in the Universe was around 23\%. This uniformity and the fact that this percentage was much
greater than what could be created in the cores of stars pointed to a
cosmic origin, as suggested earlier by Gamow. Hoyle showed that such a
percentage of Helium was indeed predicted to be synthesised in the
early stages of the Big Bang.  Subsequent calculations by Fowler and
Wagoner showed that Big Bang nucleosynthesis also produced traces of
other light elements, which were very difficult to form inside
stars. The predicted abundances matched observations very well. The
status of the Big Bang as our best theory for the origin and evolution
of the cosmos was secured, however, by the serendipitous discovery of
the CMB by Penzias and Wilson in 1964. While preparing the 20-foot
horn-shaped antenna at Bell Laboratories to perform some
radio-astronomical observations, they discovered an excess of
radiation at a temperature of around 3 Kelvin, wherever they pointed
the telescope in the sky. After exhaustive efforts to find the source
of this emission, which even included scraping out bird droppings from
the inside of the antenna, Penzias and Wilson realised that the signal
must be the CMB predicted by Alpher and Hermann.
Indeed, this oldest light in the Universe has since proven to be a
great gift to cosmology, since it provides an early-childhood snapshot
of the Universe when it was just a tiny fraction of its current
age. After the discovery of the CMB, and especially when its spectrum
was measured to be precisely that of thermal radiation from a black body, most
cosmologists were persuaded that some version of the Big Bang
scenario must have taken place.

\section{Cosmic structure and dark matter}

Since the 1960s, most work in cosmology has been in the context of the
Big Bang model, and devoted in particular to understanding how
large-scale structure in the Universe, such as galaxies and clusters,
form in the context of the Big Bang model. This has led to many
surprises and, at times, considerable scepticism in the standard Big
Bang theory, which has had to evolve considerably to match
increasingly accurate observations. Most notably, it has proven
necessary to postulate the existence of a new form of matter that
interacts only gravitationally with normal baryonic matter and is
hence invisible: dark matter.

The earliest evidence suggesting the existence of dark matter came
instead from observations of the distributions of velocities of the
galaxies within galaxy clusters. In 1933, Zwicky noticed that outer
members of the Coma cluster are moving far too quickly to be merely
tracing the gravitational potential of the visible cluster mass. In
order to make the observed velocities consistent with the virial
theorem, one needed to postulate that the cluster also contained
additional matter, which could not be seen.  Observations by Babcock
in 1939 showed that this was also the case on the scale of the
individual galaxy Andromeda. More extensive observations by Rubin and
Ford during the 1960s and 1970s of the rotation curves of numerous
edge-on spiral galaxies observed clearly showed that the (near)
circular velocities of their constituent stars as a function of
distance from the centre of the galaxy remain approximately constant
out to the observable extent of the galactic disc. This is in stark
contrast to the expected fall-off in circular velocities expected from
the visible matter contained within the stars' orbits, and suggests
the presence of a large dark matter ``halo'' in which the visible galaxy
is embedded.

The main argument for dark matter comes, however, from the central
problem for the original Big Bang model to explain the formation of
galaxies. As early as the 1930s and 1940s, Lema\^itre, Tolman and
Lifshitz all independently showed that density perturbations in an
expanding universe grow quite slowly under their own self-gravity,
with their density contrast relative to the background increasing only
in proportion to the growth of the overall scale factor of the
universe. Assuming galaxies evolved from infinitesimal density
fluctuations in the very early Universe, they inferred that galaxies
could not have formed by the current epoch, which is clearly contrary
to observations, which in the 1950s and 1960s began to uncover the
large-scale distribution of galaxies and clusters in the Universe
through the work of Neyman, Abell and Zwicky.  Moreover, in the 1960s and
1970s, a number of theoretical cosmologists, including Harrison,
Zel'dovich, Peebles and Silk, had shown that to form these galaxies
and clusters, the small density perturbations from which they evolved
should leave imprints in the temperature distribution of the CMB
across the sky.  By 1980, however, the predicted amplitude of these
temperature fluctuations exceeded observational limits on anisotropies
in the CMB and clearly a fundamental change was needed in our
understanding of the formation of structure in the Universe.  This led the theoretical cosmologist Jim Peebles to suggest that the
Universe might be dominated by a hitherto unknown form of matter, now
called dark matter, that interacts only very weakly with normal
(baryonic) matter, of which we and everything we see around us is
comprised. This allows for dark matter fluctuations to form, into
which normal matter can later ``fall'', without imprinting excess
temperature variations in the CMB. 

\begin{figure}
\centerline{\includegraphics[width=5cm]{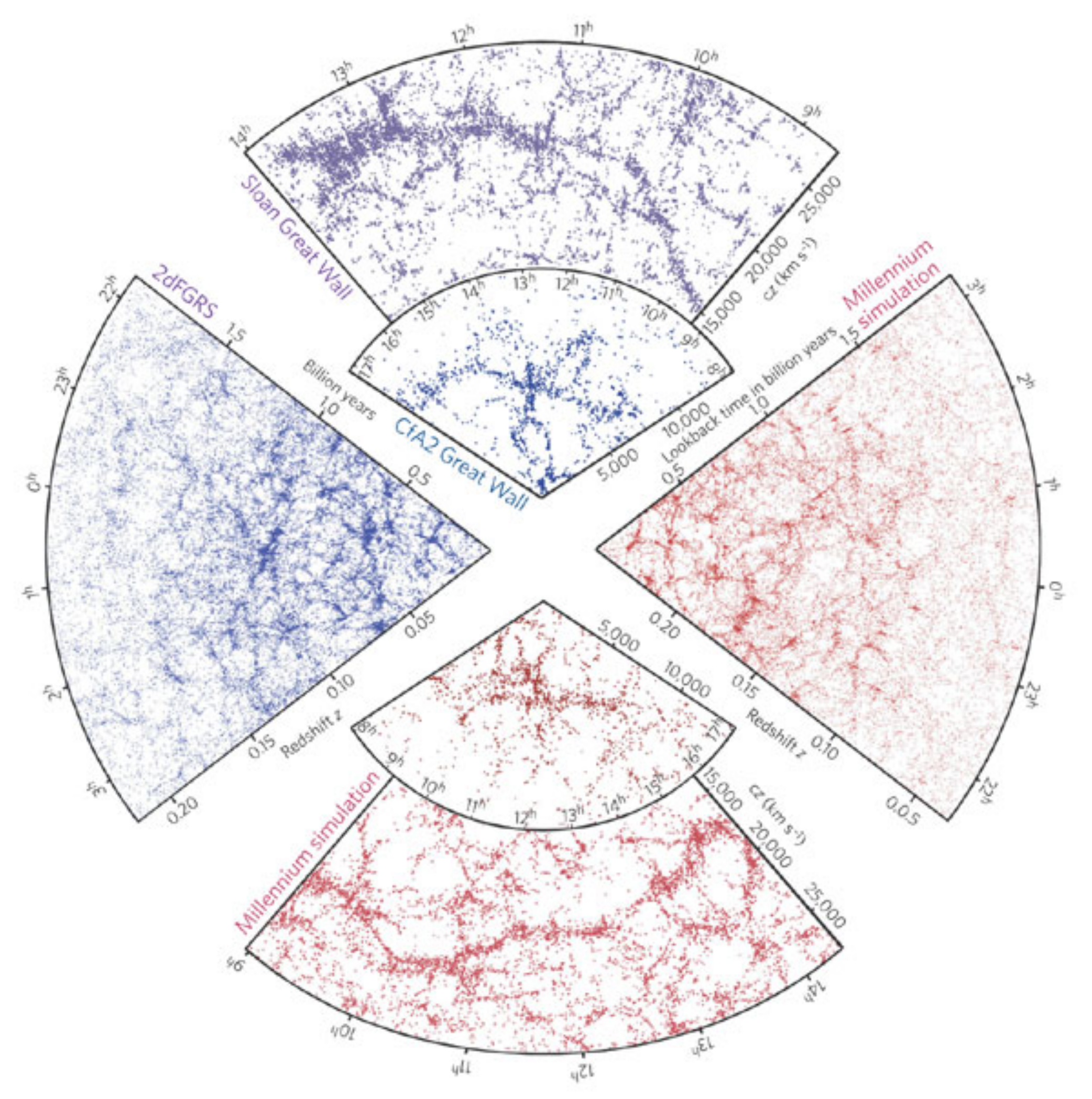}}
\caption{Top panel present the CfA2 ``Great Wall'',
    centered on the Coma cluster and to the left is one-half of the
    2dFGRS, which represents real measurements. Bottom and right panels
    present simulations constructed using ``the Millennium'' simulation and
    uses geometries and magnitude limits matching corresponding
    surveys. Credit: \cite{2006Natur.440.1137S}.}\label{f:DM}
\end{figure}

Detailed numerical simulations showed this model to be remarkably
successful in accounting for the large-scale distribution of structure
in the Universe (Figure~\ref{f:DM}).  On large scales, galaxies are collected into
clusters, clusters are part of superclusters, and superclusters are
arranged into large-scale sheets, filaments and voids. According to
cosmological ``N-body'' simulations (e.g. \citealt{1996ApJ...462..563N};
\citealt{2005Natur.435..629S}; \citealt{2007ApJ...657..262D}; \citealt{2008MNRAS.391.1685S};
\citealt{2008Natur.454..735D}), the formation of the observed large-scale
structure of luminous matter could only have taken place in the
presence of a substantial amount of dark matter. In addition, most of
the dark matter has to be both cold and non-dissipative to enable the
production of the observed structures. ``Cold'' in this context means
that it moves non-relativistically, and thus has a short
free-streaming length (for example, smaller than the size of a gas
cloud undergoing gravitational collapse). Being cold implies that the
dark matter can gravitationally aggregate on small scales and hence
seed the formation of galaxies, as mentioned above, while being
non-dissipative prevents it from cooling and collapsing with the
luminous matter and overproducing galactic discs.

Most importantly, when the Cosmic Background Explorer satellite
detected anisotropies in the CMB for the first time in 1992, they were
found to be at a level of about one part in 100,000 relative to the
2.73 Kelvin background, which is consistent with structure formation
in the cold dark matter (CDM) scenario.  During the following decade,
the CMB anisotropies were measured with increasing accuracy by a large
number of ground-based and balloon-borne experiments, by the Wilkinson Microwave Anisotropy
Probe (WMAP)
satellite over the period 2002-2009, and most recently by the Planck
satellite, which completed its observations in 2013. All these
observations remain consistent with the CDM model. Moreover, the power
in the CMB anisotropies measured on different angular scales is
consistent with the observed large-scale correlations in the
distribution of galaxies. This observation of Baryon Acoustic
Oscillations (\citealt{2006astro.ph..9591A}) strongly supports the idea that cosmic
structure formed from the passive gravitational collapse of primordial
density perturbations, the imprints of which we see in the CMB.

\enlargethispage{0.3cm}
Indeed, CMB observations can be combined with independent measurements
of $\Omega_\textrm{b,0}$ (the present-day baryonic matter density) and
$\Omega_\textrm{m,0}$ (the present-day total matter density) from Big
Bang nucleosynthesis and large-scale structure observations,
respectively, to provide very strong evidence for the existence of
dark matter.  Using the latest Planck results, one obtains posterior
mean values and 68 per cent credible intervals of $\Omega_\textrm{m,0} =
0.314 \pm 0.020$, $\Omega_\textrm{b,0}h^2= 0.02207 \pm 0.00033$ ,
$\Omega_\textrm{dm,0} h^2= 0.1196 \pm 0.0031$, indicating that dark
matter must be predominantly non-baryonic.\footnote{The present-day density parameter for the $i$th component is
  defined as $\Omega_\text{i,0}=8\pi G\rho_{i,0}/(3H_0^2)$, where
  $\rho_{i,0}$ is its physical density and $H_0\equiv 100 h$ is the
  present-day value of the Hubble parameter, which is estimated to be
  $H_0 = 67.3 \pm 1.2$ from combined cosmological probes (\citealt{2013arXiv1303.5076P}).}

\section{Gravitational lensing}

A striking visual representation of the presence of dark matter and a
beautiful illustration of general relativity is provided by
gravitational lensing.  General relativity predicts that light rays
are bent around massive bodies or, more generally, undergo deflections
when they traverse a region in which the gravitational field is
inhomogeneous.  In this manner, light from background objects can be
``lensed'' by massive objects in the foreground as the path of light
passing through the gravitational field of foreground object is
bent. The deflection of light is not just a relativistic effect, but
is also predicted by Newton's theory of gravitation, when one
considers light to be made up of a stream of particles. The Newtonian
approach does, however, predict only one-half of the deflection
predicted by general relativity. In the latter theory, a light beam
that just grazes the surface of the Sun suffers a deflection of 1.75
arcseconds, whereas Newton's theory predicts just 0.87 arcseconds.
Indeed, as I mentioned previously, the observational confirmation of
the larger value by Eddington in 1919 was a key factor in leading the
scientific community to accept Einstein's description of gravity in
terms of general relativity.

\begin{figure}
\centerline{\includegraphics[width=9cm]{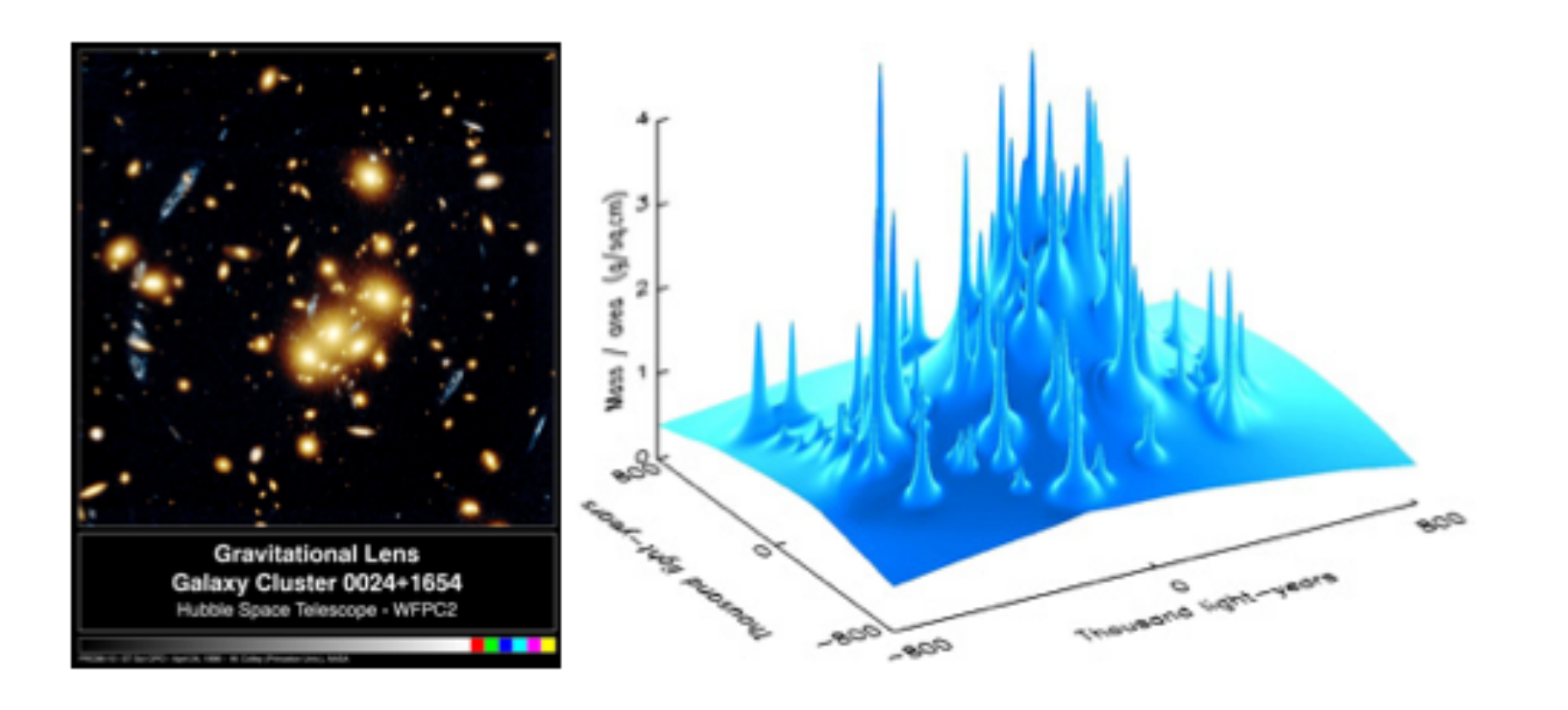}}
\caption{Presented in the left panel is an image of a gravitationally lensed cluster, while on the right is the mass map of the foreground cluster. Credit: Greg Kochanski, Ian
  Dell'Antonio, and Tony Tyson (Bell Labs, Lucent
  Technologies).}\label{f:gravlens}
\end{figure}

Very massive astronomical objects, which lie at large distances from
the Earth, can exert such a strong gravitational effect on light rays
that pass near them that a single background source can to observed as
multiples images. This phenomenon was predicted very early on in the
study of gravitational lensing, but was only observed for the first
time in 1979 \citep{1979Natur.279..381W}, since when gravitational
lensing has become a major area of research in astrophysics. Systems
which have been observed to contain multiple images are many tens in
number. Of greatest interest to astrophysics is that the analysis of
the distribution and shape of these multiple images can be used to
derive an accurate estimate of the mass distribution in the lensing
object. In particular, when light from distant galaxies is lensed
by a cluster, one can see evidence of significant gravitational
lensing, far more than can result from the observed distribution of
luminous matter in the foreground cluster, thereby implying the
presence of dark matter (e.g. \citealt{1998ApJ...498L.107T};
\citealt{2007Natur.445..286M}).  Figure~\ref{f:gravlens} shows an
example of such a cluster.

In most cases, however, the effects of gravitational lensing are far
more subtle. Typically, the size and shape of background objects are
only very slightly changed.  The nature of this change is very
difficult to determine for individual objects, since one does not know
a priori its unlensed intensity distribution. One therefore has to
average the effect of a large number of background objects to obtain a
statistical measure of this weak-lensing signal. Indeed, such
observations can be used to investigate the nature of cosmic
structures, provided one has a large collection of lensed background
objects to analyse.

\section{SNe, universal acceleration and dark energy}

\begin{figure}
\centerline{\includegraphics[width=8cm]{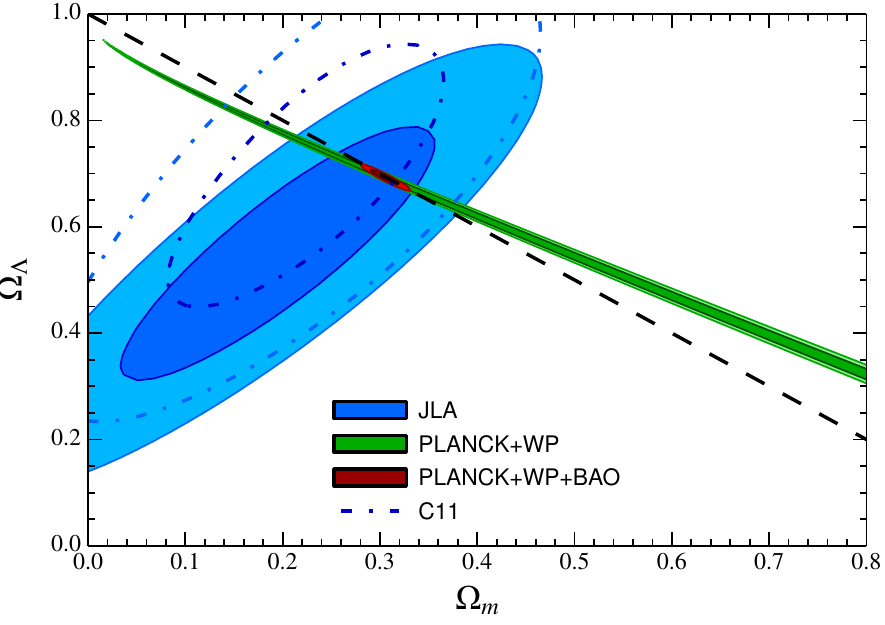}}
\caption{68\% and 95\% confidence contours for the present-day
  cosmological density parameters $\Omega_\textrm{m,0}$ and
  $\Omega_{\Lambda,0}$. Labels for the various data-sets correspond to
  the \cite{2014arXiv1401.4064B} SNIa compilation (JLA), the
  \citet{2011ApJS..192....1C} SNIa compilation (C11), the combination
  of Planck temperature and WMAP polarization
  measurements of the CMB fluctuation (PLANCK+WP), and a combination
  of measurements of the baryon acoustic oscillations scale (BAO). The
  black dashed line corresponds to a flat universe. Credit:
  \cite{2014arXiv1401.4064B}. }\label{f:Ia}
\end{figure}

In the last two decades, there has been an unexpected twist in the story of
cosmology. In 1998, measurements of the redshift--magnitude relation
for SNIa, which can be used as ``standard candles'' in cosmology,
indicated that, when the Universe was around half of its present age,
its expansion underwent a transition from a decelerating phase into an
accelerating one, which continues to the current epoch.  This came as
a complete surprise, as it was thought that the expansion should
decelerate as a result of the attractive gravitational force between
all objects slowing down the expansion.  To explain an accelerating
universe, one has to posit some additional component of the universe,
known generically as ``dark energy'', which has a large negative
pressure and thus leads to a gravitational repulsion. Amazingly, the
simplest form for such a component is provided precisely by the
additional cosmological constant term (or $\Lambda$-term) that
Einstein included in his equations of general relativity when trying
to build a static universe model, but then rejected as his ``biggest
blunder'' when he learned that the Universe is expanding. The
resulting ``$\Lambda$CDM'' scenario is our best current cosmological
model, which describes all existing observations.  Indeed, results
from WMAP, Planck and other CMB observations, combined with galaxy
surveys of large-scale structure are all consistent with a
$\Lambda$CDM model, known as the ``concordance cosmology'', in which the
total mass/energy density budget of the Universe at the present time
is comprised of approximately: 73\% dark energy, 23\% dark matter and
4\% ordinary matter (see Figure \ref{f:Ia}), and in which structure
forms from the passive gravitational evolution of scale-invariant
perturbations generated (somehow) in the very early Universe.
\section{Inflation, uncertainty and the future}

There remain numerous open questions in cosmology. Indeed, many
cosmologists view the current standard model of cosmology with
considerable scepticism. In addition to the unknown physical nature of
both dark matter and dark energy, which supposedly dominate our
Universe, our understanding of fundamental physics is only sufficient
to project back to around one ten-billionth of a second after the Big
Bang, at which epoch the typical densities and energies of particles
are at the limit of what can be reached in the latest particle physics
experiments, such as the Large Hadron Collider (e.g. \citet{2009arXiv0901.0512T, 2011arXiv1101.0593L}). At earlier times, the
physics of the Big Bang is subject to considerable speculation and
doubt. The most popular current model, known as the theory of
inflation, proposes that, almost immediately following the Big Bang,
the Universe underwent a short period of exponential expansion,
growing in size by a factor of $10^{27}$ in just $10^{-33}$ seconds,
during which microscopic quantum fluctuations in the matter fields at
the time were stretched to macroscopic scales to generate the seeds of
structure formation. Indeed, this mechanism for the origin for all the
structure in the Universe was proposed by Guth, Linde and Starobinsky,
amongst others, in the early 1980s. The inflationary model also solves
a number of other problems, such as explaining why the Universe is so
homogeneous and isotropic on the largest scales. Very recent support
for the inflationary paradigm has potentially been provided by the
BICEP2 experiment (\citealt{2014PhRvL.112x1101A, 2014arXiv1403.4302B}), which claims to have observed
polarised emission from CMB anisotropies that is consistent with the
presence of primordial gravitational waves, which are also predicted
to be produced (almost exclusively) by inflation. There are currently,
however, some concerns regarding the interpretation of the BICEP2
results, and further experimental verification is required.

Although an attractive proposal, inflation does, however, have some
theoretical problems of its own. In particular, determining the
initial conditions for inflation is both conceptually and technically
very demanding, and it may be the case that producing a period of
inflation that is consistent with observations requires an
unacceptable level of fine-tuning in the theory, but this is far from
certain. Thus, cosmology now finds itself again in a period rich in
alternative models, the development of which is driven by scepticism
in our existing description of the Universe. Only time will tell
whether (another) revolution in our thinking is required, but, based
on the experience of the last one hundred years, it seems very likely.


\chapter{Statistical methods}
\label{ch:ST}

\epigraph{`There are three kinds of lies: lies, damned lies, and statistics.'}{\textit{Benjamin Disraeli}}

There are two ways to define probability. One of them is Frequentist,
which postulates that probability is ``the ratio of the times the
event occurs in a test series to the total number of trials in the
series'' \citep{1995hep.ph...12295D}, or the ``frequencies of outcomes
in random experiments'' \citep{2003itil.book.....M}, and the other is
Bayesian, which postulates that probability is ``a measure of the
degree of belief that an event will occur''
\citep{1995hep.ph...12295D}. In my research, I choose to follow the
Bayesian interpretation of probability, since it provides the only
self-consistent extension of Boolean algebra to propositions that are
not simply true or false, but are associated with a degree of belief
(defined to lie between 0 and 1) (\citealt{1946AmJPh..14....1C}).

On a more practical note, the recent development of efficient sampling
algorithms makes the implementation of Bayesian methodology more
straightforward and reliable than its Frequentist counterpart. The
central reason for this lies in what Bayesians and Frequentists
consider to be the most important properties of probability
distributions. In contrasting the two statistical schools, much
attention is usually paid to the issue of priors. It is indeed true
that Bayesians consider the posterior probability, namely the product
of the likelihood and the prior, as the primary distribution for
inference. By contrast, the Frequentists tend to eschew the notion of
priors and concentrate on the likelihood alone. This difference is,
however, often overstated and leads practically to very little
difference in the final inferences. A far more profound practical
difference between the schools is that Bayesians consider probability
mass as most important, whereas Frequentists consider the point-value
of the probability as primary. Thus, given a probability distribution
(either the posterior or the likelihood), a Bayesian integrates under
the distribution to identify regions of the parameter space containing
the largest integrated probability, whereas the Frequentist aims to
identify the point(s) in the parameter space with the largest
probability value, at least when adopting the most commonly-used
maximum-likelihood or maximum a posteriori estimators. Using a
thermodynamic analogy, Bayesians focus on heat whereas Frequentists
concentrate on temperature. This difference between the schools is far
more fundamental than the use of priors, and has immediate practical
consequences.  By obtaining Monte Carlo samples from the distribution,
the construction of marginal probability distributions, as preferred
by the Bayesian, is far more easily and reliably achieved than the
construction of profile likelihoods, as preferred by the
Frequentist. This provides another, very practical reason for choosing
the Bayesian approach.

\section{Bayesian statistics}
In general, the probability $\Pr(A | B)$ is the degree of
belief, given $B$, of the truth of $A$. \emph{Bayes' theorem} can be used to change the order of the conditioning,
\begin{equation} \Pr(A|B) =
\frac{\Pr(B|A)\Pr(A)}{\Pr(B)}.
\end{equation}
In many cases, one wants to use a set of observations or data to infer values of parameters within a given model.  Given a model or hypotheses $H$
with a set of $N$ free parameters $\mathbf{\Theta} = \lbrac\Theta_i\rbrac$, together with a data-set $\mathbf{D}$,  Bayes' theorem implies that one can write

\begin{equation} \label{eq:bayes} \Pr( \mathbf{ \Theta} | \mathbf{D},H) = \frac{\Pr(\mathbf{D}
|\mathbf{\Theta},H)\Pr(\mathbf{\Theta}|H)}
{\Pr(\mathbf{D}|H)}, 
\end{equation}
where $\Pr(\mathbf{\Theta}|\mathbf{D}, H)$ is the posterior
probability (density) of the parameters $\mbTh$, $\pi(\mathbf{\Theta}) \equiv \Pr(\mathbf{\Theta}|H) $ is
the prior probability (density), $\mathcal{L}(\mathbf{\Theta}) \equiv \Pr(\mathbf{D}|\mathbf{\Theta},H )$ is the probability (density) of the data $\mathbf{D}$, for assumed
parameter values $\mbTh$ (called the likelihood when considered as a function of $\mbTh$), while $\Pr(\mathbf{D}|H)$ is the
\emph{Bayesian evidence}, 
\begin{eqnarray}
\mathcal{Z} \equiv \Pr(\mathbf{D}|H)  = \int \Pr(\mathbf{D}|\mbTh, H) \Pr(\mathbf{\Theta}|H)\mathrm{d}^N\mathbf{\Theta}  
= \int{\mathcal{L}(\mathbf{\Theta})\pi(\mathbf{\Theta})}\mathrm{d}^N\mathbf{\Theta}.
\label{eq:3}
\end{eqnarray}
Note that this makes the posterior in Eq.~\ref{eq:bayes} normalised to unity over the space of parameters.

In parameter estimation, the complete Bayesian inference is embodied
in the posterior distribution of the parameter values. This may be
used to obtain joint constraints on all parameters
simultaneously or constraints on individual parameters by the process
of integrating out (or marginalising over) all the other parameters.
In practice, the posterior distribution is explored by drawing samples
from it using (most often) standard Markov chain Monte Carlo (MCMC) sampling techniques. Once
they have reached equilibrium, such methods produce a set of samples
whose density is proportional to the posterior. In the entire process,
one need not calculate the evidence to normalise the posterior, since
the evidence does not depend on the parameters.

By contrast, the evidence is the key quantity of interest for the
problem of model selection. Since the evidence may be considered as
the average of the likelihood over the prior, it provides a natural
means of applying Occam's razor. If a model has a highly-peaked
likelihood, but there exist large regions of the parameter space that
are disfavoured, since the likelihood is low there, then the evidence
of the model will be small. Large evidence values occur for models for which a large fraction of the allowed parameter space is
likely. Thus, one can decide which of two models $H_{0}$ and $H_{1}$ 
is preferred by the data $\mathbf{D}$ by calculating the ratio of posterior probabilities
\begin{equation}
\frac{\Pr(H_{1}|\mathbf{D})}{\Pr(H_{0}|\mathbf{D})}
  = \frac{\Pr(\mathbf{D}|H_{1})\Pr(H_{1})}{\Pr(\mathbf{D}| H_{0})\Pr(H_{0})}
  = \frac{\mathcal{Z}_1}{\mathcal{Z}_0} \frac{\Pr(H_{1})}{\Pr(H_{0})},
\label{eq:3.1}
\end{equation}
where $\Pr(H_{1})/\Pr(H_{0})$ is the {\em a priori} probability ratio for
the two models.  In most problems this prior ratio is set to unity,
but there are cases, most notably in object detection, where one must
set this prior ratio quite carefully to offset the ``look elsewhere''
effect. One uses Jeffreys' scale given in Table \ref{tab:Jeffreys} to
interpret the ratio in Eq.~\ref{eq:3.1}.\footnote{Throughout this work,
  $\log x$ (i.e. without any subscript) denotes the natural logarithm
  of $x$, which is also commonly denoted by $\ln x$.} In practice, the evidence may also be evaluated using MCMC
sampling methods, although the standard technique of thermodynamic
integration \citep{Ruanaidh} typically requires about an order of
magnitude more samples than needed for parameter estimation. 

\subsection[Check for inconsistency between data-sets: the $\mathcal{R}$-test ]{Check for inconsistency between data-sets: \\the $\mathcal{R}$-test }\label{sec:method:Rtest}

A useful application of Bayesian model selection is in determining
whether different data-sets are mutually consistent. In principle, one
should always check that this is the case before performing a joint
analysis using them, although in practice such a check is not often
undertaken. Indeed, the vast majority of analyses in SN cosmology do
not test whether different surveys are mutually consistent before
combining them in a joint analysis to determine cosmological
parameters.

Adopting a Bayesian model selection approach, we denote by $H_0$ the
(null) hypothesis that the data-sets are mutually consistent. In this
case, one would expect each data-set to prefer broadly the same
region(s) of the model parameter space. Under the (alternative)
hypothesis $H_{1}$ that the data-sets are mutually inconsistent, one or more of them favour a different region (or
regions) of the parameter space. Simply performing a joint in this
case could lead to erroneous results (see, for example, Appendix A in
\citealt{2008JHEP...10..064F} for a demonstration).

In order to determine which one of these hypotheses is favoured by the
data, one can perform Bayesian model selection between $H_{0}$ and
$H_{1}$. Using Eq.~\ref{eq:3.1} and assuming that hypotheses $H_0$ and
$H_1$ are equally likely {\em a priori}, this can be achieved by calculating
\begin{equation}
 \mathcal{R} = \frac{\Pr(\mathbf{D}|H_{0})}{\Pr(\mathbf{D}|H_{1})}
  = \frac{\Pr(\mathbf{D}|H_{0})}{\prod_{i}\Pr(D_i|H_{1})}.
\label{eq:R-test}
\end{equation}
Here the numerator represents the standard joint analysis of all the
data-sets $\mathbf{D} = \{D_1, D_2, \cdots, D_n\}$, whereas the
denominator in the final expression represents the case in which each
data-set is analysed separately. It is worth noting, however, that the
second equality in Eq.~\ref{eq:R-test} is valid only when one allows
for potential inconsistencies in the preferred values of the {\em
  full} set of model parameters. Nevertheless, there are often
situations where one is interested only in potential inconsistencies
in the preferred values of some {\em subset} of the model
parameters. In such cases, one must use only the first equality in
Eq.~\ref{eq:R-test} and calculate the denominator
$\Pr(\mathbf{D}|H_{1})$ by performing a joint analysis of $\mathbf{D}$
in which each data-set is assigned its own ``private copy'' of only
those parameters in the subset of interest. It is clear from
Eq.~\ref{eq:R-test} that an $\mathcal{R}$-value larger than unity (or,
equivalently, a positive $\log\mathcal{R}$-value) indicates that the
(null) hypothesis $H_0$, that all the data-sets are mutually
consistent, is favoured. Otherwise, the (alternative) hypothesis $H_1$
is preferred, indicating some inconsistency between the data-sets.
One uses Jeffreys' scale given in Table \ref{tab:Jeffreys} to
interpret the value of $\mathcal{R}$.

\begin{table}
\begin{center}
\begin{tabular}{c|c|c|c}
\hline\hline
$\log(\text{odds})$ & odds & $\Pr(H_1 | \mathbf{D})$ & Interpretation \\ 
\hline
$<1.0$ & $\lesssim 3:1$ & $\lesssim 0.75$ & Inconclusive \\
$1.0$ & $\simeq 3:1$ &  $\simeq 0.75$ & Weak evidence \\
$2.5$ & $\simeq 12:1$ & $\simeq 0.92$ & Moderate evidence \\
$5.0$ & $\simeq 150:1$ & $ \simeq 0.993$ & Strong evidence \\ \hline\hline
\end{tabular}
\end{center}
\caption{Jeffreys' scale for the interpretation of Bayes factors and
  model probabilities. The posterior model probabilities for the
  preferred model are calculated by assuming only two competing
  hypotheses.}
\label{tab:Jeffreys}
\end{table}

\subsection[Analysis of potentially inconsistent data-sets:\\ hyper-parameters]{Analysis of potentially inconsistent data-sets: hyper-parameters}
\label{sec:method:hypPar}

One method for accommodating potentially inconsistent data-sets in a
joint analysis is to introduce hyper-parameters that effectively
assign a weight to each data-set that is determined directly by it own
statistical properties \citep{Hobson02}.  The space of hyper-parameter
weights is explored simultaneously with the space of original model
parameters to obtain a joint posterior distribution. By marginalising
over the original model parameters, one obtains the posterior
distribution of the hyper-parameter weights, which may be used to
determine if any inconsistencies exist between different
data-sets. Conversely, one can instead marginalise over the
hyper-parameters to recover the posterior distribution as a function
only of the original model parameters. Moreover, calculation of the
Bayesian evidence for the data, with and without the introduction of
hyper-parameters (which we denote by the hypotheses $H_1$ and $H_0$,
respectively), allows us to perform model comparison to determine
whether the data warrant the introduction of weights into the
analysis.

When analysing multiple data-sets jointly, inferred values of
hyper-parame\-ters which depart significantly from unity indicate the
presence of some inconsistency. Even in such cases, however, the
hyper-parameter approach allows for a robust joint analysis of the
data-sets. In particular, marginalisation over the hyper-parameters
allows for the resulting posterior distribution of the original model
parameters to broaden or even exhibit multi-modality resulting from
the preference of different data-sets for different regions of the
model parameter space (see \citealt{Hobson02} for more details).

\section{Nested sampling and the \textsc{MultiNest} algorithm}

It is computationally very demanding to evaluate the multidimensional
integral in Eq.~\ref{eq:3}. Nested sampling is a Monte Carlo
approach, introduced by \citet{skilling04}, that is designed to
calculate the evidence efficiently, and which also produces posterior
inferences as a by-product.  The method has been extended by
\citet{feroz08} and \citet{multinest}, who introduced the \textsc
      {MultiNest} algorithm, which is able to accommodate posteriors
      with multiple modes and/or large (curving) degeneracies. In the
      following description of the algorithm, I will closely follow
      the discussion given in these two papers.

\begin{figure}
\begin{center}
\subfigure{\includegraphics[width=0.4\columnwidth]{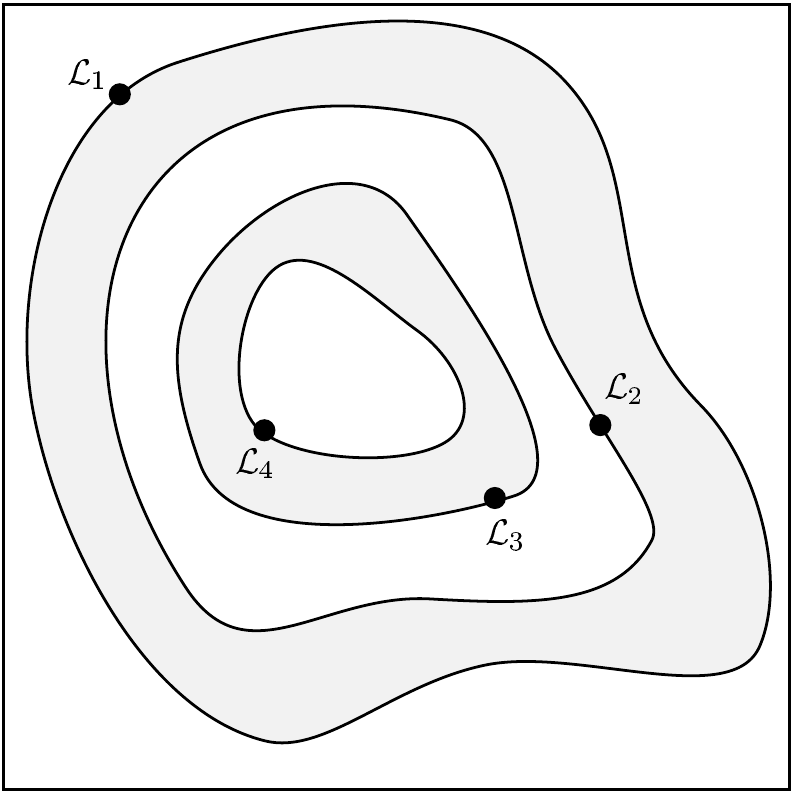}}\hspace{0.3cm}
\subfigure{\includegraphics[width=0.4\columnwidth]{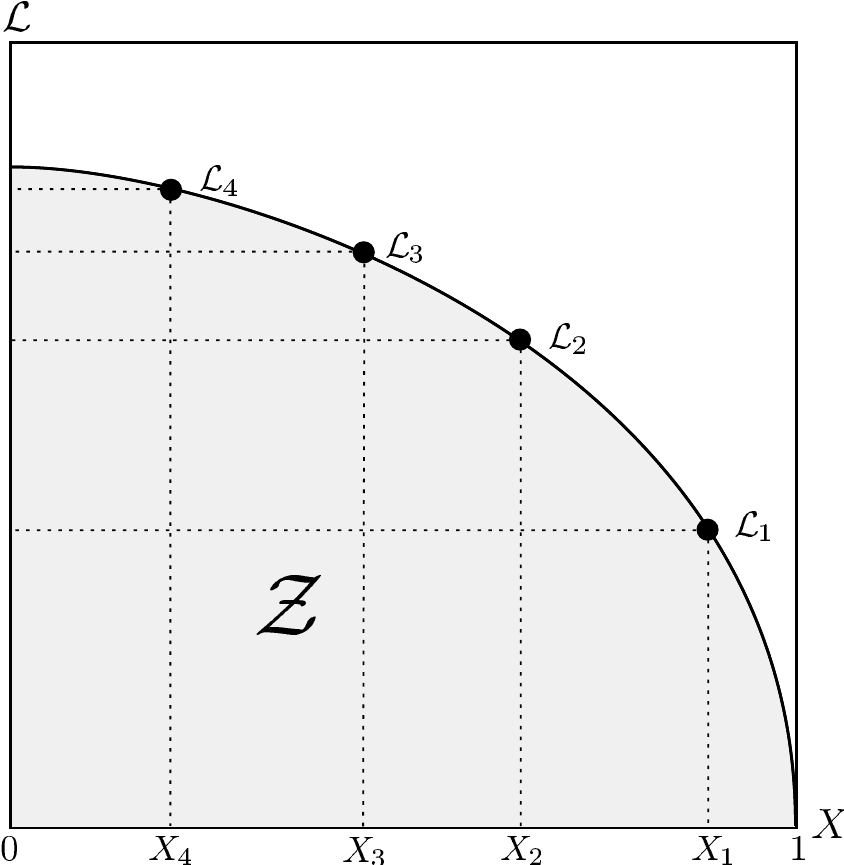}}
\caption{Left panel: An example of a two-dimensional posterior distribution. Right panel: The function $\mathcal{L}(X)$. The prior volumes
  $X_{i}$ are associated with each likelihood
  $\mathcal{L}_{i}$. Credit: \citet{feroz08}.}
\label{fig:NS}
\end{center}
\end{figure}

The key innovation in nested sampling is that the multi-dimensional
evidence integral is transformed into a one-dimensional integral.  To
perform this transformation one first defines the prior volume $X$ via
the differential relationship $\mathrm{d}X = \pi(\mathbf{\Theta})\mathrm{d}^N
\mathbf{\Theta}$. Thus one can write $X$ as
\begin{equation}
X(\lambda) = \int_{\mathcal{L}\left(\mathbf{\Theta}\right) > \lambda}
\pi(\mathbf{\Theta}) \mathrm{d}^N\mathbf{\Theta},
\label{eq:Xdef}
\end{equation}
where the domain of integration comprises the region(s) of the
parameter space that lies within the iso-likelihood contour
$\mathcal{L}(\mathbf{\Theta}) = \lambda$. Thus, one can write the 
evidence integral, Eq.~\ref{eq:3}, as:
\begin{equation}
\mathcal{Z}=\int_{0}^{1}{\mathcal{L}(X)}\mathrm{d}X,
\label{eq:nested}
\end{equation}
where $\mathcal{L}(X)$, which is the inverse of Eq.~\ref{eq:Xdef},
is a monotonically decreasing function of $X$. Thus, the evidence can
evaluated using one-dimensional numerical quadrature integration
methods, provided one can evaluate
the likelihoods $\mathcal{L}_{i}=\mathcal{L}(X_{i})$, where $X_{i}$ is
a sequence of decreasing values,
\begin{equation}
0<X_{M}<\cdots <X_{2}<X_{1}< X_{0}=1,
\end{equation}
as shown schematically in Figure~\ref{fig:NS}. In particular, one can
write the evidence as the weighted sum
\begin{equation}
\mathcal{Z}={\textstyle {\displaystyle \sum_{i=1}^{M}}\mathcal{L}_{i}w_{i}},
\label{eq:NS_sum}
\end{equation}
where the weights $w_{i}$ for the simple trapezium rule are given by
\begin{equation}
w_{i}=\frac{1}{2}(X_{i-1}-X_{i+1}).
\end{equation}
 Figure~\ref{fig:NS} gives an illustration
of this process for a two-dimensional posterior.

To perform the sum in Eq.~\ref{eq:NS_sum} one begins at iteration
$i=0$ by drawing $N$ samples from the prior distribution
$\pi(\mathbf{\Theta})$. These samples constitute the set of so-called
``live'' or ``active'' points in the nested sampling process.  At this
initial stage the volume of the prior is unity, namely $X_{0} =
1$. One then calculates the likelihood of each of the active points
and determines which of them has the lowest value (which I denote by
$\mathcal{L}_{0}$); this point is then discarded from the active
set. In order to maintain the number of active points, one then
replaces the discarded point with a new point that is again drawn from
the prior, but is now required to lie within the iso-likelihood
contour $\mathcal{L}=\mathcal{L}_{0}$. The prior volume contained
within this contour is not known precisely, since it depends on the
points in the original active set. Nonetheless, one may show that the
ratio $t = X_{1}/ X_{0}$ of the new and original prior volumes is
distributed as $\Pr(t) = Nt^{N-1}$. As the process continues, the
iterative discarding and replacement of the point with the lowest
likelihood results in the iso-likelihood contour shrinking and the
live points being constrained to ever smaller prior volumes and higher
likelihood regions. One may show that after $i$ iterations the prior
volume is $X_{i} \approx \exp(-i/N)$. The process is usually concluded
by imposing some criterion on the accuracy to which the evidence has
been calculated.

Although nested sampling is designed primarily to calculate the
evidence, a happy consequence of the method is that the final set of
active points, together with the ``historic'' set of discarded points
produced during the iterations, can be used to obtain posterior
inferences on the parameters. Indeed, one may show that
posterior-weighted samples are obtained by assigning each point the weight
\begin{equation}
p_{i}=\frac{\mathcal{L}_{i}w_{i}}{\mathcal{Z}}.
\label{eq:12}
\end{equation}
Once can then use these samples in the same way as samples obtained
from a standard MCMC method to calculate parameter means, standard
deviations, and covariances, or even to construct their marginalised
posterior distributions.

\begin{figure}
\begin{center}
\subfigure{\includegraphics[width=0.45\columnwidth,height=0.35\columnwidth]{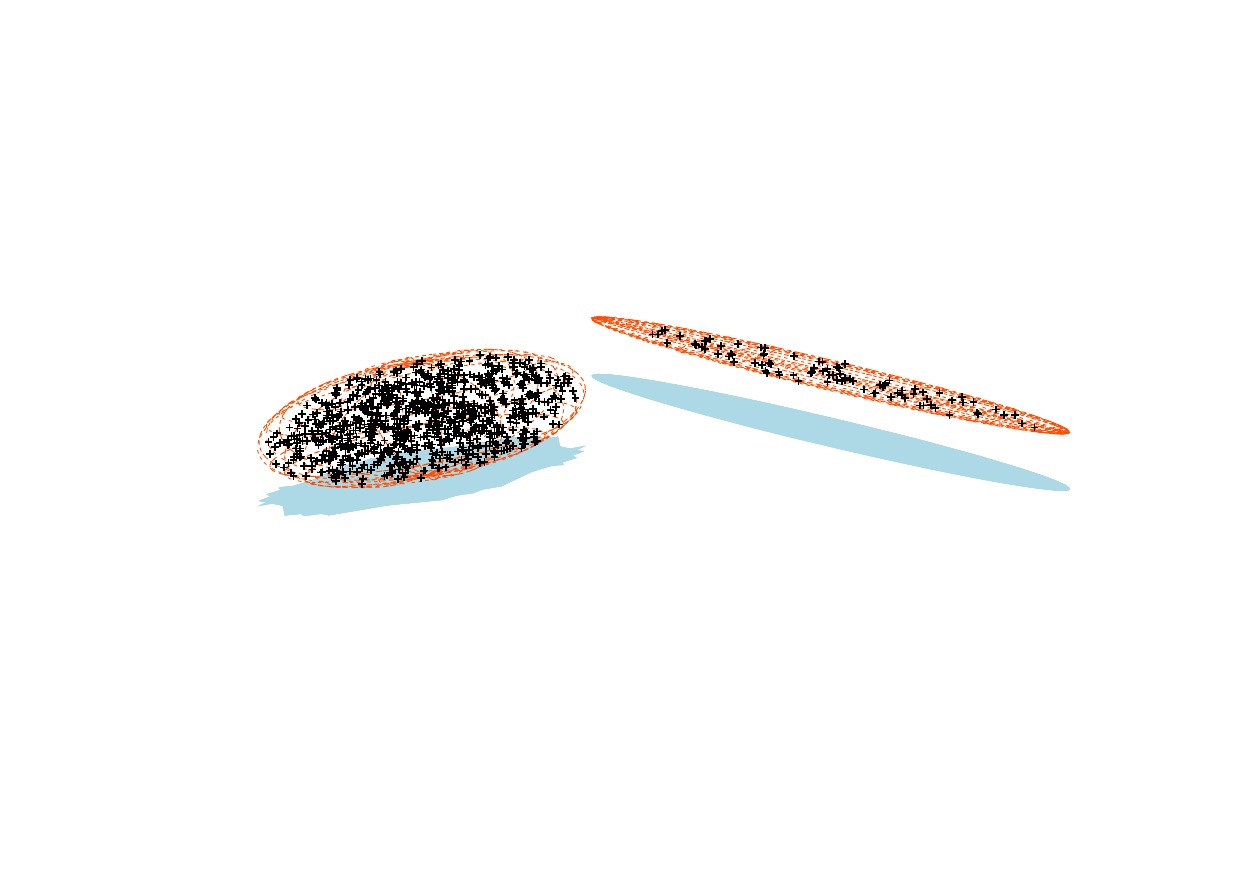}}
\setcounter{subfigure}{1}
\subfigure{\includegraphics[width=0.45\columnwidth,height=0.35\columnwidth]{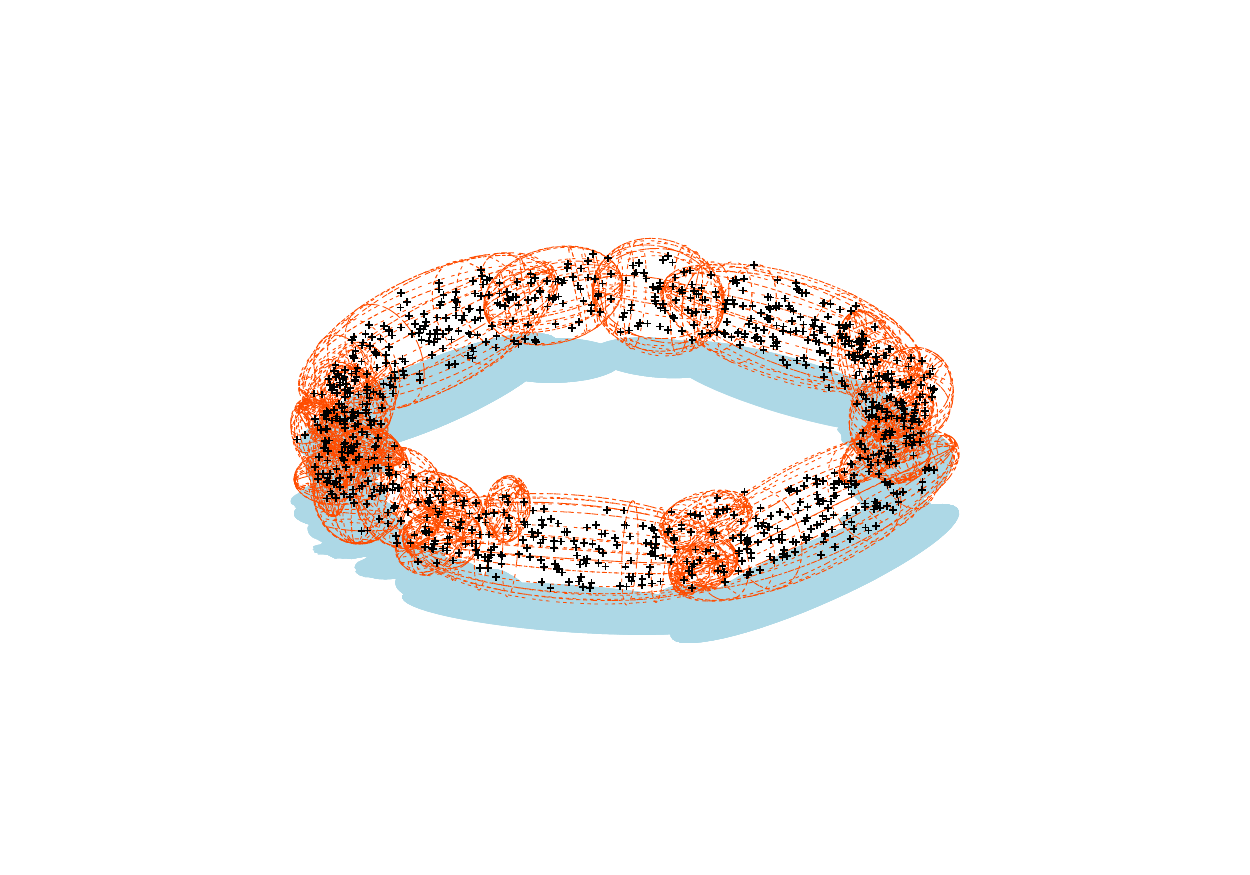}}
\setcounter{subfigure}{2}
\caption{The result of applying the ellipsoidal decomposition
  algorithm in {\sc MultiNest} to a set of 1000 points sampled from:
  two non-intersecting ellipsoids (left panel); and a torus (right panel). Credit: \citet{feroz08}.}
\label{fig:dino}
\end{center}
\end{figure}

Nested sampling has been described as only a ``meta-algorithm'', since
it leaves unanswered the key question of how, at each iteration $i$,
to draw the required replacement point from the prior
within the iso-likelihood contour $\mathcal{L}=\mathcal{L}_i$. The
{\sc MultiNest} algorithm \citep{feroz08,multinest} performs this task
using rejection sampling from a multi-ellipsoidal bound tailored to
the current active point set.  At each iteration, one performs an
expectation-maximisation process to determine the set of (possibly
overlapping) ellipsoids that encloses the set of $N$ live points in
the minimum volume, subject to the lower limit of the expected prior
volume $X_{i} = \exp(-i/N)$. The new replacement point is then drawn
uniformly from the region enclosed by these ellipsoids.

This ellipsoidal decomposition is very flexible and is able to
accommodate both multimodal structure and degeneracy lines in the
target posterior distribution. In particular, for posteriors that
contain well-defined and well-separated modes, the ellipsoidal
decomposition allows one to identify the modes and evolve the nested
sampling process in each mode separately. In essence, modes are
identified as separate entities if there exist ellipsoid(s) set that
do not overlap with any others.  An illustration of the ellipsoidal
decomposition performed by {\sc MultiNest} is given in
Figure~\ref{fig:dino}. More recently, further developments of {\sc MultiNest} has been made to enable even more accurate evaluation of the evidence \cite{2013arXiv1306.2144F}.

\section{Machine-learning and neural networks}
\label{NN}

In addition to performing parameter estimation and model selection
using Bayes\-ian methods, in this thesis I also use
machine-learning techniques, particularly for the photometric
classification of SNe into their respective types (see Section~\ref{ch:paper3} and Papers III and V). From the numerous approaches to
machine-learning, I use neural networks (NNs).  In particular, I
employ the {\sc SkyNet} package, which is a generic NN training
algorithm (\citealt{2013AAS...22143101G, 2014MNRAS.441.1741G}).

An artificial NN is a mathematical model loosely based
on the structure of the brain. A NN consists of groups of nodes that
are connected to one another by directional links that are assigned
particular weights. Using these links, each node processes information
it receives and then passes the result to other nodes.  A great short
introduction to NNs is given by \cite{2003itil.book.....M}.

\begin{figure}
\begin{center}
\includegraphics[width=0.4\columnwidth]{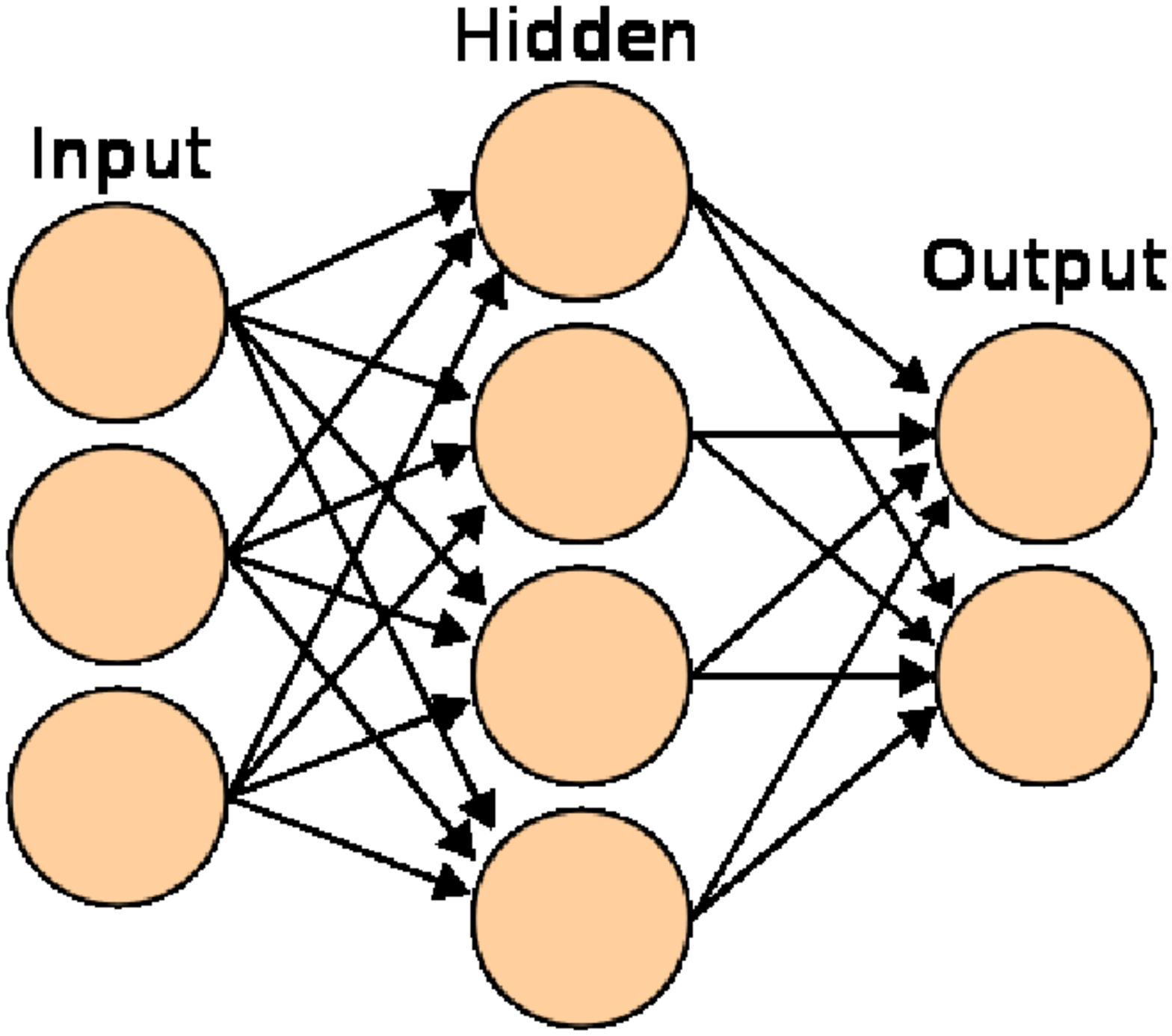}
\caption{A 3-layer NN with 3 inputs, 4 hidden nodes, and 2 outputs. Image courtesy of Wikimedia Commons.}
\label{fig:ANN}
\end{center}
\end{figure}

In this thesis, I focus entirely on the simplest form of NNs, which are
known as feed-forward networks. In particular, I will consider only
3-layer networks,  which consist of a layer of input nodes, connected
to a ``hidden'' layer, which itself is then connected to an output layer
(see Figure~\ref{fig:ANN}).
Each node (or perceptron) in the network
maps an input vector ${\bf x} \in \Re^n$ to a scalar output $f({\bf
  x};{\bf w},\theta)$ given by
\begin{equation}
\label{eq:perceptron}
f({\bf x};{\bf w},\theta) = \theta + \sum_{i=1}^n {w_{i} x_{i}},
\end{equation}
where $\{w_{i}\}$ and $\theta$ are, respectively,
the ``weights'' and ``bias'' of the perceptron. Thus, 
for a 3-layer NN, the outputs of the nodes in the hidden and output
layers are \citep{2012MNRAS.421..169G}
\begin{eqnarray}
\label{eq:hiddenformula}
\textrm{hidden layer:} \; h_{j} = g^{(1)}(f^{(1)}_{j}); \; f^{(1)}_{j} = \theta^{(1)}_{j} + \sum_{l} {w^{(1)}_{jl}x_{l}} ,\\
\label{eq:outputformula}
\textrm{output layer:} \; y_{i} = g^{(2)}(f^{(2)}_{i}); \; f^{(2)}_{i} = \theta^{(2)}_{i} + \sum_{j} {w^{(2)}_{ij}h_{j}} ,
\end{eqnarray}
where the indices $l$, $j$ and $i$ indicate, respectively, input,
hidden and output nodes, and $g^{(1)}$ and $g^{(2)}$ are called
activation functions. It may be shown that, for the NN to operate
correctly, these functions must obey certain requirements, namely that they are
smooth, monotonic and bounded. I follow \cite{2012MNRAS.421..169G} and use
$g^{(1)}(x)=\tanh(x)$ and $g^{(2)}(x)=x$.

When one ``trains'' a NN, one determines the values of the weights and
biasses that optimise the accuracy of the mapping from the input nodes
to the output nodes. The existence of a suitable mapping is guaranteed
by the ``universal approximation theorem''~\citep{UnivApprox}. As one
increases the number of hidden nodes, the accuracy of the mapping
typically increases, but so does the possibility of overfitting the
training data. Nonetheless, the ability of the network to learn
complicated mappings may be compromised if the number of hidden nodes
is too low.  There is therefore a balance between these competing
factors and the optimal number of hidden nodes is best determined by
comparing the fitting error and correlations of NNs with different
numbers of such nodes trained on the same data.

In training a NN, we wish to find the optimal set of network weights
and biasses (which together we call the network parameters ${\bf a}$)
that maximise the accuracy of the predicted outputs. However, one must
be careful to avoid overfitting to the training data at the expense of
making predictions for input values the network has not been trained
on. The general procedure for training a NN is to present it with a
set of input and outputs (or targets) $ \mathcal{D} = \{{\bf x}^{(i)},
{\bf t}^{(i)}\}$. Typically around 75\% of the set should be used for
actual NN training, while the remainder is used as a validation set of
data to determine convergence and avoid overfitting.

To train the network, one optimises the probability of reproducing the
known training data outputs with respect to the network
parameters. For problems of regression (fitting the model to a
function), this yields a log-likelihood for ${\bf a}$ in the form of a
standard $\chi^2$ misfit function, given by
\begin{equation}
\mathcal{L}({\bf a}) = -\sum_{j=1}^{n_{\rm out}} \log\sigma_j-\frac{1}{2}
 \sum_{i=1}^{n_{\rm t}}\sum_{j=1}^{n_{\rm out}} \left[
\frac{t_{j}^{(i)}-y_j({\bf x}^{(i)}; {\bf a})}{\sigma_j}
\right]^2,
\end{equation}
where $n_{\rm t}$ is the number of training data, $n_{\rm out}$ is the number
of network outputs, and $y_j({\bf x}^{(i)}; {\bf a})$ are the NN's
predicted outputs for the input vector ${\bf x}^{(i)}$ and network
parameters ${\bf a}$. The values $\sigma_j$ are hyper-parameters of the
NN model that describe the standard deviation of each of the outputs.

For a classification network that aims to learn the probabilities that
a set of inputs belongs to a set of output classes, the outputs of the
network are {\em softmaxed} to become probabilities,
\begin{equation}
p_{j} = \frac{e^{y_{j}}}{\sum_{j^{\prime}} e^{y_{j^{\prime}}}}.
\label{softmax}
\end{equation}
The classification likelihood is then given by the cross-entropy function
\begin{equation}
\mathcal{L}({\bf a}) = \sum_{i=1}^{n_{\rm t}}\sum_{j=1}^{n_{\rm c}} t_{j}^{(i)} \log 
p_{j}({\bf x}^{(i)}; {\bf a}).
\label{ClassNN-Like}
\end{equation}
In this scenario, the true and predicted output values are
probabilities. In the true outputs, all are zero except for the
correct output class, which has a value of one.


\chapter{SNe: from sky to catalogue}
\label{ch:SN_Clas}
\epigraph{`When he shall die,\\
Take him and cut him out in little stars,\\
And he will make the face of heaven so fine\\
That all the world will be in love with night\\
And pay no worship to the garish sun.'}{\textit{ William Shakespeare}}

Observations of SNe have been recorded since ancient times, transient
objects which flare brightly and appear as ``new stars'' against the
unchanging background static stars, only to fade after about a month
or so. The term \qu{nova} was first used by Tycho Brahe (1546-1601) to
describe a \qu{new star} which appeared in the constellation of
Cassiopeia on 11th November 1572, which he observed from Herrevads
kloster, Sweden. Much later Fritz Zwicky and Walter Baade
differentiated between \qu{two well-defined types of new stars or
  novae which might be distinguished as common novae and super-novae}
\citep{1934PNAS...20..254B, 1940RvMP...12...66Z}.  Nevertheless, only
since the late 1990s, in the era of charge-coupled device (CCD)
telescopes, have SNe made their contribution to cosmology. After the
great success of the Calan--Tololo survey, it became clear that one
can successfully standardise multi-band SN data, based on their
reproducible luminosities. The story of realisable ``standard
cosmological candles'' had begun.

In this chapter, I discuss SN discovery, classification and the study of
progenitor models. I give a short historical overview of past SN
surveys, summarise the current state of the field and give an outlook
on future surveys. I also summarise the techniques for
standardising SNIa, and their associated shortcomings.

\section{Observational techniques}

Since most of the energy output of a SN is in visible light, the main
methods for collecting information about them is through photometric
imaging and spectroscopy. Most of the studied SNe are usually
observed using both these techniques, so it is very useful to remember
the rough rule-of-thumb: if the SN has been detected/measured with a
telescope of diameter $D$, then spectroscopic follow-up will typically
require a telescope with diameter $2D$.

\subsection{Photometry}
 \begin{figure}
\begin{center}
\includegraphics[width=8cm]{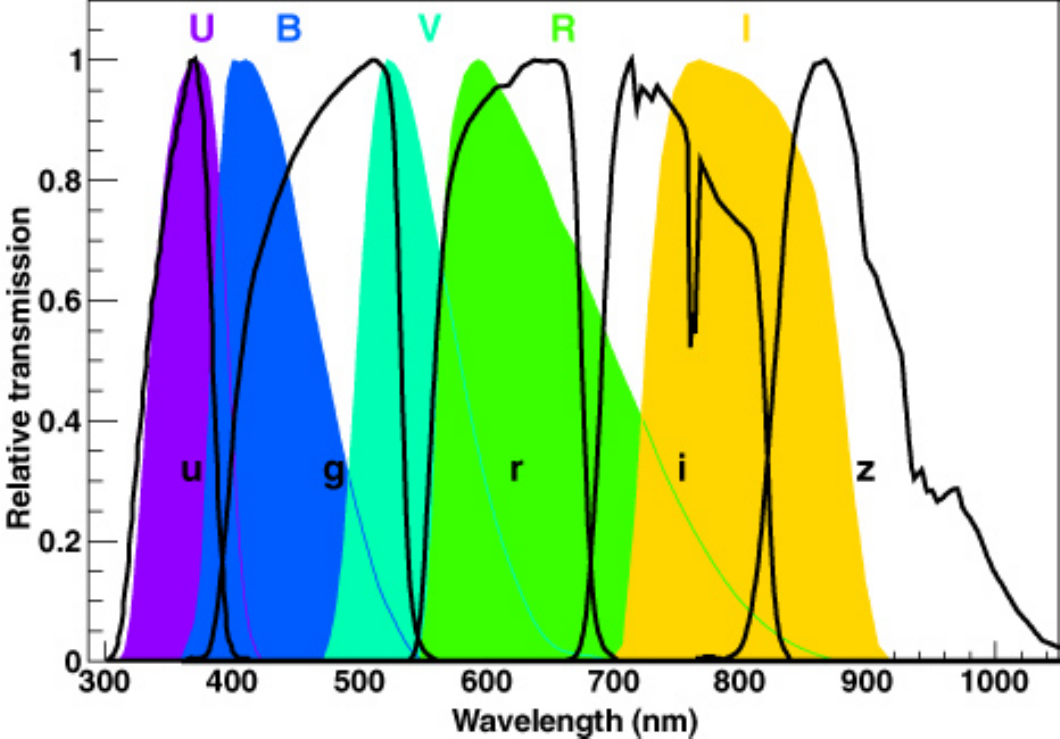}
\caption{Shapes of the commonly used photometric bands in the visible light,
  arbitrarily normalised. U, B, V, R, I refer to standard bands from
  \cite{1990PASP..102.1181B}, and u, g, r, i, z to the SDSS bands from
  \cite{1996AJ....111.1748F}.  Credit:
  \cite{2012arXiv1211.2590A}.}\label{fig:filters}
\end{center}
\end{figure}

In Figure \ref{fig:filters}, one can see two different sets of filters
commonly used in SN imaging. UBVRI is a ``standard'' set
\citep{1990PASP..102.1181B}, and u, g, r, i, z is a set of specially
designed filters used by the Sloan Digital Sky Survey (SDSS, \citealt{1996AJ....111.1748F}). Nowadays, imaging in the visible range
is made with a silicon CCD with red cutoff $\sim 1.1 \mu \text{m}$ and $10^6$
pixels per device.

A standard imaging observation is a two step process: (i) several
minutes of integration, (ii) $\sim$ 1 minute of read out. The image
resolution for ground-based telescopes is below 1 arcsecond
full-width-half-maximum and almost one order of magnitude finer in
space.

\subsection{Spectroscopy}

For spectroscopic observations, one most frequently uses fiber optics
\citep{2006astro.ph..2326G} or systems of lenses
\citep{2000INGN....2...11D}, which are positioned as a narrow slit in
the image plane and disperse the light in the perpendicular direction.
Modern multi-object spectroscopy instruments can collect data from up to $1000$ objects simultaneously. As in photometric observations,
putting spectroscopic instruments in space can be very advantageous,
since one can use a slit-less spectrometer down to very low sky
brightnesses.

\subsection{Near-infrared observations}

Very promising observations of SNe in the near infrared have been made
using low band-gap pixelised semiconductor devices, coupled to
integrated readout electronics \citep{1996NewA....1..177H}. Making
this kind of observation from the ground has problems since: (i) the
atmospheric glow rises with wavelength; (ii) there are numerous
absorption lines; and (iii) both these effects are not constant in
time. Near infrared observations also have a problem with ionizing
radiation affecting the sensors.

\section{Classification}
\label{ch:SN_Clas:class}

When describing the use of SNe for cosmological parameter inference
one usually means SNIa. However, SNe occur in a very broad range of
classes, which is still continuing to expand as SNe with previously unseen properties are
discovered.

\begin{figure}
\begin{center}
\captionsetup[figure]{skip=-10pt}
\includegraphics[width=11cm,height=5cm]{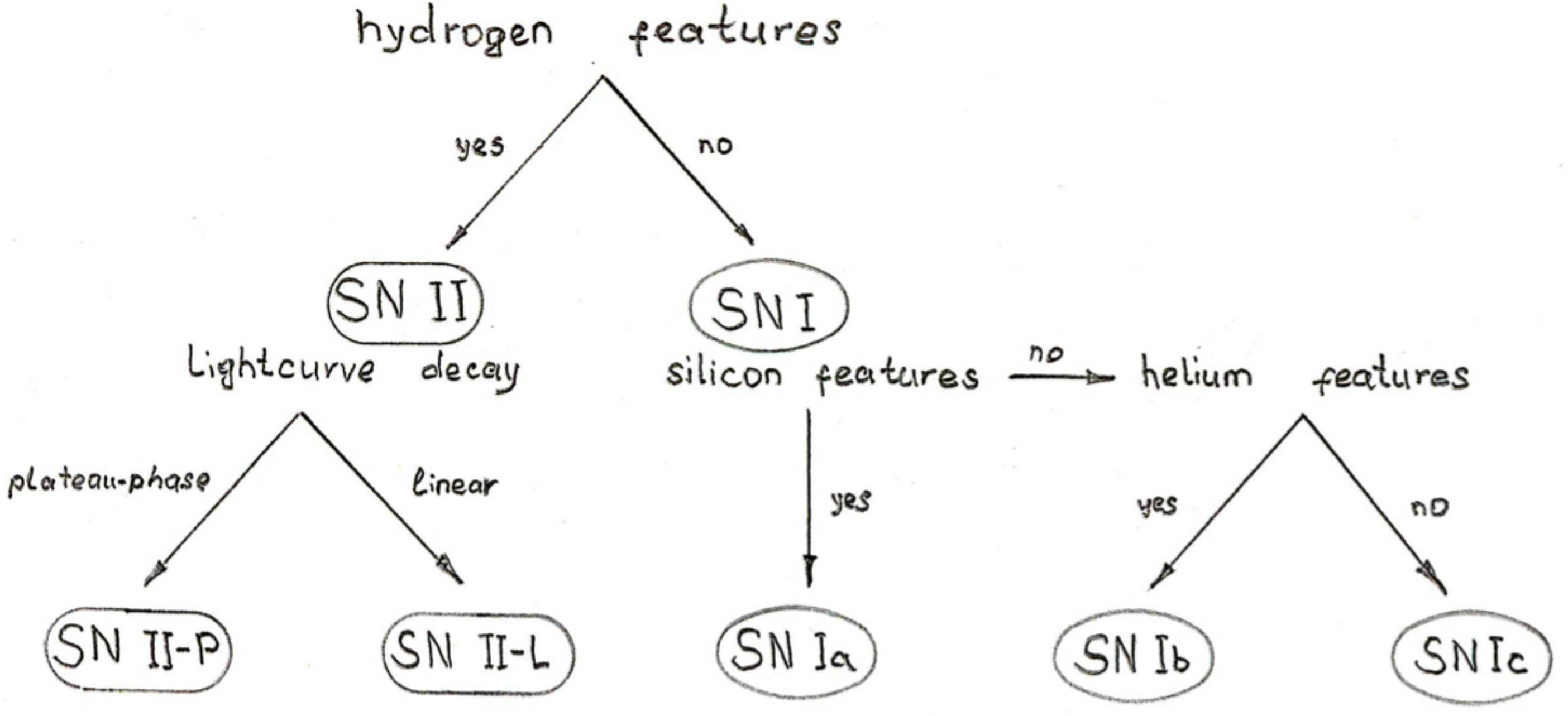}
\caption{SN classification scheme.}\label{fig:SNclass}
\end{center}
\end{figure}

The first of Baade and Zwicky's SNe had the broad features
characteristic of fast moving ejecta, being about 100 times brighter
than regular novae and not showing evidence of hydrogen lines. This was
a detection of a ``type I'' SN. Already in 1941, a different class of
SNe had been proposed \citep{1941PASP...53..224M}. These ``type II'' SNe
were fainter than those originally discovered and had a hydrogen line
in their spectra. In 1985 another unusual SN was detected, this time
it was characterised by the absence both of a hydrogen and silicon
line. This prompted a separation into subtypes within ``type I'' SNe:
SNIa were defined as events that have silicon and no hydrogen; type Ib
display neither silicon nor hydrogen, but have a strong helium line;
and type Ic have none of these lines. Figure \ref{fig:SNclass} shows
the classification scheme for the main SN types.  There is a trend to
allocate each somewhat unusual SN event to a new pigeonhole. This
queerness can be a small variation in the spectra or lightcurves.
This should in most cases be resisted, except if there are clear
physical grounds for doing so. Nonetheless, this is sometimes the
case. For example, recently a new class of superluminous SNe (SLSNe)
have come to light. These SNe are tens to hundreds of times more
luminous than “ordinary” SNe, and themselves divide into three
subclasses: SLSN-R, SLSN-I and SLSN-II, according to
\cite{2012Sci...337..927G}.

Since SNIa are the main focus of this work, I point out another way of
distinguishing between SNIa and non-Ia SNe. SNIa are thermonuclear
explosions (see Section \ref{Explos_models}), while all the other
types are core-collapse explosions.

As mentioned above, the silicon line is a signature of a ``standard
candle'', which means that in order to distinguish these events from
the variety of SN explosions we do need to have spectroscopic
measurements. From Figure \ref{fig:SNclass} one can see that, indeed,
most of the classification is performed on the basis of the
presence/absence of some element's (spectroscopic) lines, but, by
contrast, separation within type II is based on the shape of SN
lightcurves. Since spectroscopic observations are very ``expensive'' it
would be of great benefit to devise a SN classification method based
purely on photometric data (from which one, of course, cannot identify
the Si absorption line).
Many techniques targeted at SN photometric classification have been
developed, mostly based on some form of template fitting
(\citealt{2002PASP..114..833P}; \citealt{2006AJ....132..756J};
\citealt{2006AJ....131..960S}; \citealt{2007AJ....134.1285P};
\citealt{2007ApJ...659..530K}; \citealt{2007PhRvD..75j3508K};
\citealt{2009ApJ...707.1064R}; \citealt{2010ApJ...709.1420G};
\citealt{2010ApJ...723..398F}).  In such methods, the lightcurves in
different filters for the SN under consideration are compared with
those from SNe whose types are well established. Usually, composite
templates are constructed for each class, using high signal-to-noise
observations of lightcurves of well-studied SNe (see
\citealt{2002PASP..114..803N}), or spectral energy distribution models
of SNe. Such methods can produce good results, but the final
classification rates are very sensitive to the characteristics of the
templates used. One of the best-known approaches of this type is
PSNID \citep{2008AJ....135..348S,2011ApJ...738..162S}, 
which avoids many of the difficulties
encountered by simpler methods.

To address the issue of sensitivity to the templates used,
\cite{Newling2011} instead fit a parametrised functional form to the
SN lightcurves. These post-processed data are then used in either a
kernel density estimation method or a ``boosting'' machine learning
algorithm, as discussed by \cite{Newling2011}, to assign a probability to
each classification output, rather than simply assigning a specific SN
type.  More recently, \cite{2012MNRAS.419.1121R} and
\cite{2012arXiv1201.6676I} have introduced methods for SN photometric
classification that do not include on any form of template fitting,
but instead employ a mixture of dimensional reduction of the SN data
coupled with a machine learning algorithm. \cite{2012MNRAS.419.1121R}
proposed a method that uses a semi-supervised learning approach
applied to a database of SNe: first, a low-dimensional representation
of each SN is constructed from a simultaneous analysis of all the
lightcurves in the database. A classification model is then built in
this low-dimensional ``feature space'' by learning from a set of
spectroscopically confirmed training samples. This is subsequently
used to estimate the type of each unknown SN.

Subsequently, \cite{2012arXiv1201.6676I} proposed the use of
Kernel Principal Component Analysis as a tool to find a suitable
low-dimensional representation of SN lightcurves. In constructing
this representation, only a spectroscopically confirmed sample of SNe
is used. Each unlabeled lightcurve is then projected into this space
and a $k$-nearest neighbour algorithm performs the classification.

During my PhD studies, I also developed methods of photometric SNe
classification. In Paper III, I introduce an algorithm of SN
classification between SN Ia and non-Ia (see Section \ref{ch:paper3})
which does not involve templates. In Paper V, I further improve this
method by including a HNN technique.

\section{Searching for SNe}
Having discussed the main observational techniques used for SN
measurements and giving a brief description how to distinguish between
different SN types, I now describe of how SN surveys are
undertaken. Modern surveys typically require the following steps in
order to observe and classify SNe: (i) finding the events; (ii)
identifying the nature of the events using spectroscopic follow-up;
(iii) measuring the lightcurve of interesting events in as many bands
(see Figure \ref{fig:filters}) as possible.

\subsection{Finding events}
The main technique used to search for SNe is image subtraction. Since
SNe are transient events one can subtract search images taken at
different times to make non-variable objects disappear.  This method
was proposed in \cite{1987Msngr..47...46H} and first applied to real
observations in \cite{1989Natur.339..523N}.  By taking images obtained
at the rate of about twice or thrice per month, one is very likely to detect SNe
at their rising epoch, which is often early enough to trigger
spectroscopic and photometric follow-up observations.  This type of
search greatly benefits from imaging as large a field of view as
possible.
  
 Since it is often impossible to have photometric follow-up for all detected
 objects, one can implement a rolling search method, which consists of
 repeatedly imaging the same sky patch, and using the image sequence
 not only for SN detection, but also for measuring their
 lightcurves. This not only saves on photometric follow-up, but also
 has the advantage that several SNe might be observed in the same
 field of view and produces deep images in long-duration surveys.
 This technique was very successfully implemented in ground-based
 surveys, see Section \ref{ss:SnHistory}.

\subsection{Follow-up observations}

\textit{\large{Spectroscopic follow-up}:} As was mentioned
previously, spectroscopic follow-up typically requires a telescope of
twice the diameter of the one used for imaging. In terms of frequency
resolution, since SN spectra do not have narrow lines then
$\lambda/\delta\lambda \sim 100$ is sufficient for classification
purposes, but $\lambda/\delta\lambda \sim 1000$ is required to measure
the redshift sufficiently accurately, which is another important role
of spectroscopic follow-up. These observational requirements mean that
only a small fraction of SNe will have their spectra measured. Thus,
it is common during the image subtraction stage to attempt a
pre-classification to identify the most interesting candidates.  

The most challenging part of SN spectroscopy is to perform host
galaxy subtraction. Since the SNe are point sources, whereas their
host galaxies are extended, the fraction of host galaxy light mixed
with SN light will increase with redshift. For photometry this
problem is solved by making image subtractions, but this solution is
infeasible for spectroscopy. One of the current approaches is simply
to ignore the problem and make a selection against SNe with bright
host galaxies. Another way to address the problem is to use libraries
of observed SNe and galaxies to synthesize the observed spectra
\citep{2005ApJ...634.1190H}. 
Finally, an unavoidable downside of spectroscopy is its
``single-mindedness'', owing to the small field of view of most
instruments.

\textit{\large{Photometric follow-up}:} In contrast to spectroscopic
observations, photometric follow-up does not require large
observational facilities. The only constraint is that the images have
enough stars to serve as photometric and geometric
anchors. Astronomical photometry requires measurements of the source
of interest together with some standard stars using the same
instrument. With more and more rolling search surveys, photometric
follow-up will become increasingly unnecessary, especially with most
interest focussing on high-$z$ SNe, but it will still play a
major role for searches for nearby SNe.

\subsection{Surveys}

Since the early 1990s, many independent teams have explored the sky for
SNe. Some of them have targeted low redshifts, and others have looked
as deep as possible with present technologies. And the SN search
quest continues \ldots

 In this section I will describe the major SNIa surveys of the past,
 present and future, and discuss their main observational techniques,
 the time of data collection and the number of SNe observed.

\subsubsection{History of pioneering SN surveys}
\label{ss:SnHistory}

The oldest SN survey is the Calan--Tololo
survey \citep{1995AJ....109....1H,1996astro.ph..9062H}, which observed a sample of
``nearby'' SNe at redshifts below roughly $0.1$. Photometry and spectroscopy
for these SNe were measured on relatively small telescopes, of 1 m and
2 m in diameter, respectively.

After the great success of Calan--Tololo, in the mid-90s two teams
started a search for high-$z$ SNe in order to use them as standard
candles: the Supernova Cosmology Project (SCP) and the high-$z$ team
(HZT). By 1995 the projects were already yielding their first
promising results, and both teams consequently received plenty of
observational time, including photometric follow-up with the Hubble Space
Telescope (HST). The results from each team were published in
\cite{1998AJ....116.1009R} and \cite{1999ApJ...517..565P}, with 10 and
42 distant SNe respectively, and they came to the same conclusion: that the
expansion of the Universe is accelerating. Such a surprising and
profound discovery was obviously a very strong reason to continue the
search for further distant SNe.

The accuracy of the colour measurement was, however, a major problem
in these works, and so measuring accurate lightcurves was essential in
order to improve high-$z$ SN results. From 1997, HST started a new
program on photometric follow-up of ground-based searches, followed in
2002 by an independent program for finding faint high-$z$ events using
the Advanced Camera for Surveys \citep{2003astro.ph..9368K,
  2004ApJ...607..665R, 2006astro.ph.11572R}. This search allowed the
collection of good SN data, with events up to $z \sim 1.0$.

\subsubsection{Second generation of SN surveys}

\begin{table}
\begin{center}
\begin{tabular}{l|ccc}
\hline\hline
Name   &  ESSENCE & SNLS &  SDSS  \\
\hline
Imager  & \begin{tabular}{c}   CTIO 4-m \\on the 0.36~${\rm deg^2}$ \\ Mosaic-II \end{tabular}&      \begin{tabular}{c} CFHT 3.6-m \\ equipped with\\ 1~${\rm deg^2}$ Megacam \end{tabular} &  \begin{tabular}{c}   SDSS 2.5-m \\with its  1.52\\ ${\rm
  deg^2}$ camera \end{tabular}  \\
  \hline
  Bands & $ R, I$ & $ g,r,i, z$ & $ u,g,r,i,z$\\  
  \hline
   $z$ range &   [0.3,0.7] & [0.2,1.0] &  [0.1,0.4]\\ 
  \hline
  Location & Northern Chile & Hawaii & Apache Point, US\\ 
  \hline
  Monitored & 36 points & 4 points &300 ${\rm deg^2}$  \\
   regions &  (i.e. $\sim 10~ {\rm deg^2}$)  &  (i.e. $\sim 10~ {\rm deg^2}$)  & equatorial
stripe\\  
 \hline
   Frequency&   4th night & 4th to 5th night &  2nd night\\ 
  \hline
   Period &  \begin{tabular}{c}  for a 3-month \\per seasons\\ 2003-2008 \end{tabular} &\begin{tabular}{c} as long as
points\\ remained visible \\ 2003-2008 \end{tabular} & \begin{tabular}{c} 3 months\\ per year\\2005-2007 years  \end{tabular}\\ 
\hline
   $N_\text{SN events}$ &  $\sim$ 100 & $\sim$ 250 &  $\sim$ 370\\ 
  \hline\hline

\end{tabular}
\caption{Summary of second generation SN surveys.
\label{tab:rollingsearch}}
\end{center}
\end{table}
Within a decade of the success of Calan--Tololo, a few low-$z$ SN
surveys had targeted nearby sky for SNe: CfA \citep{1999AJ....117..707R,
  2006AJ....131..527J, 2009ApJ...700..331H}, the Carnegie Supernova
Project (CSP; \citealt{2009AAS...21442704C}) and the Lick Observatory Supernovae
Search (LOSS; \citealt{1999astro.ph.12336L}), which provides events both
inside and outside the Hubble flow. The latest completed nearby SN survey is
SNFactory \citep{2009EAS....36...11C, 2009AAS...21348902T,
  2009A&A...500L..17B}. The data from all these nearby searches now
provides an excellent low-redshift ``anchor'' for cosmological studies
using high-$z$ SNe.

The second generation of high-$z$ surveys include: ESSENCE
\citep{2007astro.ph..1043M}, SNLS \citep{2005astro.ph.10447A} and
SDSS \citep{2008AJ....136.2306H}, each of which used the rolling
search technique. Their goal was to increase both the number and
quality of well-measured high-$z$ SNIa. Indeed, many lightcurves for
high-$z$ SNe have been collected, but unfortunately not all detected
SNe have spectroscopic follow-up observations and measured redshifts, since using 4-m
and 8-m class telescopes for all of them was not feasible. A summary
of these surveys is presented in Table \ref{tab:rollingsearch}.

\subsubsection{Joint analysis}
\label{JA}
To place tight constraints on cosmological parameters, one needs a set
of SNe at a range of redshifts, with coherent distance estimates and a
good understanding of all systematic errors. This is usually achieved
by the analysis of SN compilations from different surveys, since
covering a large redshift range generally requires different
instruments. Any new compilation also typically contains new events
and state-of-the-art techniques for photometric calibration, and delivers
correlated uncertainties.

 The most commonly used compilation is ``Union'', which has had a few
 generations: Union \citep{2008ApJ...686..749K}, Union2 \citep{
   2010ApJ...716..712A} and the latest Union2.1 \citep{
   2012ApJ...746...85S}, and was very successful in making SN data
 more accessible for cosmological analysis outside of the SN community,
 where most of the users are primarily interested in cosmology
 constraints, and not so interested in parameters associated with the
 SNe themselves.  In addition to the ``Union'' compilation, one should
 also mention the most recent compilation, published in the latest
 SDSS data release paper by \cite{2014arXiv1401.4064B}. Indeed, the
 cosmological constraints derived from this latest survey were shown
 in Figure~\ref{f:Ia} in the Introduction.

One does, however, have to be very careful when combining data-sets
together. The joint analysis of combined surveys comes at a price,
since one must first check that the individual SN surveys produce
results that are mutually consistent. If this is not the case, any
results derived from their combination may be misleading. In Paper IV,
I present a method to perform this task and apply it to
existing compilations.

\subsubsection{Current and future SN surveys}

The intermediate Palomar Transient Factory (iPTF) is one of the nearby
SN surveys currently in operation.  Built as a continuation of the
Palomar Transient Factory, it has now collected data for about~$ 2200$
SNe, out of which about~$1400$ are SNIa. The observations are made in the
R- and g-band and most of them have spectroscopic follow-ups. In 2016,
iPTF will be transformed into the Zwicky Transient Factory
(ZTF). Using a reworked version of the same telescope as iPTF, ZTF
will use a new camera: the world's largest in field-of-view at nearly 50 ${\rm
  deg^2}$. This new camera will enable a full scan of the visible sky
every night.

CSP II is an another nearby SNe search started in 2011 and anticipated
to operate for five years. The difference between this survey and iPTF
is that, rather than performing optical observations, it collects data
in the near infrared and, together with time-series spectroscopy, it
tries to achieve a distance precision of 1-2$\%$ to build a definitive
low-redshift reference for future rest-frame infrared observations of
distant SNIa.

The Pan-STARRS (Panoramic Survey Telescope and Rapid Response System)
constitutes a new generation of rolling search surveys.  Designed in
the University of Hawaii's Institute for Astronomy, it has a
wide-field camera that can make images of the whole sky every four
nights. So far, only 1.5 years of results have been obtained
\citep{2013arXiv1310.3828R, 2013arXiv1310.3824S}, which has resulted in
146 spectroscopically confirmed SNIa at $0.03<z<0.65$.

Another high-$z$ survey currently in operation is DES, which saw
first-light in 2012 and will continue for five years.  This survey
operates on the Blanco 4-meter telescope in the Chilean Andes, with a
570-Megapixel digital camera, DECam.  DES surveys a large swathe of
the southern sky and will provide deep images of it. DES plans to
obtain well sampled lightcurves for more than several thousand
SNe. Unfortunately, DES does not have a spectroscopic follow-up
program, and so will have to rely on photometric classification
methods to determine which SNe are of Type Ia.

Finally, I would like to mention that the Large Synoptic Survey Telescope will begin operations in 2019 and
will photograph the entire available sky every few nights and have
data of outstanding quality. If all goes well, science observations
will commence in 2021.

Several mission concepts to measure SNe from space have also been
developed; unfortunately none of them passed the selection
processes. Working from space is crucial at $z \gtrsim 1$, because
reliable distances should then be measured in the near infrared, in
order to allow a direct comparison with nearby events measured in blue
bands.

\section{SNIa}
\label{Ia_main}

Since SNIa are those used in cosmology, they have become the most
studied type of SNe. In this Section, I briefly describe the physics of
SNIa explosions and ways in which they can be standardised for use in
cosmology.
 
\subsection{Explosion models}
\label{Explos_models}

The physics of a SNIa explosion, one of the most energetic events in
the Universe, is still unclear.  The most popular hypothesis is that
the SNIa is a thermonuclear explosion in carbon–oxygen white dwarfs in
close binaries \citep{1960ApJ...132..565H,
  1997Sci...276.1378N}. \cite{2004NewAR..48..605T} show that the observed amount of
energy in SNIa explosions is approximately the amount of energy that
would be produced in the conversion of carbon and oxygen into iron. In
order for this process to occur, white dwarfs must be close to the
Chandrasekhar mass, so that carbon ignition can start. A realistic way
for a white dwarf to grow to the Chandrasekhar mass is through mass
transfer within a close binary. Unfortunately, the nature of the stars
that can be the donors is not clear. Also, no progenitor system before a
SN explosion has been conclusively identified. Another problem with
this model is that there is observational evidence for systems where
the progenitor has a mass much lower (e.g.~\citealt{2009AJ....138..376F})
or higher (e.g.~\citealt{2006Natur.443..308H}) than the standard
Chandrasekhar mass. Leading models for the SNIa progenitor are
single-degenerate \citep{1973ApJ...186.1007W, 1982ApJ...253..798N} and
double-degenerate \citep{1981NInfo..49....3T,
  1984ApJS...54..335I,1984ApJ...277..355W}. There are many good
reviews of this topic, see e.g.~\cite{2012NewAR..56..122W,
  2000astro.ph..6305H}.

 The nature of SNIa explosions is interesting not only from the point
 of view of stellar and galaxy evolution, but also for cosmological
 studies. Early studies of nearby SNe and numerical simulations of
 their explosion models have been used to derive the value of $H_0$
 \citep{1992ApJ...392...35B,1996astro.ph..2025H,2004astro.ph.10686S}.
 Unfortunately, this is the only example when explosion models have
 been used to infer cosmological parameters. The reason for this is
 the complexity of the explosion and the consequent light
 production. SN lightcurve models discussed in Section \ref{ch:SALT2} can only broadly reproduce observed lightcurves. The
 same holds for the models of spectra.  Improving the models could allow
 us to use them as templates for fitting data, which would reduce the
 distance scatter, and give insights into redshift-dependent
 systematic biasses in distances.

\subsection{SN lightcurves}
\begin{figure}
\includegraphics[width=6cm,height=6cm]{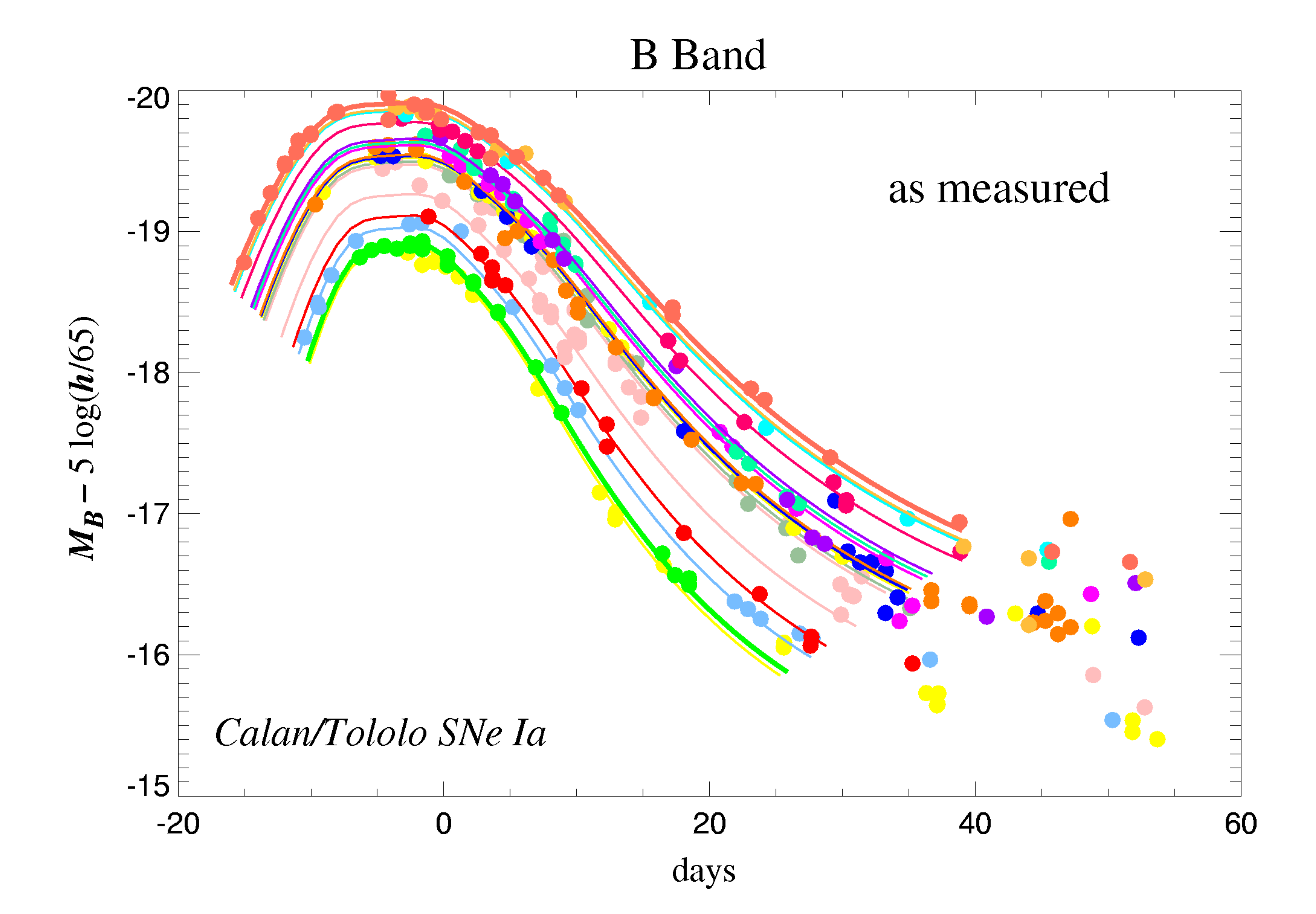}
\includegraphics[width=6cm,height=6cm]{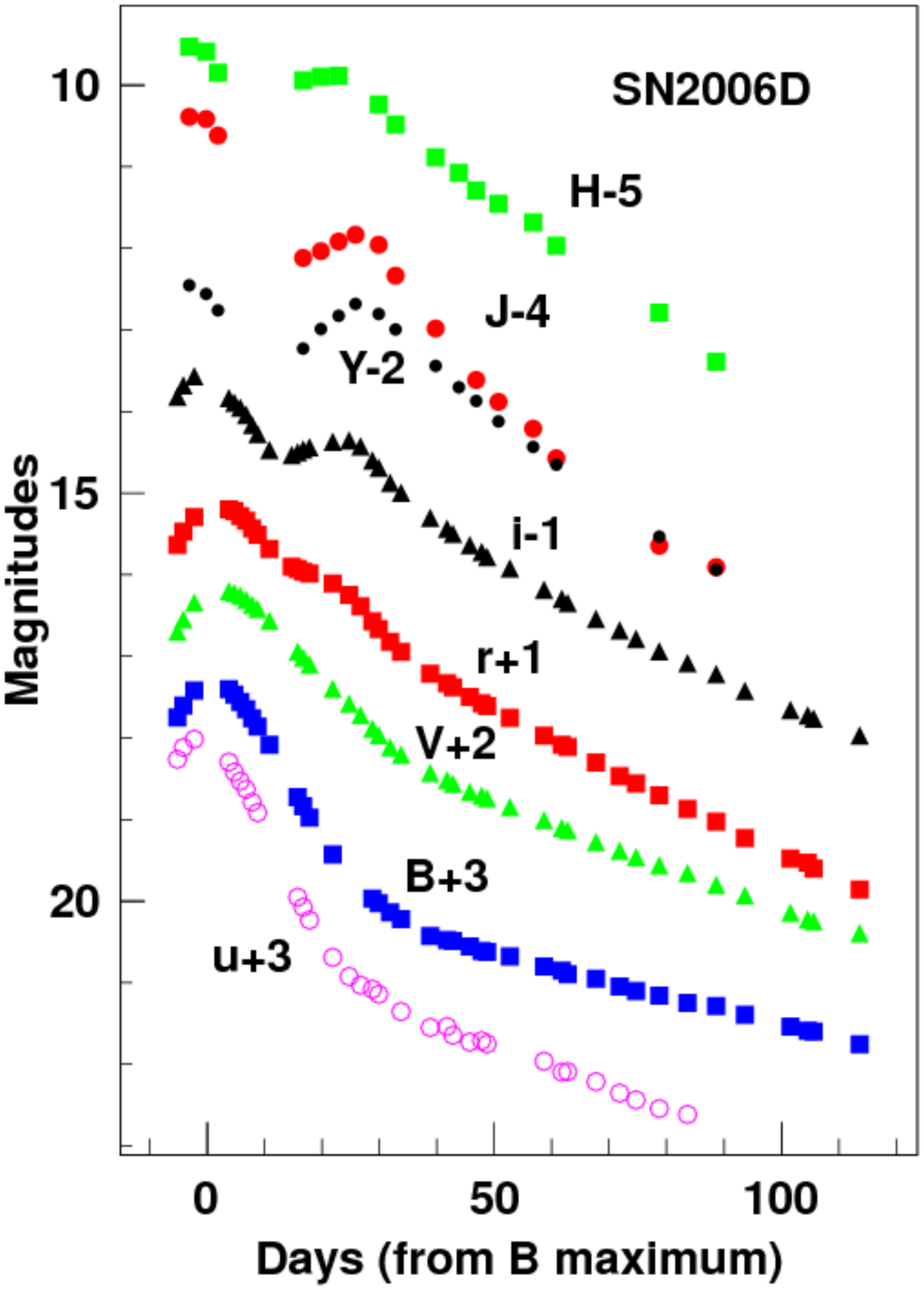}
 \captionof{figure}{Left panel: SN lightcurves in the B-band. Credit: Perlmutter (2003). Right panel: Lightcurves in different bands. Credit: \cite{2012arXiv1211.2590A}.}\label{f:SN_light}
\end{figure}

A SNIa emits most of its energy in the visible light, with some energy
in the near UV and near IR. As was noted early on, SNIa have
reproducible lightcurves \citep{1964ARA&A...2..247M}, as shown in the left
panel of Figure~\ref{f:SN_light}, but with very different behaviour in
different colours; the example of SN2006D is shown in the right panel
of Figure~\ref{f:SN_light}.

In the left panel of Figure \ref{f:SN_light} one can see that SNIa exhibit
a variability in the width of their lightcurves. Accounting for time
dilation in order to compare restframe widths does not completely
eliminate this difficulty. The problem was ``solved'' in 1993 from a
small sample of well-measured events \citep{phillips1993}.  The
Phillips relation quantifies how SNe that are intrinsically brighter
have lightcurves that decline more slowly from their maximum.  This
relation is central to the use of SNIa for measuring cosmological
distances. Moreover, in the $B$-band, this variability of lightcurves
is typically described by stretching the time axis of a single lightcurve template \citep{1997ApJ...483..565P,2001astro.ph..4382G}. The
rise and fall timescales seem to vary together for
\cite{2001astro.ph..4382G} and \cite{2006astro.ph..7363C}, while
\cite{2010ApJ...712..350H} finds them to be essentially independent.

SNIa also exhibit variability in their colours (measured, e.g. at
maximum) even at a fixed decline rate \citep{GuySullivan2010,
  2009ApJ...693..207B}. The source of colour variability, which is
unrelated to the decline rate, is still unclear. One of the options is
that SNe intrinsically have different colors
(e.g.~\citealt{2011ApJ...729...55F}), while another hypothesis relates
it to extinction by dust in the host galaxy. Most likely the truth is
a mixture of both and possibly some other astronomical reasons that we
do not yet understand.

To use SNIa for cosmological searches, their fluxes must be expressed
in the same way, and for this purposes one uses lightcurve fitters.

\subsection{Standardization of SNe}
\label{ch:StandOFsne}

Using empirical lightcurve models not only ``standardises'' SN
lightcurves, but also ``compresses'' the photometric data
characterizing an event. The aim of all the lightcurve models can be
summarised as the derivation of a distance, for which one needs a
brightness (anything that scales linearly with the observed flux),
the decline rate and colour.

In the original work of \cite{phillips1993}, the data measurements
were very well sampled, so there was no need for an explicit
model. Phillips built smooth discrete templates with
different $\Delta m_{15}$ values, where this quantity denotes the
decline rate measured as the magnitude difference between peak and 15
(restframe) days later.  \cite{1995AJ....109....1H} generalised this
method to allow it to fit much more sparse observations. This method
enjoyed broad usage and the resulting templates have been continuously
updated \citep{1999AJ....118.1766P,
  2004A&A...415..863G,2006astro.ph..3407P}. Recently, the {\sc Snoopy} model
\citep{2011AJ....141...19B} revisited the $\Delta m_{15}$ paradigm
with an extension of it into the near IR. Another method was used to
account for the decline-rate variation: the ``stretch'' paradigm
\citep{1997ApJ...483..565P} proposed to stretch the time axis of these
templates in the B and V bands. A disadvantage of all
these types of models is that they rely on lightcurve templates.

\subsubsection{Spectral Adaptive Lightcurve Template 2 (SALT2)}
\label{ch:SALT2}
\begin{figure}
\begin{center}
\includegraphics[width=10cm,height=6cm]{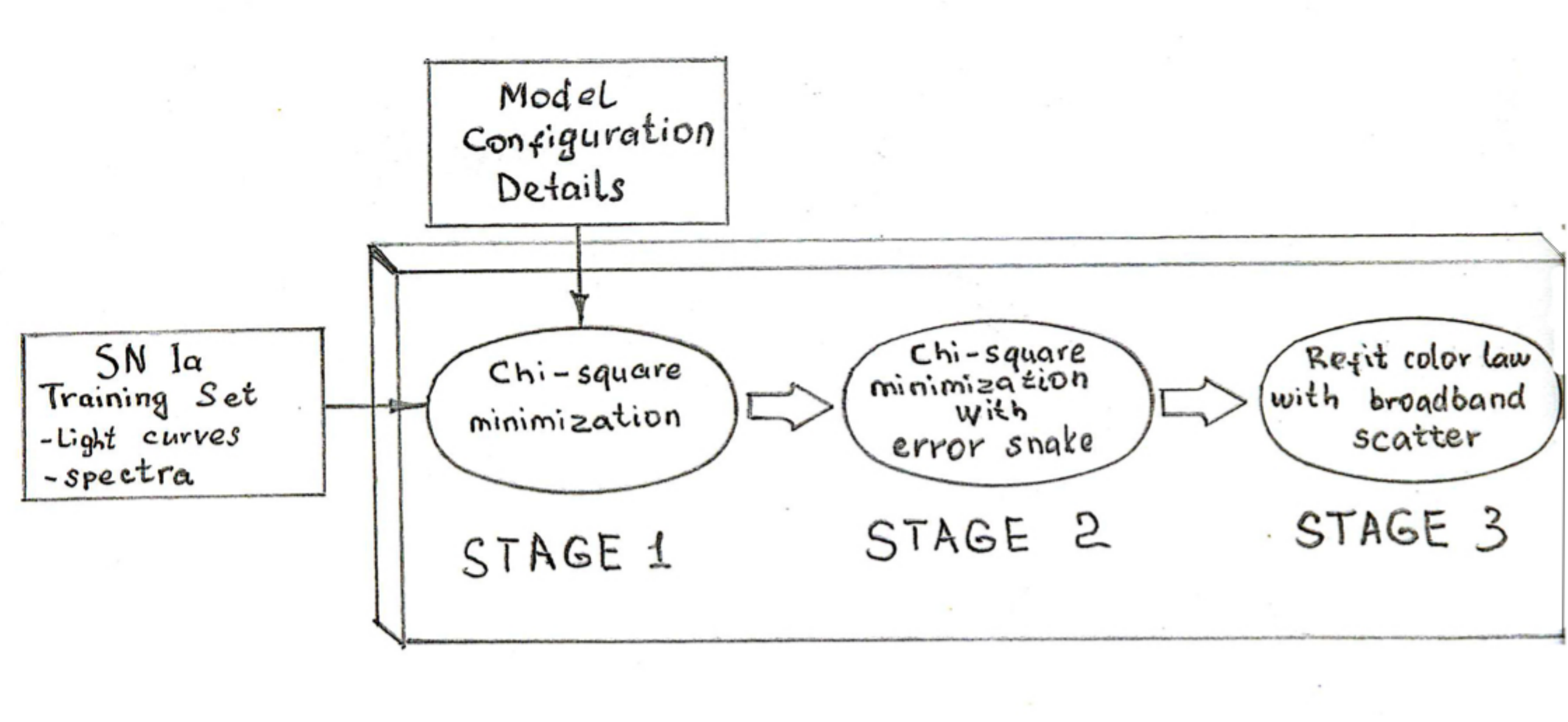}
\caption{The three stages of the SALT2 model training process.  Each
  stage calculates best-fit model parameters using successively
  improved estimates of model uncertainties.}\label{fig:SALT2}
\end{center}
\end{figure}

The most commonly used empirical lightcurve model is currently SALT2
\citep{2007A&A...466...11G,2014arXiv1401.4065M}. The main idea behind this
method is that all the measurements are fit using a function of the phase $p$ and the wavelength $\lambda$
\begin{equation}
\protect \label{eq:S2function}
F(p,\lambda)=x_0~[M_0(p,\lambda)+x_1 M_1(p,\lambda)]~\exp[c~\rm{CL}(\lambda)],
\end{equation}
where $M_0$ is the mean SNIa spectral energy distribution, $M_1$
accounts for light\-curve width variations, $\rm{CL}$ is a color law which
incorporates any wavelength-dependent color variations and is
independent of epoch, $x_1$ is the lightcurve shape parameter, $x_0$
is the overall flux scale and $c$ is the peak B--V color. The first
three of the above parameters are SN-independent and describe all
SNIa, whereas the last three parameters are unique for each SN.  No
assumptions about dust or extinction laws are made a priori.  A
cartoon schematic of the SALT2 training process is shown in Figure
\ref{fig:SALT2}.

SALT2 also allows one to accommodate the intrinsic variability of SNIa. 
To achieve this, the model includes sources of additional
uncertainty. These are a broadband magnitude scatter $k(\lambda)$,
which deals with the color law and the $c$ parameter, and a spectral
``error snake'' $S(p,\lambda)$ to account for the $x_1$ parameter. In
order to calculate and include these effects, SALT2 performs three
iterations of $\chi^2$-minimisation, as illustrated in Figure
\ref{fig:SALT2}.

As a result, for each SN, SALT2 reports best-fit values $\hat{x}_0,
\hat{x}_1, \hat{c}$, the redshift $\hat{z}$, and the covariance matrix
\begin{equation} \label{eq:def_CSALT2}
\hat{C}_{\rm{SALT2}} = \left( \begin{array}{c c c} \sigma_{x_0}^2  & \sigma_{x_0 ,  x_1} & \sigma_{x_0 ,  c}  \\ 
\sigma_{x_0,  x_1} & \sigma^2_{x_1} & \sigma_{x_1,c}  \\
\sigma_{x_0,  c} & \sigma_{x_1,c} & \sigma^2_{c}  
\end{array}\right).
\end{equation}
Let us denote the result of the SALT2 lightcurve fitting procedure
as
\begin{equation} \label{eq:dataSALT2}
D_{\text{SALT2},i} = \{ \hat{z}_i, \hat{x}_{0i}, \hat{x}_{1i},\hat{c}_i, \hat{C}_{i, \rm{SALT2}} \},
\end{equation}
where $i$ runs through the $n$ SNe in the sample. These outputs can be
used to obtain distance estimates for the SNe using Eq.~\ref{eq:muobsdef}.

\subsubsection{Other lightcurve models and distance estimators}
\label{ch:nonSALT2}

As mentioned previously, the SALT2 model is one of the most frequently
used SN lightcurve fitters. Together with SALT
\citep{2005A&A...443..781G,AstierGuy2006} and SiFTO \citep{2008ApJ...681..482C}, it
mostly uses the ``stretch'' paradigm. SALT is just a earlier version of
SALT2, whereas SiFTO models the lightcurves from spectral energy
distribution templates. SiFTO uses the fact that SN lightcurves have
different shapes in different bands (see the right panel of Figure~\ref{f:SN_light}). As in SALT2, these methods do not give direct
distant estimates, but one needs to use equations of the form of
Eq.~\ref{eq:muobsdef}. One can also perform analyses with combined
SALT2/SiFTO fits, see e.g.~\cite{GuySullivan2010}.

Another frequently used model is the Multi Light Curve Shape (MLCS,
\citealt{1996ApJ...473...88R}) model, which consists of a
one-parameter family of lightcurve shapes with standard visible bands
B, V, R and I, see Figure~\ref{fig:filters}. This complete model of SN lightcurves depends on three parameters (plus a reference date): a distance
modulus, a brightness offset, and an extinction value. Initially
trained on 12 events it was updated to about 100 events. Also, the
model has been extended towards the blue by adding the U-band. This
second version is named MLCS2k2 \citep{2007ApJ...659..122J}.

Among other methods are the Bayesian Adapted Template Match method
\citep{2003astro.ph..5008T}, the so-called ``Bailey ratio''
\citep{2009A&A...500L..17B}, the CMagic distance estimator
\citep{2003astro.ph..2341W}, the Gaussian-process regression
method \citep{2013ApJ...766...84K}, plus many more.

Unfortunately, not only is none of these methods the ``correct'' one,
but the cosmological parameter constraints one obtains using the same
data but different lightcurve fitting methods are significantly
different. Comparisons between SALT2 and MLCS2k2 were made in
\cite{2009ApJS..185...32K}, which finds that when the restframe
$U$-band is taken into account, systematically different distances are
obtained. In \cite{GuySullivan2010}, SALT2 and SiFTO are compared and
the differences obtained are acceptable if the compilation is contains
the level of a few hundred events.

\section{Type non-Ia SNe in cosmological analyses}
\label{ch:IIp_cosmology}

Usually non-Ia SNe are an unwanted contaminant in SNIa
compilations. Nevertheless, there is one type of non-Ia SNe that can
be used for cosmological analyses. Type II Plateau SNe (SNII-P)
can be used as standard candles to determine luminosity distances,
although only for smaller distances and with lower accuracy than hose of
SNIa. Despite this shortcoming, SNII-P explosions are better
understood than SNIa. Another advantage of SNII-P is that they have
been found only in late-type galaxies, whereas SNe Ia have been
observed both in late- and early-type galaxies. Consequently, one
would expect distance estimates from SNII-P to suffer less from biasses
resulting from different galactic environments. This difference in
systematic effects allows SNII-P data to complement SNIa analyses
\citep{2010ApJ...708..661D}.


\chapter{Cosmological parameter inference using SNe}
\label{ch:Cosmology}
\epigraph{`All my Bayesian friends have objected at this point that there's
  no such concept as bias in Bayesian analysis. It is true that there
  is no meaningful, exact definition of bias except in a frequentist
  sense. What I mean here is that the answer [\ldots] is usually wrong, and
  in a given direction. The dictionary calls this ``bias''.'}{Stephen Gull}

Undoubtedly the greatest accomplishment of SN studies over the past
20 years is the discovery of the current accelerating expansion of the
Universe.  This was recognised by the award of the Nobel Prize in
Physics in 2011, which was divided one-half to Saul Perlmutter, the
other half jointly to Brian P. Schmidt and Adam G. Riess.

In order to constrain cosmological parameters, one needs to make an
inference from the observational measurements. It is usual to adopt a
two-stage approach in which SN observations are first analysed using
a lightcurve fitting program, the outputs of which are then used in a
second-stage of inference for the cosmological parameters. The first
step of this process was already described in
Section~\ref{ch:StandOFsne}.  In this chapter, I outline the
commonly-used methods of cosmological inference, assessing their
advantages and disadvantages. Since the SALT2 algorithm is the most
widely used for SN fitting and I made use of it in my papers, I will
also use it here. However, the results and conclusions presented in
this Chapter are not specific to SALT2 and can be easily generalised
to the outputs of other lightcurve fitting codes like SiFTO, MLCS2k2,
etc.

\section{Comparing theory and observations}

As discussed in Section \ref{ch:SALT2}, on the observational side, for
each SN, the SALT2 algorithm reports best-fit values $\hat{m}_{B}^*,
\hat{x}_0, \hat{x}_1, \hat{c}$, the redshift $\hat{z}$, and the
covariance matrix $\hat{C}_{\rm{SALT2}} $ defined in
Eq.~\ref{eq:def_CSALT2}.  This covariance matrix does not include
covariances involving $\hat{m}_{B}^*$. We use
\begin{eqnarray}
 \sigma_{m_{B}^*,  x_1}= -\frac{5 \sigma_{x_0,  x_1}}{2 x_0  \ln10}\\
 \sigma_{m_{B}^*,  c}= -\frac{5 \sigma_{x_0,  c}}{2 x_0  \ln10}
\end{eqnarray}
to construct the covariance matrix 
\begin{equation} \label{eq:def_Cm_new}
\hat{C} = \left( \begin{array}{c c c} \sigma_{m_B^*}^2  & \sigma_{m_B^* ,  x_1} & \sigma_{m_B^* ,  c}  \\ 
\sigma_{m_B^*,  x_1} & \sigma^2_{x_1} & \sigma_{x_1,c}  \\
\sigma_{m_B^*,  c} & \sigma_{x_1,c} & \sigma^2_{c}  
\end{array} \right). 
\end{equation}
Let us denote the result of the SALT2 lightcurve fitting procedure,
 with the above rescaled covariance matrix,
as
\begin{equation} \label{eq:data}
D_i = \{ \hat{z}_i, \hat{m}_{Bi}^*, \hat{x}_{1i},\hat{c}_i, \hat{C}_i \},
\end{equation}
where $i$ runs through the $n$ SNe in the sample.

The data $D_i$ define the ``observed'' distance modulus $\mu^{\rm obs}_i$
for the $i$th SN as
\begin{equation}
\mu_i^{\rm obs} (\alpha,\beta,M_0) = \hat{m}_{B,i}^\ast - M_0 + \alpha
\hat{x}_{1,i} - \beta\hat{c}_i.
\label{eq:muobsdef}
\end{equation}
This expression contains three unknown parameters, all of which are
assumed global, i.e. having the same value for all SNIa. These
parameters are: $M_0$, the $B$-band absolute magnitude of the SN, and
$\alpha$, $\beta$, which are nuisance parameters controlling the
stretch and colour corrections.

On the theory side, in a Friedman-Robertson-Walker cosmology, if an
object has an absolute luminosity $L$ and one measures its flux to be
$F$, then its luminosity distance is given by\footnote{For simplicity, I refer here to bolometric quantities,
  i.e. integrated over all frequencies, rather than in terms of fluxes
  in specific frequency bands.}
\begin{equation}
D_\textrm{L}  \equiv \left(\frac{L}{4\pi F}\right)^{1/2}.
\label{eq:11}
\end{equation}
The distance modulus is defined by $\mu = m - M_0$, where 
$m \equiv -2.5 \log_{10} F$ is the apparent magnitude of the object and 
$M_0$ is its absolute magnitude. Hence, $\mu$ can be written as
\begin{equation}
\mu = 5 \log_{10} \left(\frac{D_\textrm{L}}{\mbox{Mpc}}\right) + 25,
\label{eq:dmod1}
\end{equation}
where the constant offset is included such that one satisfies the convention
that $\mu=0$ at $D_\textrm{L} = 10$~pc.

In terms of the cosmological parameters $\mathscr{C}=\{\Omega_\textrm{m,0},\Omega_\textrm{de,0},H_0,w\}$, an  object at a redshift $z$ has the 
luminosity distance
\begin{equation}
D_\textrm{L}(z,\mathscr{C}) = \frac{c}{H_0}\frac{(1+z)}{\sqrt{|
\Omega_{k,0}|}}
S\left(\sqrt{|\Omega_{k,0}|}I(z)\right),
\label{eq:12}
\end{equation}
where 
\begin{equation}
I(z)\!\ \equiv\!\! \int_0^z \!\!\!\frac{d\bar{z}}
{\sqrt{(1+\bar{z})^3\Omega_\textrm{m,0} \! + \! 
(1+\bar{z})^{3(1+w)}\Omega_\textrm{de,0} 
\! + \! (1+\bar{z})^2 \Omega_{k,0}}},
\end{equation}
in which (neglecting the present-day energy density in radiation)
$\Omega_{k,0} \equiv 1 - \Omega_{\textrm{m},0}-\Omega_\textrm{de,0}$
and $S(x)=x$, $\sin x$ or $\sinh x$ for a spatially-flat
($\Omega_{k,0}=0$), closed ($\Omega_{k,0}<0$) or open
($\Omega_{k,0}>0$) universe, respectively. A cosmological constant
corresponds to the special case in which the dark-energy
equation-of-state parameter has the value $w=-1$ in this case, the
present-day density parameter is usually denoted by
$\Omega_{\Lambda,0}$.

To compare the observations and theory, one could therefore identify a
set of objects ($i=1,2,\ldots,N$) whose absolute magnitudes are known
{\em a priori} (standard candles), measure their
distance moduli and redshifts, and
consider the differences (often termed Hubble diagram residuals)
\begin{equation}
\Delta\mu_i = \mu_i^\textrm{obs}(\alpha,\beta,M_0)-\mu(z_i,\mathscr{C}).
\label{eq:deltamu}
\end{equation}
These could then be used to place constraints on the cosmological
parameters $\mathscr{C}$, plus other parameters of interest, such as
$\alpha, \beta \textrm{ and } M_0$.

One should note, however, that if the standard candles all have the
same absolute magnitude $M_0$, but this value is unknown, then it is
degenerate with the Hubble constant $H_0$, as evident from
Eqs.~\ref{eq:11} and \ref{eq:12}. Perhaps more importantly, Nature has
neglected even to provide such a set of ``uncalibrated'' standard
candles. Instead, we must make do with SNIa for which the absolute
magnitudes vary, but can nonetheless be standardised, as discussed in
Section \ref{Ia_main}.

\section{Toy simulations of SN data}
\label{ch:Cosm:sim}

In order to discuss the advantages and disadvantages of different
inference methods, I apply them to toy simulated SNIa data. In these
simulations I assume that the off-diagonal elements of the covariance
matrix in Eq.~\ref{eq:def_Cm_new} are zero; this makes a negligible
difference since the off-diagonal elements are very small in practice.

To make the simulation, the procedure below is performed for each SN ($i=1,2,\ldots,N_{\rm SN}$), and the values of the various parameters used in the
simulations are given in Table~\ref{tab:cosm:sims}.

\begin{table}
\begin{center}
\begin{tabular}{l|c}
\hline\hline
Parameter   &  Value used for simulations   \\
\hline
$\sigma_{\text{int}}$  &    0.01  \\
$\alpha$  &    0.12  \\
$\beta$  &    3.2 \\
$M_0$  &    $-19.3$ \\
$\sigma_{m_b} $ &  0.05  \\
$\sigma_{x_1} $ &  0.5  \\
$\sigma_{c} $ &  0.05  \\
$\sigma_{z} $ &  0.0001  \\
$\Omega_{\rm m,0} $ &  0.3 \\
$h $ &  0.7 \\
\hline\hline
\end{tabular}
\caption{Parameter values used in the generation of simulated SNIa 
data, assuming a spatially-flat universe.
\label{tab:cosm:sims}}
\end{center}
\end{table}

\begin{enumerate} 
\item The redshift is drawn independently from $z_i \sim U(0,1)$, where $U(a,b)$ denotes a uniform distribution in the range $[a,b]$.
\item The predicted $\mu_i(z_i,\mathscr{C})$ is calculated using Eq.~\ref{eq:dmod1}.
\item The hidden variables $M_i$, $x_{1,i}$ and $c_i$ are drawn from the
  respective distributions $M_i \sim {\cal N}(M_0,\sigma_{\rm int}^2)$, $x_{1,i} \sim U(-5.0,3.0)$ and $c_i \sim U(-0.2,0.3)$, where ${\cal N}(\mu,\sigma^2)$ denotes a normal (Gaussian) distribution with mean $\mu$ and variance $\sigma^2$.
\item The value of $m^\ast_{B,i}$ is calculated using the Phillips relation
$m^\ast_{B,i} = \mu(z_i,{\mathscr{C}})+M_i-\alpha
  x_{1,i} + \beta c_i$.
\item The simulated observational data are obtained by drawing
  independently from the distributions $\hat{z}_i \sim {\cal
    N}(z_i,\sigma^2_{z,i})$, $\hat{m}_{B,i}^\ast \sim {\cal
    N}(m_{B,i}^\ast,\sigma_{m^\ast_{B,i}}^2)$, $\hat{x}_{1,i} \sim
  {\cal N}(x_{1,i},\sigma_{x_{1,i}}^2)$ and $\hat{c}_i \sim {\cal
    N}(c_i,\sigma_{c,i}^2)$.
\end{enumerate}

For my simulations, I generated 100 independent data-sets, each one containing 200 SNe. In my analyses, I assume $h=0.7$ (as is
established to a few per cent accuracy by a number of cosmological
probes) and vary $\Omega_{\rm m,0}, M_0, \alpha$ and $\beta$.

\section{Naive definition of the likelihood }
\label{ch:Cosmology:naive_like}

Assuming that the Hubble diagram residuals are Gaussian-distributed
one often defines the likelihood as

\begin{equation}
\mathcal{L} = \prod\limits_{i=1}^{N_{\rm SN}}{\frac{1}{\sqrt{2\pi}\sigma_i(
\alpha,\beta,\sigma_\text{int})}}\exp \left\{- \frac{[\mu_i^\text{obs}(\alpha,\beta,M_0)-\mu_i(\mathscr{C})]^2}{2\sigma_i^2(\alpha,\beta,\sigma_\text{int})}\right\}.
\label{eq:norm_like}
\end{equation}
In this expression, I have included the dependence on (only)
the parameters to be fitted. Here the
total dispersion $\sigma_{i}^2$ is given by:
\be
\sigma_{i}^2 = (\sigma_{\mu,i}^z)^2 + \sigma_\text{int}^2 
+ \sigma_{\text{fit},i}^2(\alpha,\beta).
\label{eq:sigmadef}
\ee
The three components arise as follows:
\begin{enumerate}
\item Uncertainties in the peculiar velocity and/or the spectroscopic
  measurements of either the host galaxy or SNIa itself lead to an
  uncertainty $\sigma_{z,i}$ in its estimated redshift, which in turn
  induces an error $\sigma_{\mu,i}^z$ in the distance modulus.
\item Even after correction for stretch and colour, there remains some
  global variation in the SNIa absolute magnitudes. The quantity
  $\sigma_\text{int}$ contains all of these intrinsic dispersion
  errors.
\item The uncertainty in the fitting of the parameters by SALT2 is given by
\be
\sigma_{\text{fit},i}^2 = \boldsymbol{\psi}^{\rm t} \hat{C}_i  \boldsymbol{\psi},
\label{eq:sigmafitdef}
\ee
where the transposed vector $\boldsymbol{\psi}^{\rm t} = \left(1, \alpha, -\beta
\right)$ and $\hat{C}_i $ is the covariance matrix given in Eq.~\ref{eq:def_Cm_new}.
\end{enumerate}
Additional errors, such as those due to lensing or Milky Way dust extinction can also be added in at this stage, but I do not consider such errors here.
 
For convenience, one often writes Eq.~\ref{eq:norm_like} in the form:
\begin{equation}
-\ln(L) = \mbox{const} + \sum_{i=1}^{N_{\rm SN}}\frac{[\mu_i^\text{obs}-\mu_i]^2}{2 \sigma_i^2} + \sum_{i=1}^{N_{\rm SN}} \ln(\sigma_i).
\label{eq:norm_like_long}
\end{equation}

\begin{figure}
\centerline{\includegraphics[width=8cm]{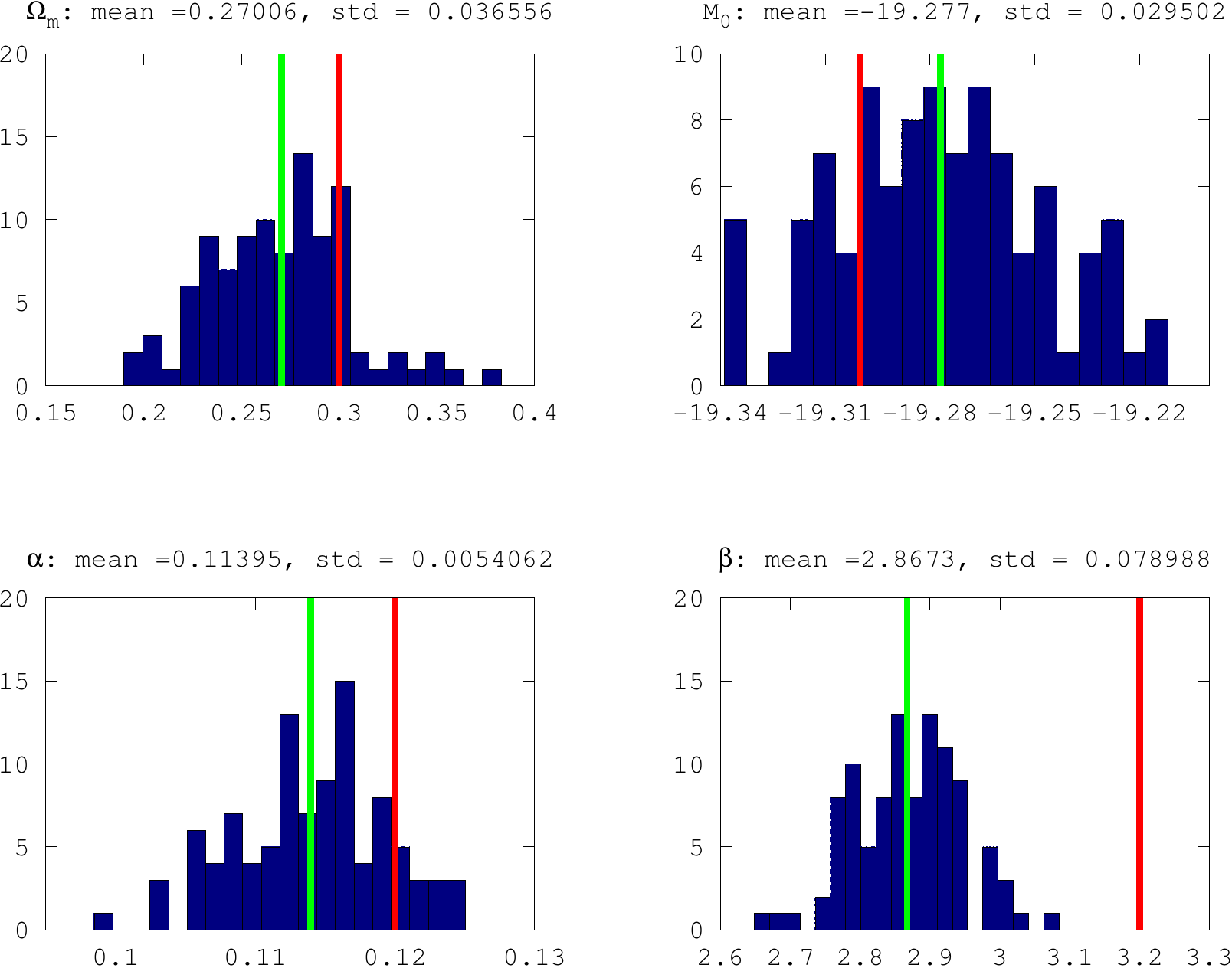}}
\caption{Histograms showing the distribution of the point estimates
  of the parameters $\Omega_{\rm m,0}$, $M_0$, $\alpha$ and $\beta$. The
  green vertical lines show the mean values of the point estimates,
  and the solid red vertical lines show the true values of the 
  parameters used to simulate the data. The data are analysed using the
  naive likelihood method.}\label{f:naive_large}
\end{figure}

As shown in Figure \ref{f:naive_large}, maximising this function unfortunately leads to
biased results for some of the parameters. This is particularly acute for the $\beta$ parameter. This effect stops us using the likelihood
in Eq.~\ref{eq:norm_like}, although it has been employed and investigated in some
previous analyses \citep{2005physics..11182D}. 

\begin{figure}
\centerline{\includegraphics[width=8cm]{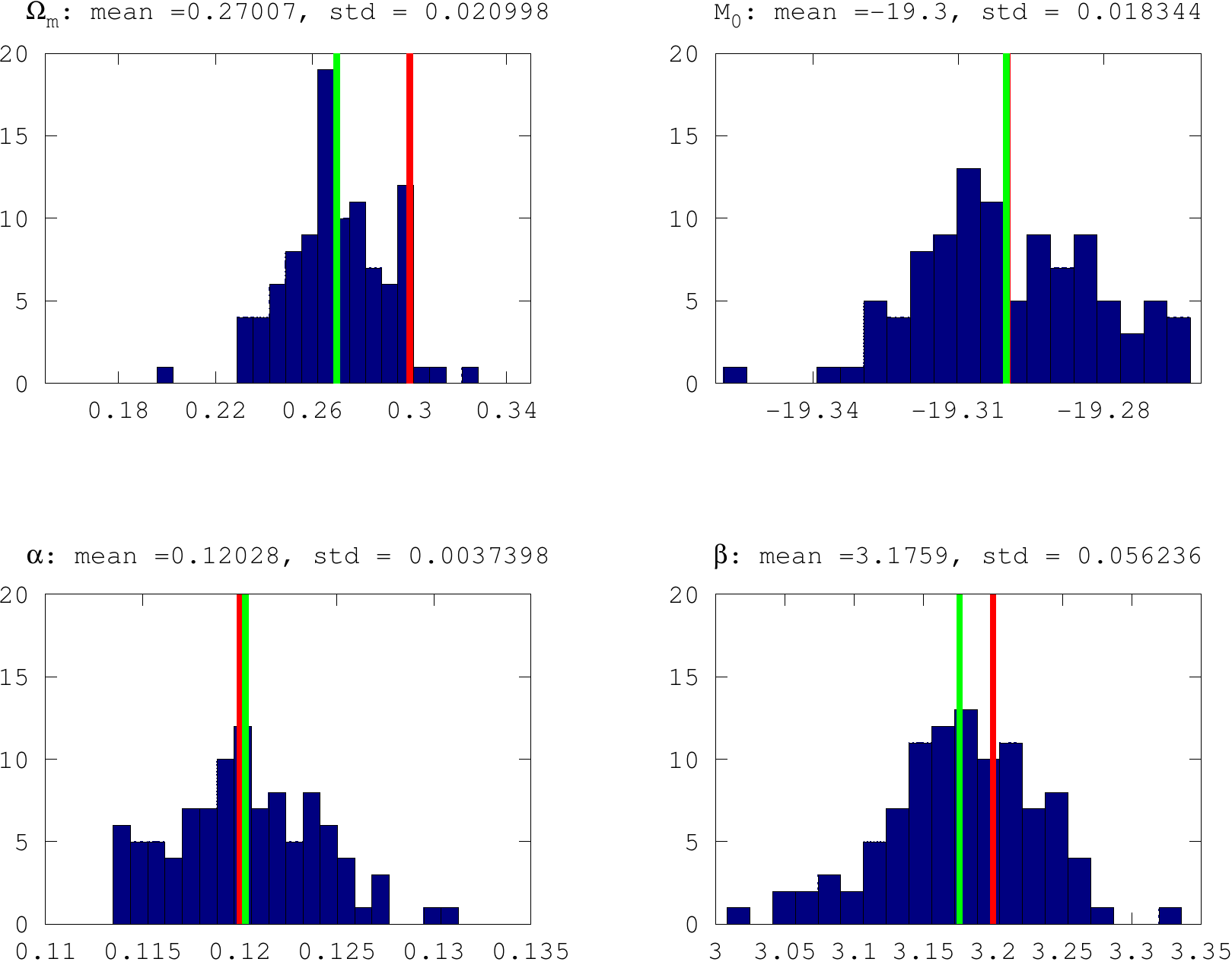}}
\caption{The same as Figure \ref{f:naive_large}, but for simulations
  generated using the values $\sigma_{x_1} = 0.1$ and $\sigma_{c} = 0.01$.}\label{f:naive_small}
\end{figure}

It is worth pointing out that, in the case when the error bars on the
color and stretch parameters are small, the biasses ``disappear'', as one
can see in Figure \ref{f:naive_small}.

\cite{Gull} identified the origin of the problem, albeit in the
conceptually more straightforward context of fitting a straight line
\be
y = ax +b,
\label{eq:Gull_line}
\ee
to a set of data $(x_i,y_i)$ where the $x_i$ and $y_i$ both have an
associated measurement error and $a, b$ are the parameters of the
model. One can see that this linear toy model is very similar to the
relation between the parameters in Eq.~\ref{eq:muobsdef}, where the
parameters $a, b$ of the linear model correspond to the parameters
$\alpha, \beta$. I will return to Gull's findings in Section
\ref{ch:Cosmology:BHM}, but I first describe the more pragmatic,
empirical approach adopted by the SN community to address the problem
of bias associated with the use of Eq.~\ref{eq:muobsdef}.

\section{The standard $\chi ^2$-method }
\label{ch:Cosmology:chi2}

\begin{figure}
\centerline{\includegraphics[width=9cm]{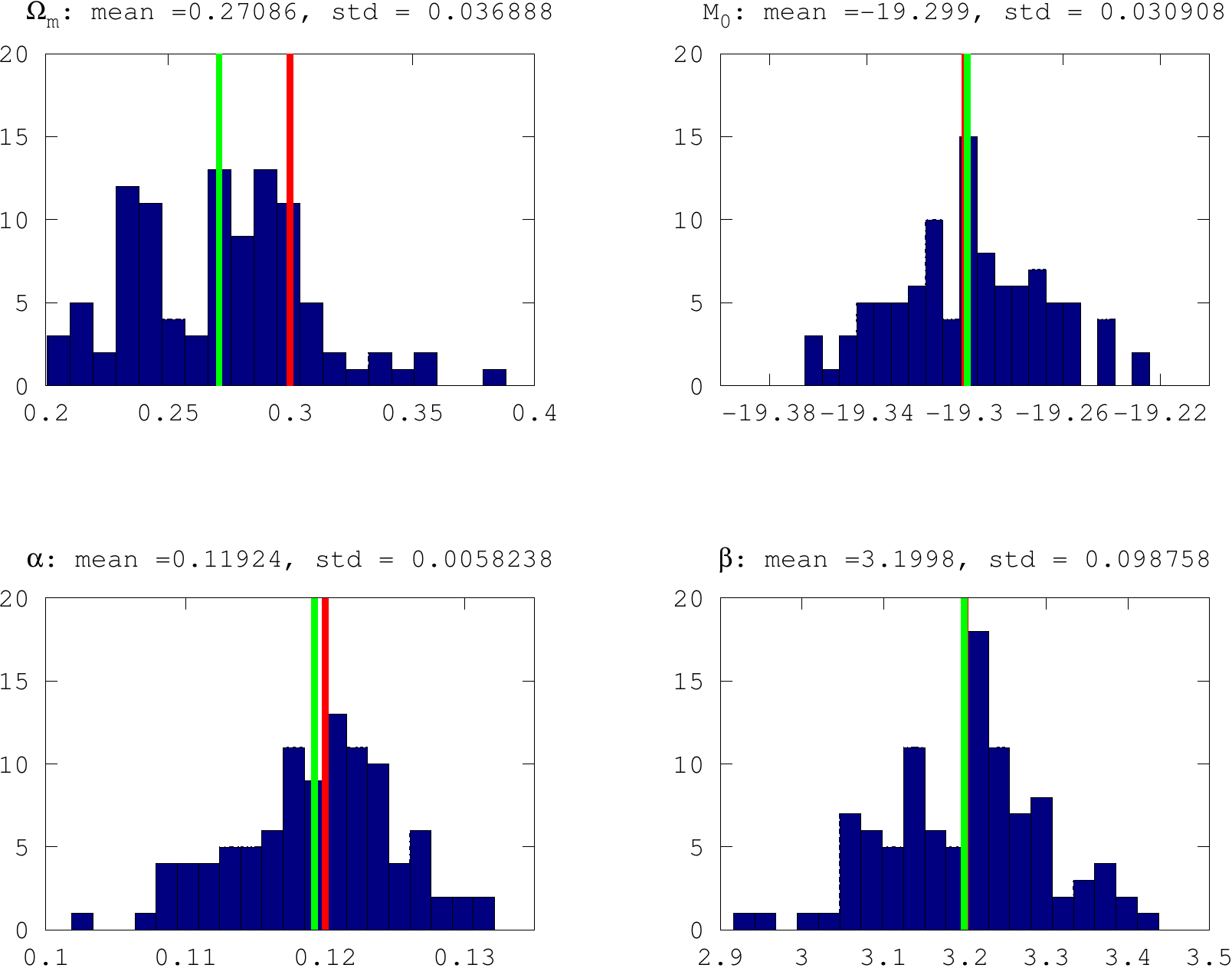}}
\caption{The same as Figure \ref{f:naive_large}, but for simulations analysed with the $\chi^2$-method.}\label{f:chi2_large}
\end{figure}

The precise methods used for the estimation of cosmological parameters
differ between SN consortia, but the main elements are common to all.
Central to the approach is the $\chi^2$-statistic, which is defined as
\be \label{eq:chisq}
\chi^2(\mathscr{C},\alpha,\beta, M_0,\sigma_\text{int}) = \sum_{i=1}^{N_{\rm SN}} 
\frac{[\mu_i^\text{obs}(\alpha,\beta,M_0)-\mu_i(\mathscr{C})]^2}{\sigma_i^2(\alpha,\beta,\sigma_\text{int})},
\ee
where $\sigma_{i}^2 $ is that given in Eq.~\ref{eq:sigmadef}. 

We see that this expression is merely the second term on the
right-hand side of Eq.~\ref{eq:norm_like_long}. The full expression
for the (minus) log-likelihood on the right-hand side of
Eq.~\ref{eq:norm_like_long} contains two competing terms: minimisation
of the second term ($\chi^2$) would favor large values of $\sigma_{i}$,
whereas minimisation of the third term would favor small values of
$\sigma_{i}$. In Eq.~\ref{eq:chisq} this last term is simply ignored.
Somewhat surprisingly, it has been established through simulation that
discarding this term removes the bias in the estimated parameters,
provided $\sigma_\text{int}$ is held fixed. It is fair to say that the
fundamental reason for this ``magical'' removal of the bias is not
well understood, even by those members of the SN community that use
the method on a regular basis.

Typically, the $\chi ^2$-function (Eq.~\ref{eq:chisq}) is minimized
simultaneously with respect to the cosmological parameters
$\mathscr{C}$ and the global SNIa nuisance parameters $\alpha$,
$\beta$ and $M_0$. This minimisation can, however, be performed using different
search algorithms (e.g.~MCMC techniques or grid searches) and the treatment
of $M_0$ (which is degenerate with $H_0$), in particular whether this
parameter is marginalised over analytically or numerically.

There remains, however, the issue of determining an appropriate value
for $\sigma_\text{int}$, which is usually performed as follows.  Once
the minimum value of $\chi ^2$ has been obtained, the value of
$\sigma_\text{int}$ is estimated by requiring that $\chi^2_{\rm
  min}/N_{\rm dof} \sim 1$; this process is usually iterated until
convergence is obtained.

The $\chi^2$-method has been fully tested and proven to be satisfactory for
cosmological parameter inference. Results from the toy example are
shown in Figure \ref{f:chi2_large}.  Nevertheless, the method does have a few problems:
\begin{enumerate}
\item The use of $\chi^2$ in Eq.~\ref{eq:chisq} is not statistically
  well-motivated, but is based only on empirical evidence and
  experience \citep{Gull}.

\item The global parameters $\alpha, \beta$ appear in both the
  numerator and denominator, since they act as both range and location
  parameters. Thus, the errors on $\alpha, \beta$ are not
  Gaussian. The informal test which states that $\chi^2_{\rm
    min}/N_{\rm dof} \sim1$ for a good fit model only holds in the
  Gaussian case. Hence its use cannot be justified here.

\item Since I use an unnormalised likelihood (without the second term
  on the right-hand side of Eq.~\ref{eq:norm_like_long}), one can not
  calculate the Bayesian evidence, and hence model selection is not
  possible. One can look on this problem differently: every model will
  fit the data equally well, since by construction $\sigma_{\rm{int}}$
  is determined by demanding that $\chi^2_{\rm min}/N_{\rm dof} \sim1$.

\item This method obtains only a best value for $\sigma_{\rm{int}}$ without
  any indication of the error on that value. 
\end{enumerate}

\section{BHM}
\label{ch:Cosmology:BHM}

\cite{Gull} solved the problem of fitting a straight line to data with
errors in both $x$ and $y$. The abstract of this paper is reproduced bellow:
\begin{quotation}
 `A Bayesian solution is presented to the problem of straight-line
 fitting when both variables $x$ and $y$ are subject to error. The
 solution, which is fully symmetric with respect to $x$ and $y$, contains
 a very surprising feature: it requires an informative prior for the
 distribution of sample positions. An uninformative prior leads to a
 bias in the estimated slope.'
\end{quotation}
In essence, this bias is the same as that which stopped us from using
the likelihood in Eq.~\ref{eq:norm_like} for estimating cosmological
parameters. 

Gull's proposed solution has two main steps:
\begin{enumerate}
\item Introduction of ``hidden variables'', which are the ``true''
  values of the measured quantities; these variables are nuisance
  parameters and will be integrated away in the end.
\item The imposition of an informative prior on these variables, which contains hyper-parameters that are allowed to vary simultaneously with the other parameters, and are then marginalised over. 
\end{enumerate}
These ideas were implemented in the context of SN cosmological
analyses in the BHM \citep{march11}. 
\begin{figure}
\centerline{\includegraphics[width=9cm]{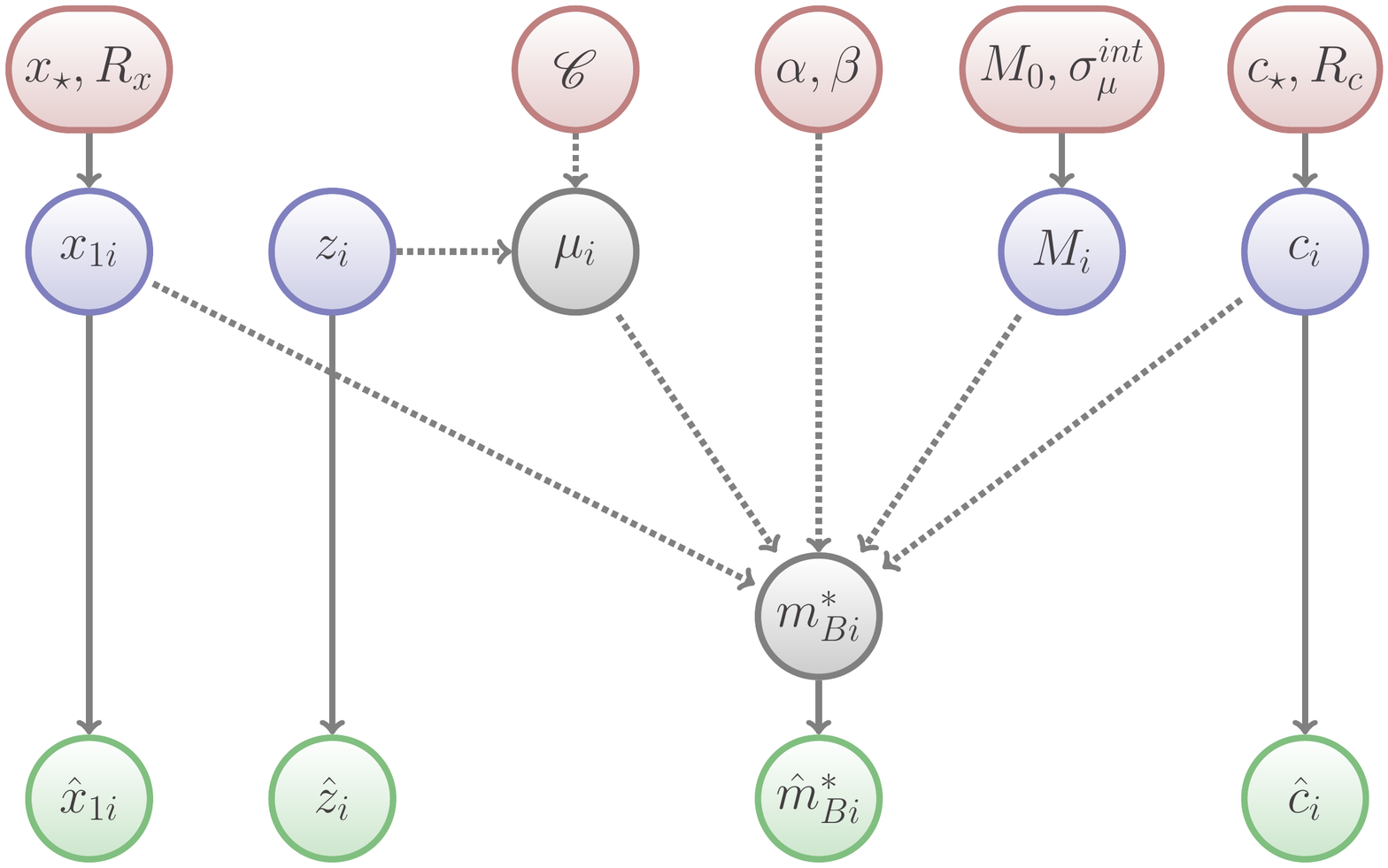}}
\caption{Graphical network showing the deterministic (dashed) and
  probabilistic (solid) connections between variables in the
  BHM. Variables of interest are in red, hidden (unobserved) variables
  are in blue and observed data (denoted by hats) are in
  green. Credit: \cite{march11}.}\label{f:BHM}
\end{figure}

In this method one considers a probability for the full catalogue of
SNe {\em simultaneously}, since the form of the BHM likelihood for a set
of SN observations ($i=1,2,\ldots, N)$ is {\em not} simply the
product of the likelihoods for each individual SN. 

One begins by considering the probability 
\begin{equation}
P \equiv \Pr(\hat{\mathbf{m}}_{B}^\ast,\hat{\mathbf{x}}_{1},\hat{\mathbf{c}},
\hat{\mathbf{z}},\mathbf{m}^\ast_{B},\mathbf{x}_{1},\mathbf{c},\mathbf{z},\mathbf{M}
|\mathscr{C},\alpha,\beta,
\mathbf{\sigma_\text{int}},\widehat{\mathbf{C}},\boldsymbol{\sigma}_{z}),
\label{eq:b1}
\end{equation}
where boldface symbols denote vectors containing the corresponding
parameters for each SN ($i=1,2,\ldots, N)$. I denote the parameters of
the model to be fitted by $\boldsymbol{\phi} =
\{\mathscr{C},\alpha,\beta,\sigma_\text{int}\}$, and those assumed
known by
$\boldsymbol{\psi}=\{\widehat{\mathbf{C}},\boldsymbol{\sigma}_{z}\}$.

The BHM likelihood function described in \cite{march11} is obtained by
marginalising Eq.~\ref{eq:b1} over the hidden variables, together with
other nuisance parameters describing the SN population. One can see
the relation between different types of variables in this model in
Figure \ref{f:BHM}. Thus, one obtains ${\cal
  L}_\text{BHM}(\boldsymbol{\phi}) \equiv
\Pr(\hat{\mathbf{m}}_{B}^\ast,\hat{\mathbf{x}}_{1},\hat{\mathbf{c}},\hat{\mathbf{z}}
|\boldsymbol{\phi},\boldsymbol{\psi})$ by marginalising as follows:

\begin{small}
\begin{equation}
{\cal L}_\text{BHM}(\boldsymbol{\phi})
=
\int\!\!\cdots\!\!\int {\rm d}\mathbf{m}_B^\ast\,{\rm d}\mathbf{x}_{1}\,{\rm d}\mathbf{c}\,{\rm d}\mathbf{z}\,{\rm d}\mathbf{M}\,
\Pr(\hat{\mathbf{m}}_{B}^\ast,\hat{\mathbf{x}}_{1},\hat{\mathbf{c}},
\hat{\mathbf{z}},\mathbf{m}^\ast_{B},\mathbf{x}_{1},\mathbf{c},\mathbf{z},\mathbf{M}
|\boldsymbol{\phi},\boldsymbol{\psi}). 
\label{eq:b1a}
\end{equation}
\end{small}

To perform this marginalisation, one assumes that: (i) the measured
redshift $\hat{z}_i$ is independent of $\hat{m}_{B,i}^\ast$,
$\hat{x}_{1,i}$ and $\hat{c}_i$; (ii) the true redshift $z_i$ is
independent of $M_i$, $x_i$, $c_i$, and (iii) the exact relationship
$\mu(z_i,\mathscr{C}) = m_{B,i}^\ast - M_i + \alpha x_{1,i} - \beta
c_i$ between the hidden variables holds. This enables one to write Eq.~\ref{eq:b1} as
\begin{eqnarray}
P = 
\Pr(\hat{\mathbf{m}}_{B}^\ast,\hat{\mathbf{x}}_{1},\hat{\mathbf{c}}|\mathbf{m}^\ast_{B},\mathbf{x}_{1},\mathbf{c},\widehat{\mathbf{C}})
\,\Pr(\hat{\mathbf{z}}|\mathbf{z},\boldsymbol{\sigma}_{z})\,
\Pr(\mathbf{M},\mathbf{x}_{1},\mathbf{c}|\sigma_\text{int})\,
\nonumber \\
\Pr(\mathbf{z})\,
\delta[\mathbf{m}_{B}^\ast - \mathbf{M} + 
\alpha \mathbf{x}_{1} -
\beta \mathbf{c} - \mathbf{\mu}(\mathbf{z},\mathscr{C})].
\label{eq:b2}
\end{eqnarray} 
The presence of the delta-function allows one to perform the integral
over $\mathbf{m}^\ast_B$ immediately. Moreover, the first two
probability distributions on the right-hand side are simply the
product of the corresponding distributions for each SN separately,
namely
\begin{align}
\Pr(\hat{\mathbf{m}}_{B}^\ast,\hat{\mathbf{x}}_{1},\hat{\mathbf{c}}|\mathbf{m}^\ast_{B},\mathbf{x}_{1},\mathbf{c},\widehat{\mathbf{C}})
& = \prod_{i=1}^N \frac{
\exp\left[-\tfrac{1}{2}(\hat{\mathbf{v}}-\mathbf{v})^{\rm t}
  \widehat{C}_i^{-1} (\hat{\mathbf{v}}-\mathbf{v}) \right]}
{|2\pi \widehat{C}_i|^{1/2}},\\ 
\Pr(\hat{\mathbf{z}}|\mathbf{z},\boldsymbol{\sigma}_{z})
& = \prod_{i=1}^N \frac{\exp\left[-(\hat{z}_i-z_i)^2/(2\sigma_{z,i}^2)\right]}
{(2\pi\sigma_{z,i}^2)^{1/2}},
\end{align}
where the vector $\mathbf{v}$ is defined by $\mathbf{v} = (\mathbf{m}_{B}^\ast,\mathbf{x}_{1},\mathbf{c})^\text{t}$ and similarly for $\hat{\mathbf{v}}$.

A key difference between the BHM and the standard $\chi^2$-method is
the specification of the prior
$\Pr(\mathbf{M},\mathbf{x}_{1},\mathbf{c}|\sigma_\text{int})$ on the
right-hand side of Eq.~\ref{eq:b2}. A common choice is to consider the
SNe as true standard candles by assigning $\Pr(M_i)=\delta(M_i-M_0)$,
while adopting a uniform normalised top hat distribution on each
remaining hidden variable $c_i$ and $x_{1,i}$
\citep{2005physics..11182D}, but this just leads to the naive
likelihood discussed in Section \ref{ch:Cosmology:naive_like}, with
$\sigma_\text{int}$ set to zero. I therefore impose a different prior
here.  It is again assumed separable in the sense that
$\Pr(\mathbf{M},\mathbf{x}_{1},\mathbf{c}|\sigma_\text{int})=
\Pr(\mathbf{M}|\sigma_\text{int})\Pr(\mathbf{x}_{1})\Pr(\mathbf{c})$,
but each of the distributions on the right-hand side does not
factorise into terms corresponding to individual SNe. This occurs
since the BHM introduces and marginalises over additional nuisance
hyper-parameters $M_0$, $x_\ast$, $R_x$, $ c_\ast$ and $R_c$
associated with the SN population, and described below. In particular,
one writes
\begin{align}
\Pr(\mathbf{M}|\sigma_\text{int}) & 
= \int \rm{d}M_0 \,\Pr(M_0)
\,\prod_{i=1}^N \Pr(M_i|M_0,\sigma_\text{int}), \label{eq:b4} \\
\Pr(\mathbf{x}_{1}) & 
= \int\!\!\!\int {\rm d}x_\ast \,{\rm d}R_x\, 
\Pr(x_\ast) \Pr(R_x) \,\prod_{i=1}^N \Pr(x_{1,i}|x_\ast,R_x), \label{eq:b5}\\
\Pr(\mathbf{c}) & 
= \int\!\!\!\int {\rm d}c_\ast\, {\rm d}R_c\, 
\Pr(c_\ast) \Pr(R_c) \,\prod_{i=1}^N \Pr(c_{i}|c_\ast,R_c), \label{eq:b6}
\end{align}
in which it is assumed that a number of probability distributions are
separable.  The prior distribution of each of the hidden variables
$M_i$, $x_{1,i}$, $c_i$, is assumed to be Gaussian, so that
\begin{align}
\Pr(M_i|M_0,\sigma_\text{int}) 
& = {\cal N}(M_0,\sigma_{\rm int}^2), \\ 
\Pr(x_{1,i}|x_\ast,R_x) 
& = {\cal N}(x_\ast,R_x^2), \\
\Pr(c_i|c_\ast,R_c) 
& = {\cal  N}(c_\ast,R_c^2). 
\end{align}
\begin{figure}
\centerline{
\includegraphics[width=3.7cm]{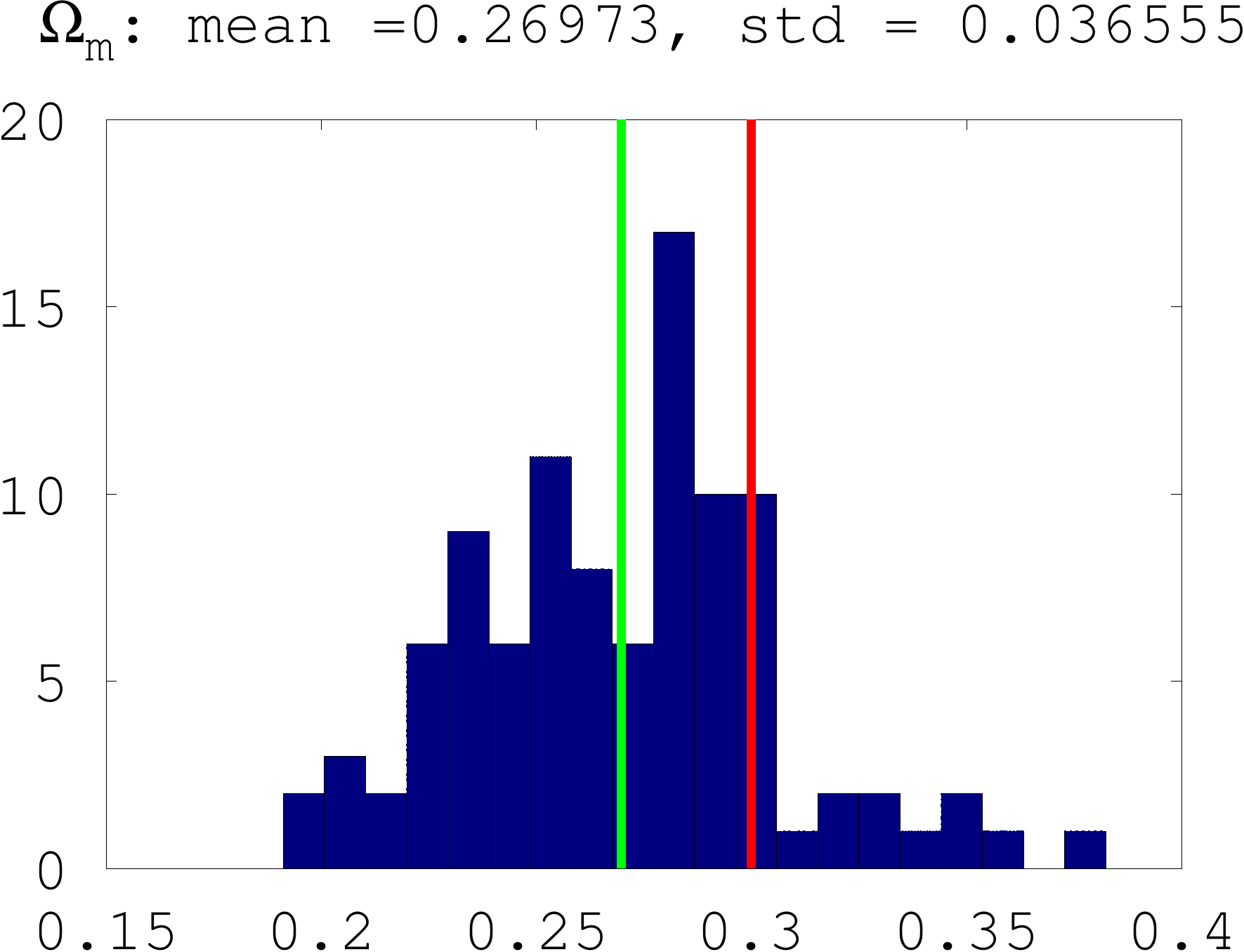}
\includegraphics[width=3.7cm]{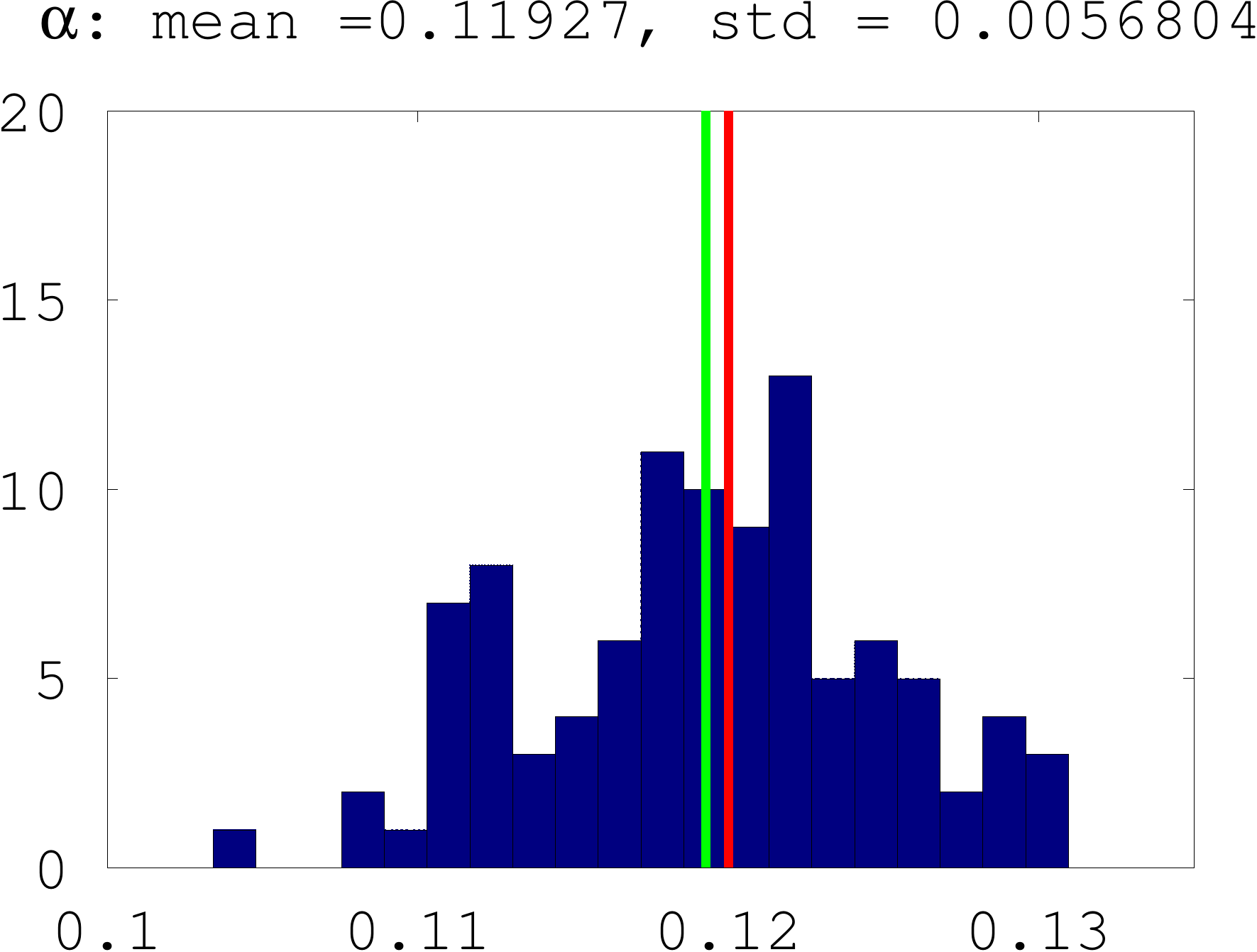}
\includegraphics[width=3.7cm]{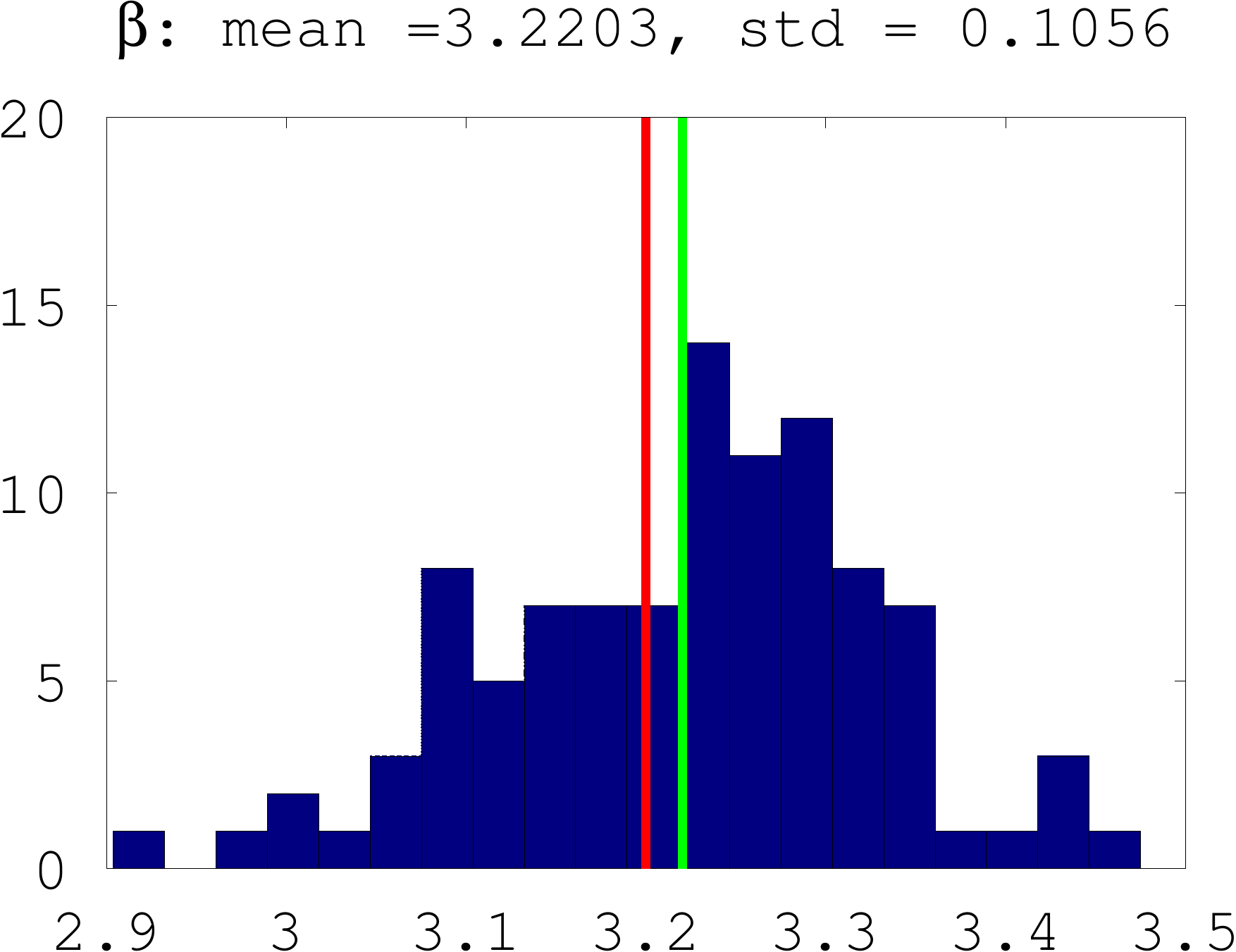}}
\caption{The same as Figure \ref{f:naive_large}, but for simulations analysed with the BHM. Note that the parameter $M_0$ is marginalised over analytically.}\label{f:BHM_large}
\end{figure}

Finally, one must also assign the priors on the nuisance
hyper-parameters $M_0$, $x_\ast$, $R_x$, $ c_\ast$, $R_c$, which are
taken to be
\begin{align}
\Pr(M_0)&={\cal N}(\bar{M}_0,\sigma_{M_0}^2),\\
\Pr(x_\ast)&={\cal N}(0,\sigma_{x_\ast}^2),\\ 
\Pr(c_\ast)&={\cal N}(0,\sigma_{c_\ast}^2), \\
\Pr(R_x)&=1/R_x,\\
\Pr(R_c)&=1/R_c,
\end{align}
where one assumes $\bar{M}_0=-19.3$~mag, $\sigma_{M_0}=2.0$~mag,
$\sigma_{x_\ast}=1$ and $\sigma_{c_\ast}=1$. The priors on $x_\ast$
and $c_\ast$ are taken as Gaussian because this is the maximum-entropy
prior on variables that can take positive and negative values, and for
which one has an expectation for the mean and variance. In this case a
mean of zero and a standard deviation of unity is appropriate.  Since
the widths $R_x$ and $R_c$ are non-negative scale parameters, we adopt
the non-informative Jeffreys prior on them.

All the necessary integrals in Eq.~\ref{eq:b1a} and Eqs.~\ref{eq:b4}-\ref{eq:b6} are Gaussian, except those over $R_x$ and
$R_c$. Thus, March et al. (2011) integrate analytically to obtain
a final expression for the likelihood in terms of an integral over
only $R_x$ and $R_c$; these parameters are therefore added to the
parameter vector $\boldsymbol{\phi}$, sampled from and marginalised out
numerically to recover ${\cal L}_\text{BHM}(\boldsymbol{\phi})$ defined in Eq.~\ref{eq:b1a}.

One can see from Figure \ref{f:BHM_large}, that the BHM shows the same level of precision for parameter estimates as the standard $\chi^2$ method,
so both methods yield unbiased parameter constraints. For the BHM, however, one
obtains two additional benefits: (i) the full probability distribution
for $\sigma_\text{int}$; and (ii) the ability to perform consistent Bayesian inference, including model selection. In Paper I, I compare the $\chi^2$-method and BHM in much more detail. I applied both of these methods to more realistic examples than the toy example used in this chapter and to real data (see Paper I and Section \ref{ch:paper1}).
\begin{figure}
\centerline{\includegraphics[width=10cm]{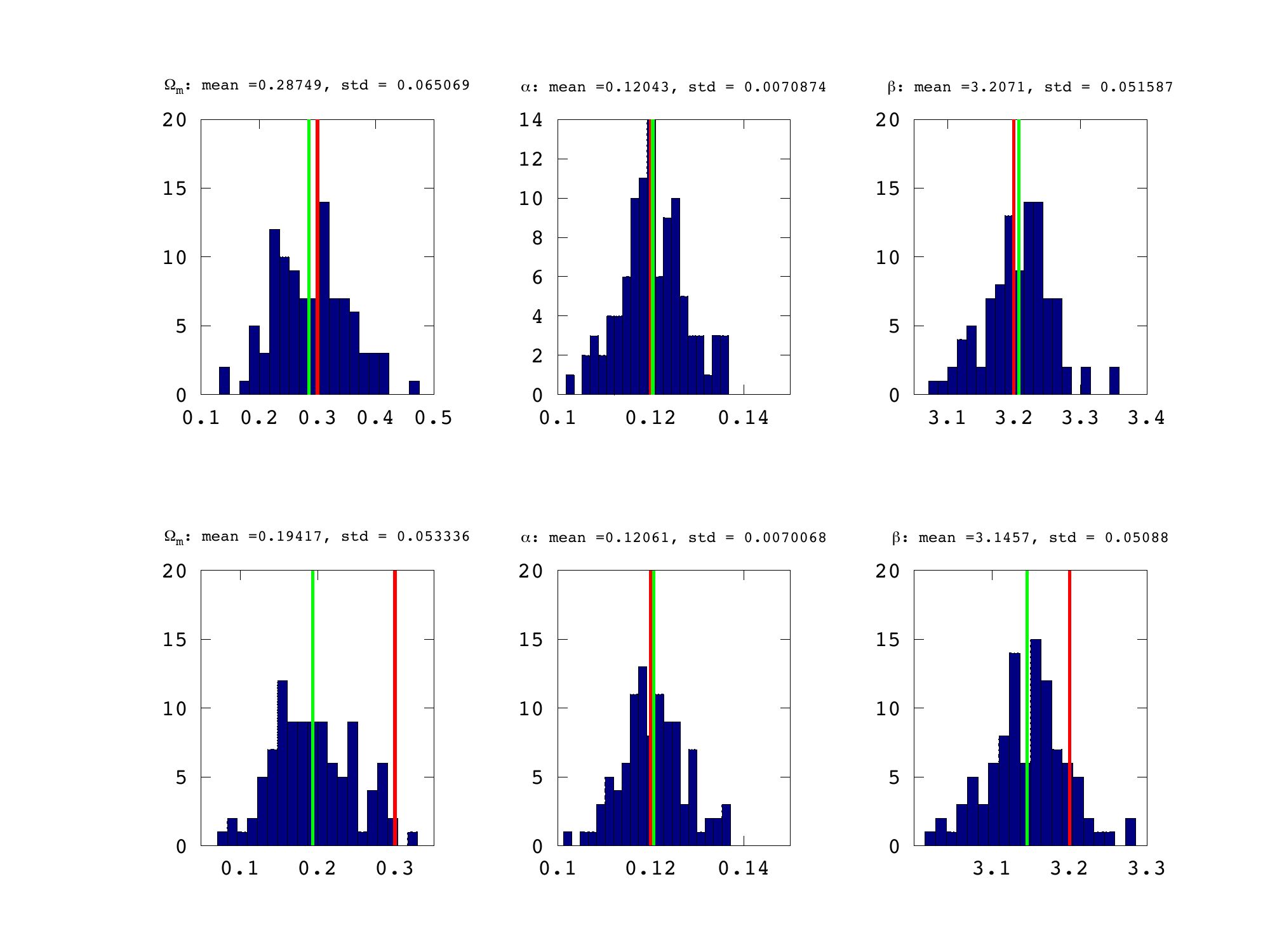}}
\caption{The same as Figure \ref{f:BHM_large}, but for SNe with $x_1(z>0.5) = x_1(z<0.5) +1$, analysed using the $\chi^2$-method (top row) and the BHM (bottom row).}\label{f:2bins}
\end{figure}

In spite of these advantages, one should also mention some
shortcomings of BHM. First, a key difference between BHM and the
standard $\chi^2$-method is that ${\cal L}_\text{BHM}$ does not depend
on $M_0$, since this parameter is marginalised out
analytically. Similarly, so too are the hidden variables $M_i$, which
correspond to the ``true'' absolute magnitude of each SN.  Consequently,
one cannot calculate the residuals in Eq.~\ref{eq:deltamu} for each SN.
This can make it difficult to compare the BHM with other methods, since
visual inspection of the residuals often plays an important role.  A
further, and perhaps more critical, shortcoming of the BHM is the
assumption that the hidden variables $c_i$ and $x_{1,i}$ do not depend
on redshift; this is a very crude approximation that is almost
certainly not true. In the event that these quantities do evolve with
redshift, one may show using simulations that the BHM can begin to produce
biased parameter constraints, whereas the standard $\chi^2$-method is
more robust to this effect, even though the formulation does not
explicitly allow for this eventuality; this is illustrated in Figure
\ref{f:2bins}.

\section{Generalised Bayesian Likelihood}
\label{ch:Cosmology:GBL}

The above shortcomings of the BHM result from the assumption of simple
(redshift-independent) Gaussian priors on the hidden variables and
nuisance parameters, which is necessary in order to marginalise over
them analytically.  Consequently, it is of interest to retain the
overall structure of the BHM, but to replace the analytical
marginalisation with a numerical one. This results in a larger
computational burden, but allows for the imposition of more realistic
priors on the hidden variables, including redshift dependence, and
provides access to best-fit values (indeed full marginal posterior
distributions) of the hidden variables.

It therefore seems natural to improve upon the BHM as follows, to produce 
a Generalised Bayesian Likelihood.
\begin{enumerate}
\item Integrate Eq.~\ref{eq:b1a} and Eqs.~\ref{eq:b4}-\ref{eq:b6} numerically by sampling from the set of
  hidden parameters $x_{1,i}$, $c_i$, $z_i$, $M_i$ $(i=1,2,\ldots, N)$
  and nuisance hyper-parameters $M_0$, $x_\ast$, $R_x$, $ c_\ast$,
  $R_c$ (in addition to the parameters of interest $\bphi$), and
  numerically (as opposed to analytically) marginalising over them.
\item Allow for more realistic priors on the hidden and nuisance
  parameters than the simple separable Gaussian forms assumed in the BHM,
  in particular including the possibility of redshift dependence of
$c_i$ and $x_{1,i}$.
\end{enumerate} 
The resulting approach does, however, carry with it a substantially
increased computational burden compared to the BHM, since the dimension of
the parameter space to be sampled is increased by $4N+5$, where $N$ is
the number of SNe being analysed. Even for existing SN data-sets,
such as Union2 with over 600 SNe, this leads to a parameter space with over
2000 dimensions. To sample from a posterior distribution of this
dimensionality and evaluate the evidence exceeds the current
capabilities of the widely used Bayesian inference code
\textsc{MultiNest}.  Therefore alternative methods are
required. Provided the posterior distribution is relatively benign,
i.e. smooth, unimodal and without very pronounced tails, then MCMC
methods, such as Gibbs or Hamiltonian sampling, can produce reliable
parameter estimates, and be combined with thermodynamic integration to
evaluate the evidence. Another possibility is to use alternative forms
of nested sampling, such as the DNest algorithm
\citep{2010ascl.soft10029B}, which is considerably less efficient than
MultiNest up to around 50 dimensions, but is capable of producing
posterior samples and evidence estimates in spaces with several
thousand dimensions. I am currently investigating these various
alternative sampling methods, and plan to present the Generalised
Bayesian Likelihood and its application to real and simulated SN data
in a forthcoming publication.


\chapter{Gravitational lensing of SNe}
\label{ch:lensing}
 
\epigraph{`Light thinks it travels faster than anything but it is
  wrong. No matter how fast light travels, it finds the darkness has
  always got there first, and is waiting for it.'}{\textit{Terry
    Pratchett}}


General relativity predicts that light rays are bent around massive
bodies or, more generally, undergo deflections when they traverse a
region in which the gravitational field is inhomogeneous. This effect is relevant to the study of SNe since the mass distribution along the line-of-sight to a SN will cause a gravitational lensing effect. The discussion in the previous chapter, and indeed in most papers on SN cosmology, ignores this effect. Nonetheless, such gravitational lensing can be used to constrain the form of the galaxy dark matter haloes along the lines-of-sight to the SNe, as I show in Paper II and Section~\ref{ch:paper2}.  In this
chapter I therefore give a brief description of some of the basic features of gravitational lensing.
 \section{Lens equation}\label{ch:lensing:theory}
 
  \begin{figure}
\includegraphics[width=13cm,height=6cm]{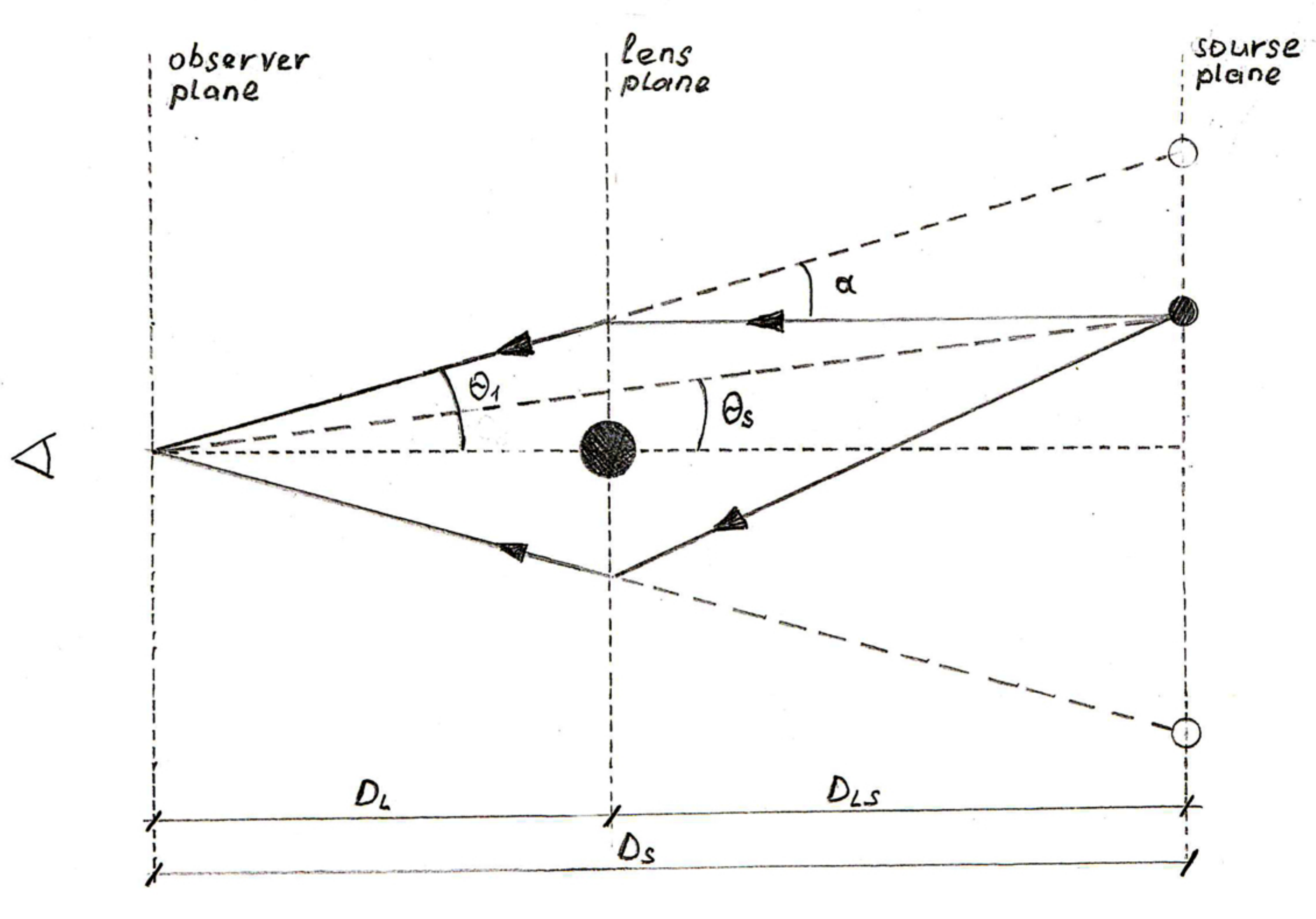}
\caption{Gravitational lens geometry.}\label{fig:lens_eq}
\end{figure}

Ray-tracing in curved spacetime, as illustrated in
Figure \ref{fig:lens_eq}, can be described by a lens equation,

 \begin{equation}
 \vec{\theta}_\text{S} =  \vec{\theta}_1 - \frac{D_{\text{LS}}}{D_\text{S}} \alpha,
 \label{eq:lens}
 \end{equation} 
where $\vec{\theta}_\text{S}$ is the angle between source and the lens that
would be observed in absence of lensing, $\vec{\theta_1}$ is the
observed angle between image and lens, and $\alpha$ is the deflection
angle; all the distances are angular-diameter distances $D_\text{A}$, since these are defined such that 
$D_\text{A} = \ell/\Delta\Theta$, where $\ell$ is the proper transverse distance and $\Delta\Theta$ is the angle it subtends at the observer.
 In
our expanding Universe, for objects at redshift $z_A$ and $z_B$, the angular-diameter distance is given by
\begin{equation}
 D_{z_A z_B} =  \frac{1}{(1+z_B) H_0} \int_{z_A}^{z_B} \frac{dz}{\sqrt{\Omega_{\text{m},0} (1+z)^3 + \Omega_{\Lambda,0}} },
 \label{eq:ADD}
 \end{equation}
 where I have set $c=1$. 
 
In the case of a point mass lens, with mass M,  Eq.~\ref{eq:lens} becomes
 \begin{equation}
 \theta_\text{S} =  \theta_1 - \frac{D_{\text{LS}}}{D_\text{S} D_\text{L}} \frac{4 G M}{\theta_1}.
 \label{eq:pointmass}
 \end{equation} 
In the very extreme case in which the source and lens are collinear, the
source will be lensed into an Einstein ring and the angular separation
is given by the Einstein angle:
\begin{equation}
 \theta_\text{E} = \sqrt{  \frac{4 G M D_{\text{LS}}}{D_\text{S} D_\text{L}}  }.
 \label{eq:Einrad}
 \end{equation} 
  Thus for each lens one may define a critical surface mass density, above which
 one obtains an Einstein ring or multiple images; this is given by
 \begin{equation}
 \Sigma_\text{c} = \frac{1}{4\pi G}  \frac{D_\text{S} D_\text{L}}{ D_{\text{LS}}} ,
 \label{eq:sigc}
 \end{equation} 
 
 Now, if one considers a more general lens than a point mass, for any
 surface density $\Sigma (\theta_1) $, which is obtained by projecting
 the matter distribution on to the lens plane, one can introduce
 the dimensionless scaled surface mass density, or convergence
 $\kappa (\vec{\theta_1)}$, given by
    \begin{equation}
\kappa (\vec{\theta}_1)= \frac{\Sigma  (\vec{{\theta}_1})}{ \Sigma_\text{c}} .
 \label{eq:conver}
 \end{equation} 
  In other words, the convergence describes the focussing by the lens
  of the light emitted by the source. This focussing causes the
  source to appear larger. According to Liouville's theorem
  (conservation of the phase-space density of the photons emitted by
  the source), the increase of size will lead to increase of brightness.
  At the same time, distortion by twisting of the light rays in the
  lens can occur. This will lead to shearing of the image shape. 

To describe both phenomena one can introduce the lens map
 \[ A = \left( \begin{array}{cc}
1-\kappa -\gamma_1 & -\gamma_2 \\ -\gamma_2 &1-\kappa
+\gamma_1 \end{array} \right),\] where $\gamma_1 = \gamma \cos(2\phi)$
 and $\gamma_2 = \gamma \sin(2\phi)$, with $\gamma$ being the
 ellipticity and $\phi$ is a position angle for an initially circular
 source that has been lensed into an ellipse.
In terms of the convergence and shear one can define the important quantity known as the magnification:
    \begin{equation}
\mu = \frac{1}{(1-\kappa )^2 -\gamma^2},
 \label{eq:magn}
 \end{equation}
 which corresponds to the ratio of the image area to the source area.
 
 \section{Types of gravitational lensing}\label{ch:lensing:types}
 
 The obvious case of interest is when the source is within the
 Einstein angle of the lens and multiple images, arcs, or even
 distinct parts of an Einstein ring appear; this is called strong
 gravitational lensing. Although predicted long before, the first
 multiple-image system was discovered by 
 \citet{1979Natur.279..381W}.

Observing multiple images of the same source, one can estimate the
lens matter density distribution from simple Euclidean space
calculations. One can also use the fact that some of the sources
vary with time, so the multiple images could also vary with time. Time delays of the multiple images are another powerful mechanism which can be used
to calculate the Hubble constant.

Strong lensing is often divided into subcategories, depending on the
angular resolution of the images. Using the point source approximation
in Eq. \ref{eq:Einrad} we can get a feeling for the amount of the
lensing in typical astronomical situations:
\begin{eqnarray}
\theta_\text{E} = 0.9 \sqrt{\left(\frac{M}{{M_{\odot}}}\right)\left(\frac{10~\text{kpc}}{ D} \right)}  ~~\text{milliarcseconds}
\nonumber \\
= 0.9 \sqrt{\left(\frac{M}{{10^{11} M_{\odot}}}\right)
\left(\frac{ \text{Gpc}}{ D}\right)}  ~~\text{arcseconds}
 \label{eq:magn}
\end{eqnarray}

The above estimates relate, respectively, to the two most commonly
occurring situations: microlensing  and macrolensing. Microlensing is interesting since lenses range from
about the mass of a planet to the mass of a star. Microlensing is
commonly used to: (i) constrain the nature of dark matter, for
example in searches for MACHOs (massive compact halo objects); (ii)
detect extrasolar planets; (iii) constrain the structure
of the Milky Way disk, and do much more.  Macrolensing, with
separations of typically arcseconds, is the range where most of the
collected images occur, so macrolensing is often a synonym for strong
lensing itself.

Depending on the alignment of the lens and source, weak lensing could
appear. Calling this lensing ``weak'' we should remember that it is not necessarily less important. Usually, weak lensing
occurs when the lens is located outside the Einstein radius, and
compared to strong gravitational lensing, results in small
magnifications and small image distortions, which makes it often
impossible to detect it without a priori knowledge of the source
properties. While strongly lensed images often can tell us about the
structure of the lens, weak gravitational lensing images allow us only to probe the
statistical properties of the matter distribution on the
line-of-sight. Nevertheless, weak lensing is one of the most common
effects observed in the Universe. At some level, all objects that emit
light and are observed at Earth are affected.

 \section{SNe through gravitational telescopes}\label{ch:lensing:examples}

Zwicky had already proposed by the 1930s to use galaxies as
gravitational lenses. However, it was not until
the late 1970s that the first gravitationally lensed objects were
detected. From the very beginning, SNe have been taken into account in
the calculation of the lensed images as being one of the background
sources \citep{1964MNRAS.128..307R}. Since that time several
applications of lensed SNe have been explored.

Unfortunately, no multiply imaged SN has yet been observed.
Nevertheless, magnification/demagnification by large-scale structure
along the line-of-sight can be used to estimate weak lensing effects.

This effect is a statistical “nuisance” in obtaining cosmological inferences
from SNIa. There are two ways to correct for it: (i) assume some
overall magnification distribution (e.g. \citealt{2005ApJ...631..678H, 2008ApJ...673..657M}) or (ii) calculate convergence along each SN line-of-sight using simplified scaling relations applied to the nearby
foreground galaxies observed (e.g. \citealt{2008A&A...487..467J, jonssonSNLS, kronborg10, 2013MNRAS.429.2392K, 2014ApJ...780...24S}). The last method can also be used to estimate
parameters of the objects on the line-of-sight, such as dark matter
halo parameters, which I have done in Paper II and Section \ref{ch:paper2}.

In recent wide-field surveys, a few SNe have been found that are
detectably magnified. Among them are three gravitationally lensed SNe
(SN CLO12Car, SN CLN12Did, and SN CLA11Tib) behind CLASH clusters
\citep{2014ApJ...786....9P}.  The reason for lensed SNe to be such
rare events is that the SN has to be precisely aligned with a
gravitational lens.


\chapter{Summary of my results}
\label{ch:Summ}

\epigraph{`In God we trust. All others must bring
  data.'}{\textit{W. Edwards Deming}}

My main research results are in the five papers included in this
thesis, which describe different aspects of how observations of
SNe can be used in astrophysics and cosmology.  In particular,
the papers focus on: the use of SNIa in
constraining the background cosmological model describing our
Universe; the use of gravitational lensing of distant SNIa by foreground cosmic structure to constrain the nature of
dark matter haloes of galaxies; the determination from photometric
lightcurves of whether or not a given SN is of type Ia, and
hence can be used in cosmological inference; checking for consistency between
different SN data-sets within large compilations. In this chapter, I will
summarise the results from my papers, give an update on the already published
results and describe findings that are yet to be published.

\section{SNe and cosmology}
\label{ch:paper1}

In Chapter \ref{ch:Cosmology}, I have already presented a general
account of different methods for cosmological inference from SNIa
data. In Paper I, I present a detailed comparison of the standard
$\chi^2$-methodology and the recently proposed BHM applied to
SNIa lightcurves fitted with the SALT2 technique.  I
described these two methods in Section \ref{ch:Cosmology:chi2} and
\ref{ch:Cosmology:BHM} respectively. Through the analysis of realistically
simulated SN data-sets, I obtain similar results in Paper I to those
obtained from the toy example described in Chapter~\ref{ch:Cosmology}.

\subsection{Biasses in the $\chi^2$-method and BHM}

\begin{figure}
\begin{center}
\includegraphics[height=4.25cm]{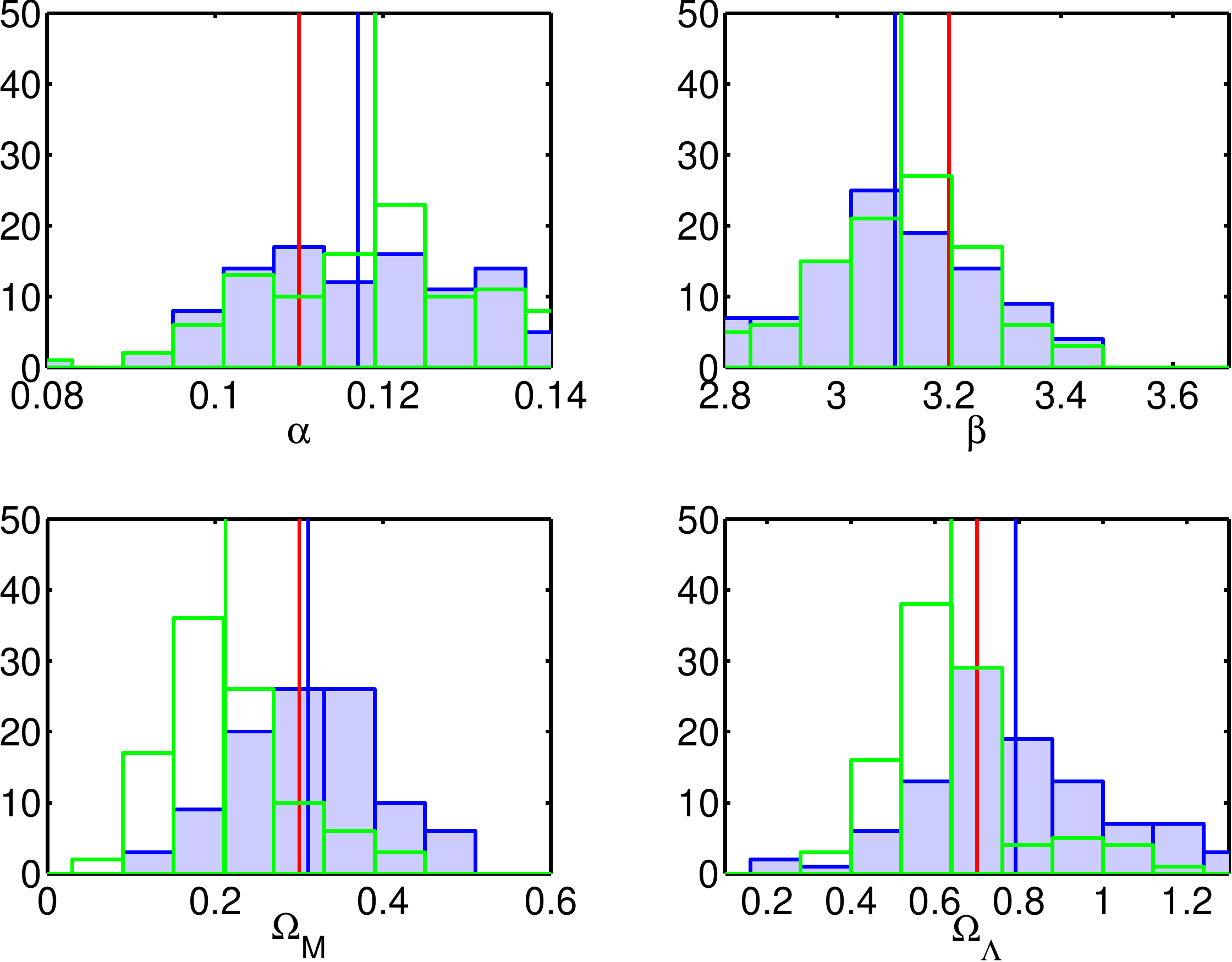}~~~~~~~~~~~~~
\includegraphics[height=4.25cm]{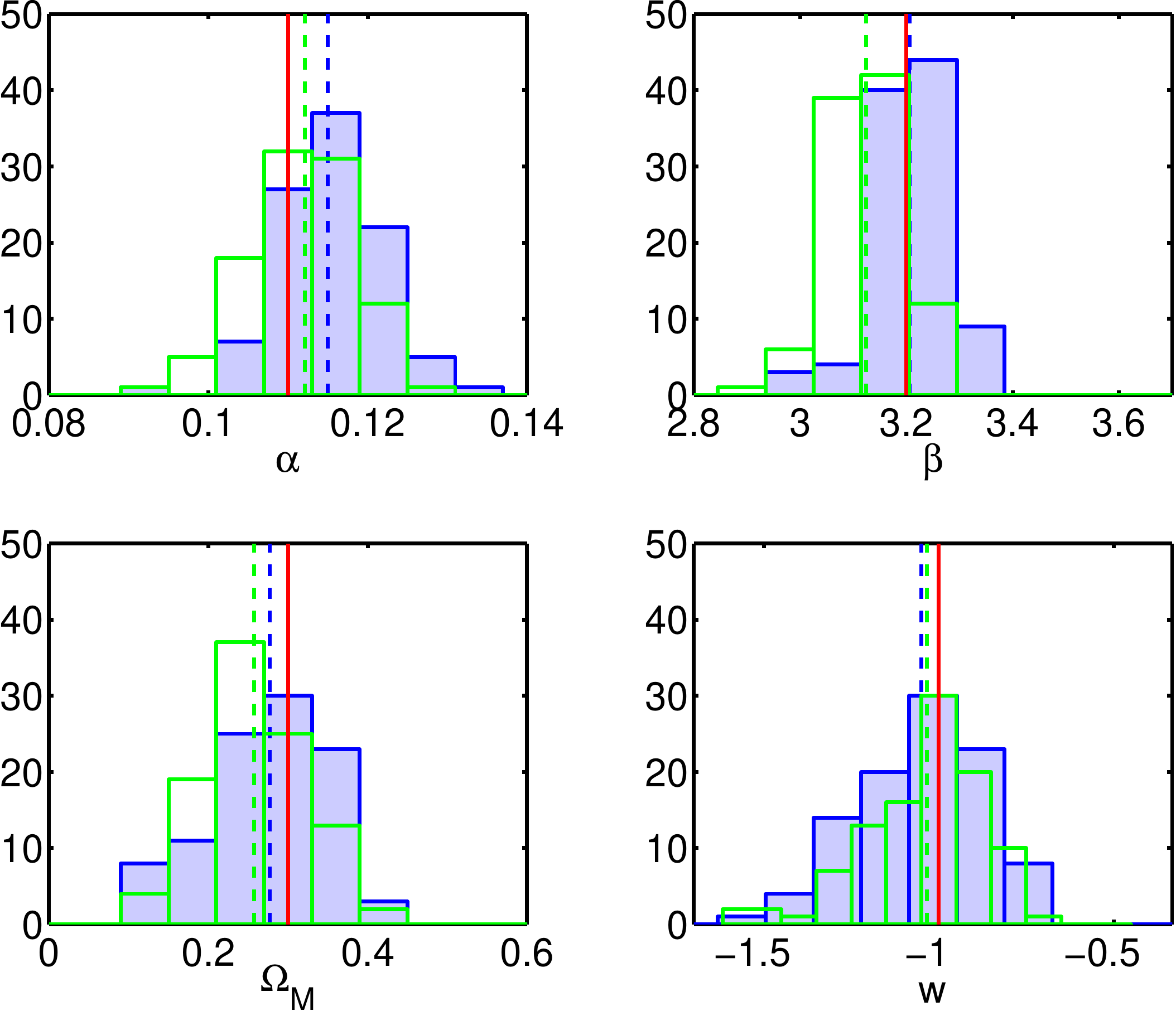}
\caption{ Left (right) panel: sampling distributions derived from a $\Lambda$CDM analysis of 100 simulated SNLS (`cosmology') data-sets.  
  The histograms show the point estimates for the SNIa global parameters $\alpha,\beta$ and the cosmological
  parameters $\Omega_{\rm m,0},\Omega_{\Lambda,0}$, inferred using the
  BHM (filled blue histograms) and the chi-square method (empty green histogram). Blue and green
  vertical lines show the mean values of the point estimates, solid red
  vertical lines show the value of the true (i.e. model input)
  parameter used to simulate the data. From Paper I. }\label{fig:lcdm-snls3a}
\end{center}
\end{figure}

In Paper I, I establish that small biasses in the recovery of
cosmological parameters occur for both the standard $\chi^2$-methodology and the BHM. These biasses are not the same for each
method, however, leading to discrepancies between their results, which
are greatest when analysing just a single survey, such as SNLS; see
the left panel in Figure \ref{fig:lcdm-snls3a}. I find that the BHM
offers a modest advantage over the $\chi^2$-method in that it produces
slightly less biassed estimates of the parameters; this is
particularly true for $\Omega_{\rm m,0}$. The biasses on the
$\Omega_{\rm m,0}$ estimates produced by the two methods are in
opposite directions, which results in approximately a $2\sigma$ discrepancy for
any given realisation of SNLS-type data. Most interestingly, I find
this to be the case for the real SNLS data-set. However, in
simulations, one finds that increasing the redshift range of the data-set reduces the discrepancy between the methods; see the right panel
of Figure \ref{fig:lcdm-snls3a}. As more higher and lower redshift SNIa
are added to the sample, the estimates of the cosmological parameters
of interest from the two methods begin to converge. Nonetheless, this
conclusion may be premature; although we generated state-of-the-art
simulations, they contained no redshift dependence of the SN
properties.

\subsection{BHM with redshift-dependent stretch and colour corrections}
\label{ch:paper1:2bins}
\begin{table*}
\begin{center}
\begin{footnotesize}
\begin{tabular}{l|rrr|rrr}
\hline\hline
Data-set & \multicolumn{3}{c}{Union2} & \multicolumn{3}{c}{S11}\\
&  $\log(\mathcal{Z}_{H_{12}})$ & $\!\!\log(\mathcal{Z}_{H_{22}})$ & $\!\!\log(\mathcal{Z}_{H_{13}})$ & $\log(\mathcal{Z}_{H_{12}})$ & $\!\!\log(\mathcal{Z}_{H_{22}})$ & $\!\!\log(\mathcal{Z}_{H_{13}})$\\
\hline
ESSENCE &  9.21 & 10.18 & 8.57 & --- & --- & --- \\
HST &  0.80 & $-0.16$ & 0.23 & $-0.30$ & $-0.63$ & $-1.07$ \\
SDSS &  $-0.25$ & $-0.77$ & $-1.13$ & $-0.33$ & $-0.61$ & $-1.27$ \\
SNLS  & 2.08 & 1.34 & 1.41 & 4.87 & 3.62 & 3.68  \\
CfA & 0.00 & $-0.20$ & $-0.95$ & $-0.01$ & $-0.32$ & $-1.02$ \\
Low-$z$ &  --- & --- & --- & 0.01 & $-0.30$ & $-0.97$ \\[1mm]
\hline\hline
\end{tabular}
\caption{$\log$-evidence values for models $H_{n_{\alpha}n_{\beta}}$
  ($n_{\alpha}$ and $n_{\beta}$ being the number of redshift bins for
  stretch and colour parameter populations respectively) relative to the model
  $H_{11}$. Errors on these $\log$-evidence values are all around
  $0.09$.}\label{tab:re}
\end{footnotesize}
\end{center}
\end{table*}

As mentioned in Chapter~\ref{ch:Cosmology}, it is possible for the
stretch and colour parameter populations to evolve with redshift; indeed some
studies have already tried to explore this possibility
(\citealt{2009ApJS..185...32K, 2011ApJS..192....1C}).
\citet{2009ApJS..185...32K}, in their Section 10.2.3, present evidence
for the redshift evolution of the color parameter $\beta$ for the
SALT2 lightcurve fitting algorithm for different combinations of
samples in the full data-set. This question has been revisited in
Section 5.7 of \citet{2011ApJS..192....1C}; they find that using later
versions of SALT2 results just in marginal evidence for the evolution
of the parameter $\beta$, but they do not discuss how this changes
with different combinations of data-sets. I revisit this question by
applying the BHM to the data from Union2 and
\citet{2011ApJ...737..102S} (hereafter S11). I divide these SNe according to the
telescope with which they have been observed; in particular I divide
the SNe into the following subsets: ESSENCE, HST, SDSS, SNLS, CfA and a
compilation of low-$z$ SNIa measurements.

In order to check whether these data-sets have redshift evolution in the
stretch and colour parameters, I modify the BHM to allow for this evolution by
introducing multiple contiguous redshift bins for the stretch and
colour parameter populations. SNe within different stretch (colour) redshift
bins are allowed to have different values of $\alpha$ ($\beta$) and
$R_x$ ($R_c$). This is achieved by allowing the priors on $\alpha$
($\beta$) and $R_x$ ($R_c$) to be completely independent in different
stretch (colour) redshift bins. The lower (upper) limits on first
(last) redshift bin are set to $z_{\rm min}$ ($z_{\rm max}$), where
$z_{\rm min}$ and $z_{\rm max}$ are the minimum and maximum SN
redshifts in the catalogue. The other end points of redshift bins are
set as free parameters which are estimated along with other BHM
parameters. The number of stretch and colour redshift bins,
represented by $n_{\alpha}$ and $n_{\beta}$, are estimated by Bayesian
model selection, done by analysing models with different values of
$n_{\alpha}$ and $n_{\beta}$, starting with $n_{\alpha} = n_{\beta} =
1$, and picking the model with the highest value for the Bayesian
evidence. We denote these models by $H_{n_{\alpha}n_{\beta}}$.

 The log-evidence values for models $H_{12}$, $H_{22}$ and $H_{13}$,
 all with respect to the base model $H_{11}$, are given in
 Table~\ref{tab:re}. It is evident from this table that $H_{12}$
 ($n_{\alpha} = 1$, $n_{\beta} = 2$) is the preferred model for most
 data-sets. Even when $H_{12}$ is not the most preferred model,
 preference for other models over it is not very strong. Thus, in
 agreement with the previous studies mentioned above, we see that our
 Bayesian model selection approach also provides evidence for some
 evolution with redshift of the $\beta$ parameter. This is an
 interesting finding in its own right, but also has important
 consequences for using the standard BHM, which assumes no redshift
 dependence for any of the parameters. Clearly, this assumption is
 broken by the real data, and so one must take care in interpreting
 results obtained using the ``vanilla'' BHM. Although the modification
 to the BHM introduced above has its uses, a more
 statistically-principled approach to extending the BHM is to allow
 the priors on the colour and stretch parameters to depend on
 redshift.  This generalised Bayesian likelihood method is discussed
 in Section~\ref{ch:Cosmology:GBL}.

\section{SNe and cosmic structure}
\label{ch:paper2}

In Paper II, I present a Bayesian statistical methodology for
constraining the properties of dark matter haloes of foreground
galaxies that intersect the lines-of-sight towards SNIa. This builds
on the BHM used in Paper I.

In this approach, the parameters of interest are those describing the
dark matter haloes assumed to exist around the known galaxies along
the lines-of-sight to the SNIa. My method yields an effective
likelihood function, which gives the probability of obtaining the
observed SNIa data (i.e.~the parameter values obtained in SALT2
lightcurve fits) as a function of these parameters. Once appropriate
priors have been placed on the parameters, the full posterior
distribution is explored using {\sc MultiNest} to obtain parameter
constraints and also calculate the Bayesian evidence for use in model
comparison.

I investigate two different models for the density profile
$\rho(\mathbf{r})$ of the dark matter halo: the truncated singular
isothermal sphere (tSIS) and the Navarro--Frenk--White (NFW) profile (\citealt{navarro97}),
both of which are widely used models in astronomy. My results for the
SIS profile are presented in Paper II. The NFW results were also
included in the first version of Paper II, but were subsequently
removed during the refereeing process, since it was necessary to add
significant further discussion of the methodology, which resulted in
the paper becoming too long. I summarise below the main finding for
both the SIS and NFW profiles; note that the full version of the
original paper is still online as the 1st arXiv version of Paper II.

\subsection{SIS model}
\label{ch:paper2:SIS}

\begin{figure}
\begin{center}
\includegraphics[width=6cm,height=7cm]{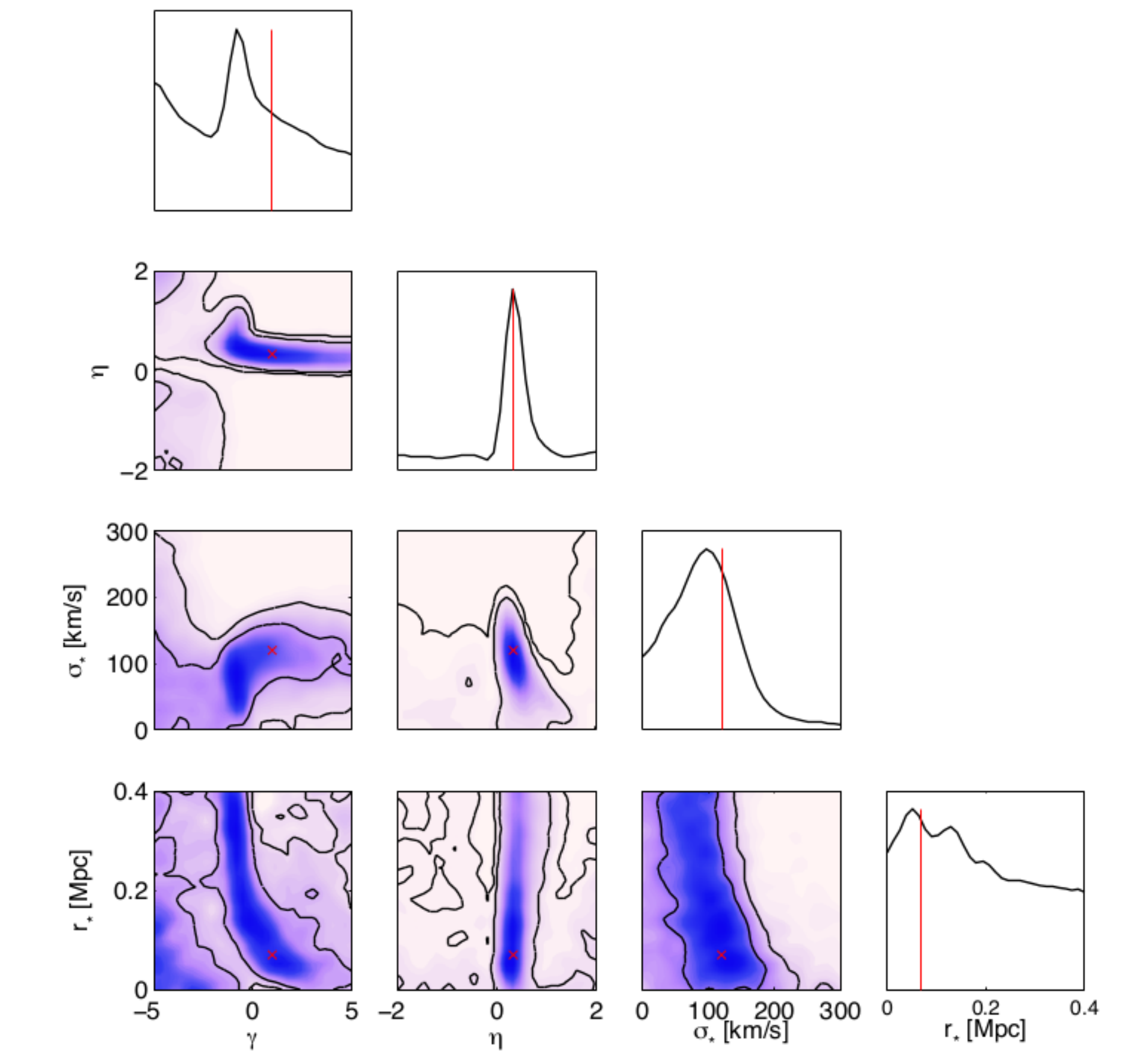}
\includegraphics[width=6cm,height=7cm]{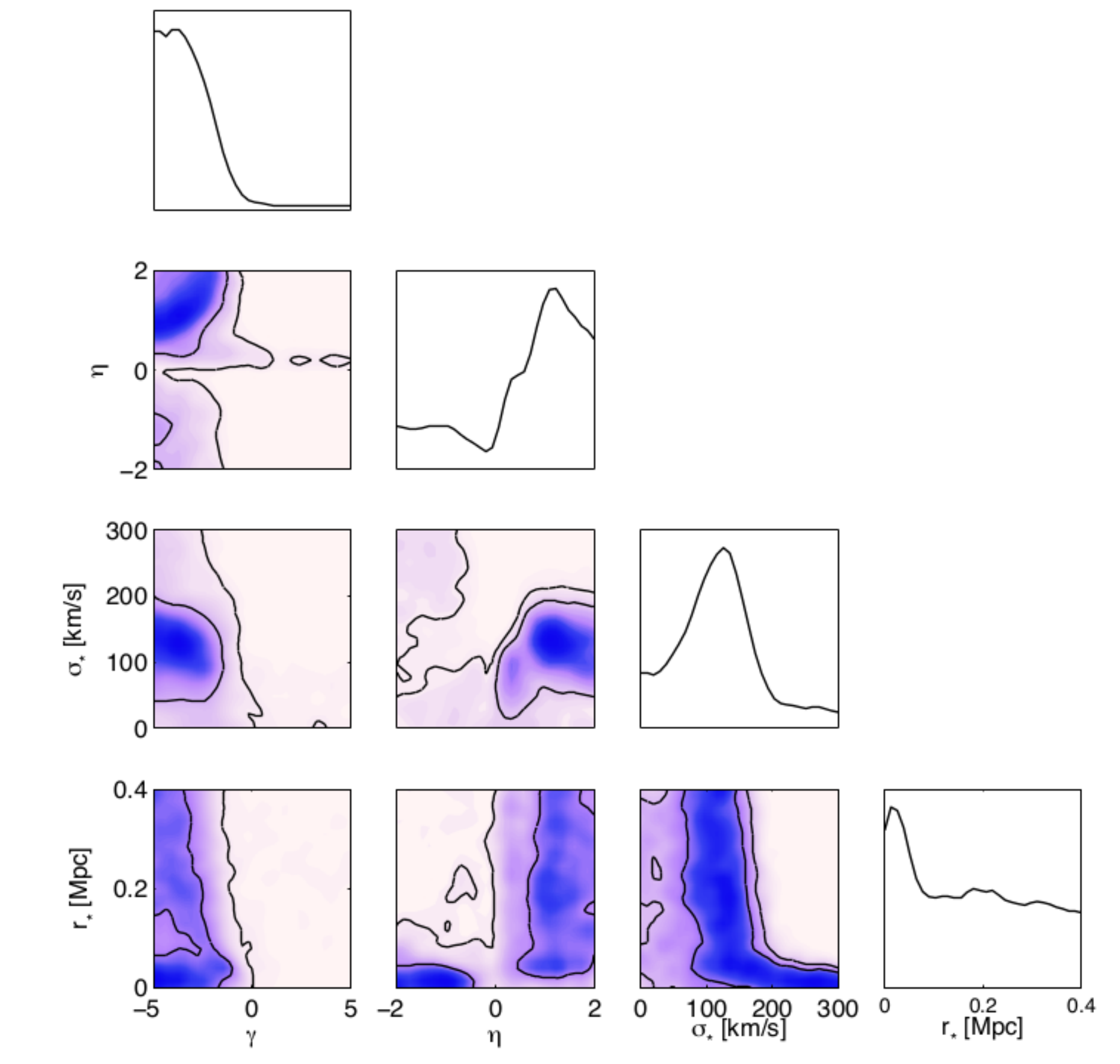}
\caption{Left (right) panel: 1D and 2D marginalised posteriors distributions for the
  parameters of the
  tSIS halo model, derived from the analysis of 500 simulated
  SNIa generated from a tSIS
  model (real SNLS data). From Paper II.}\label{fig:sim-tsis-1gal-500}
\end{center}
\end{figure}

The SIS model has a radial density profile given by
\begin{equation}
\rho(r)=\frac{\sigma^2}{2\pi G}\frac{1}{r^2},
\end{equation}
which is a function of just one single free parameter, namely the
one-dimensional velocity dispersion $\sigma$ of its constituent
particles. A disadvantage of the SIS profile is that the total mass
diverges. Consequently, I use a modified version that depends on a
second free parameter $r_{\rm t}$, which defines the radius at which
the SIS profile is truncated. For finite $r_{\rm t}$, the total mass
does not diverge.

It is straightforward to show that for an object described by a
tSIS profile, the surface density is
\begin{equation}
\Sigma(\xi)=\left\{ 
\begin{array}{ll}
\frac{\sigma^2}{\pi G\xi}\arctan\sqrt{ {r_{\rm t}^2}/{\xi^2}-1} &\mbox{ if $\xi \leq r_{\rm t}$} \\[2mm]
 0  &\mbox{ if $\xi > r_{\rm t}$}.
       \end{array}
\right.
\label{eq:sissigma}
\end{equation}
By substituting this expression into Eq.~\ref{eq:conver}, one obtains
the corresponding convergence $\kappa_{\rm gal}(\xi)$ produced by a
galaxy dark matter halo of this form.

An important complication that arises is the potential relationship
between the velocity dispersion and the galaxy luminosity. To allow
for this possibility and to determine the form of the relationship, I
assume the scaling relations:
\begin{equation}
\sigma=\sigma_{*}\left( \frac{L}{L_*} \right)^{\eta},
\label{eq:sislaw}
\end{equation}
\begin{equation}
r_{\rm t}=r_*\left(\frac{\sigma}{\sigma_*}\right)^{\gamma}=r_*\left(\frac{L}{L_*}\right)^{\eta\gamma},
\label{eq:trlaw}
\end{equation}
Thus, for the tSIS halo model, we wish to constrain the four
parameters $\mathbf{h}=\{\gamma,\eta,\sigma_\ast,r_\ast\}$.

To evaluate my methodology, I first apply it to simulated SNIa
data-sets. To this end, I generate and analyse multiple sets of
simulations, each the same size as the real SNLS data-set. My main
finding was that there is a wide variation in the significance at
which one may detect a gravitational lensing signal. This results from
the strong dependency of the gravitational lensing signal on whether
the data sample contains some SNIa that are strongly magnified. The
number of such SNIa in the sample has a marked effect on the derived
log-evidence $\Delta\log\mathcal{Z}$ relative to a model assuming no
lensing, which I find ranges from about $-1.5$ to $4.5$, with a
median value of $-0.6$. The parameter constraints derived from this
median catalogue are very broad; indeed the constraints are similar to
those obtained from a simulation containing no lensing
signal. Nonetheless, as shown in the left panel of
Figure \ref{fig:sim-tsis-1gal-500}, if one increases the number of SNe
in the data-set up to 500, the constraints become much tighter and one
is able to set limits on the parameters of the dark matter
haloes. Indeed, these constraints contain the true values input to the
simulations. Thus, provided the sample of SNe is sufficiently large,
my method is able to detect the gravitational lensing signal and
recover the correct halo parameters.

When applied to real SNLS data (consisting of 162 SNIa), the
parameters constraints are those shown in the right panel of
Figure \ref{fig:sim-tsis-1gal-500}. On performing a Bayesian model
comparison, I find that the model for a lensing signal produced by
tSIS haloes is only just preferred by 0.2 log-evidence units
relative to the no lensing model, which is similar to the uncertainty
in the evaluated evidence. Consequently, there is no support for choosing either the lensing or no lensing models. This marginal
detection is contrary to previous studies, although these earlier
works did not perform Bayesian model selection, but instead focussed
on goodness-of-fit statistics at the best-fit point in parameter
space. One can begin to reconcile these findings by noting that the
parameter constraints for the tSIS halo model (see the right
panel of Figure \ref{fig:sim-tsis-1gal-500}) do appear somewhat tighter
than those obtained for simulations of 162 SNIa without the inclusion
of a lensing signal. This does suggest a borderline detection of a
lensing signal in the real SNLS data.

\subsection{NFW model}
\label{ch:paper2:NFW}

The NFW profile has a radial density
distribution given by
\begin{equation}
\label{eq:nfw_rho}
  \rho(r)=\frac{\delta_{c}~\rho_{\rm c}}{(r/r_{\rm{s}})(1+r/r_{\rm{s}})^{2}}~,
\end{equation}
where $\rho_{\rm c} = 3H^{2}(z)/8\pi G$ and $H(z)$ are the critical
density and Hubble parameter, respectively, at the redshift, $z$, of
the halo.  The scale radius $r_{\rm s} = r_{200}/c$ is a
characteristic radius for the halo, where $r_{200}$ (the virial
radius) is the radius at which the mass density of the halo drops to
$200 \rho_{\rm c}$, the dimensionless number $c$ is the concentration
parameter, and
\begin{equation}
\delta_{c}= \frac{200}{3}\frac{c^{3}}{\ln(1+c)-c(1+c)^{-1}}
\end{equation}
is a characteristic overdensity. The profile therefore depends on two
free parameters: the virial radius $r_{200}$ and the concentration
parameter $c$. 

The mass contained within the virial radius $r_{200}$ is
\begin{equation}
M_{200} \equiv M(r \le r_{200}) = \frac{800\pi}{3}\rho_c r_{200}^3 = 
\frac{800\pi}{3} \frac{\rho_{\rm m}}{\Omega_{\rm m}}r_{200}^3,
\end{equation}
where $\rho_{\rm m}$ is the mean matter density of the universe and
$\Omega_{\rm m}$ is the matter density parameter, both evaluated at
the redshift $z$ of the halo.
\begin{figure*}
\begin{center}
\includegraphics[width=6cm,height=7cm]{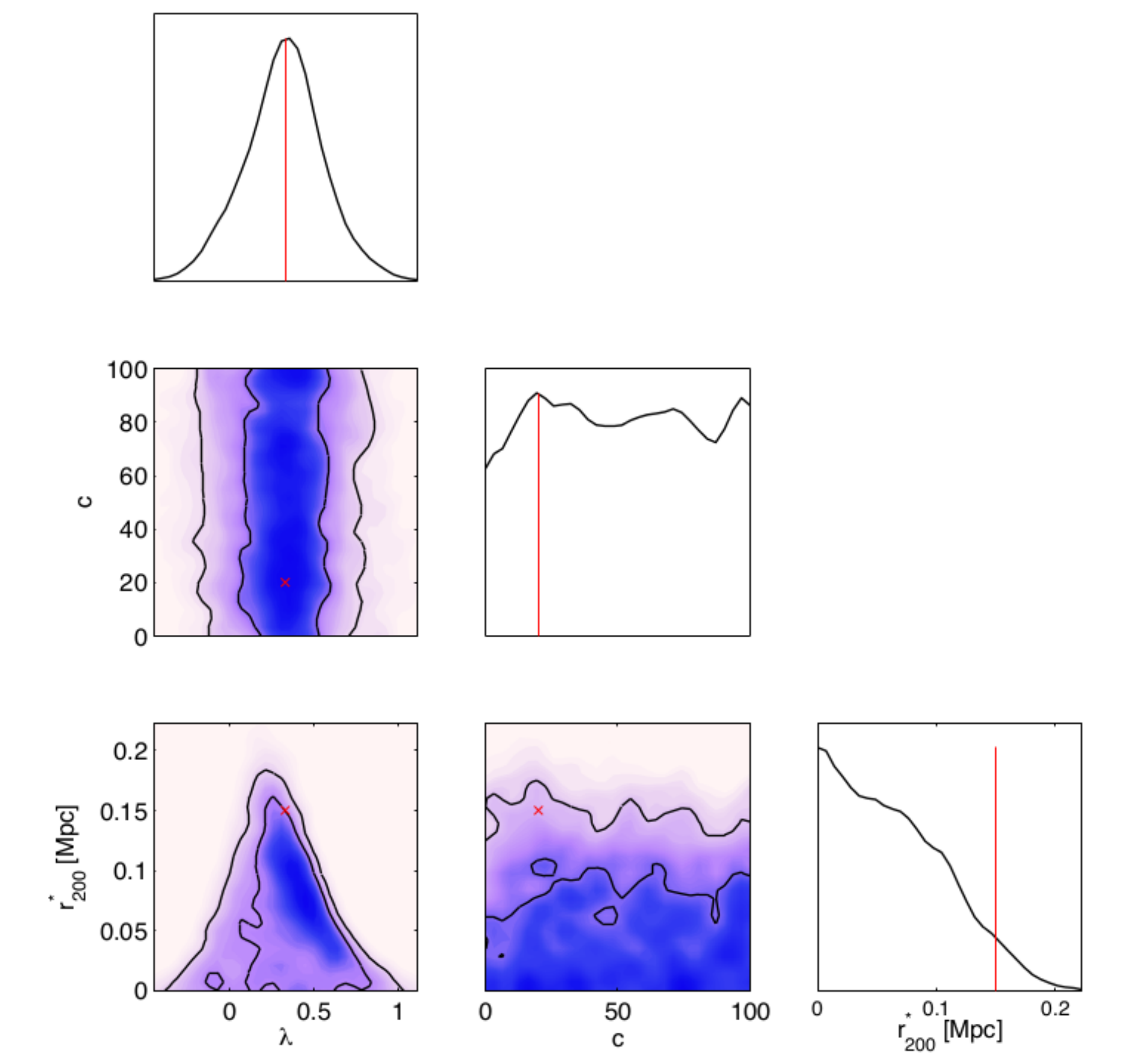}
\includegraphics[width=6cm,height=7cm]{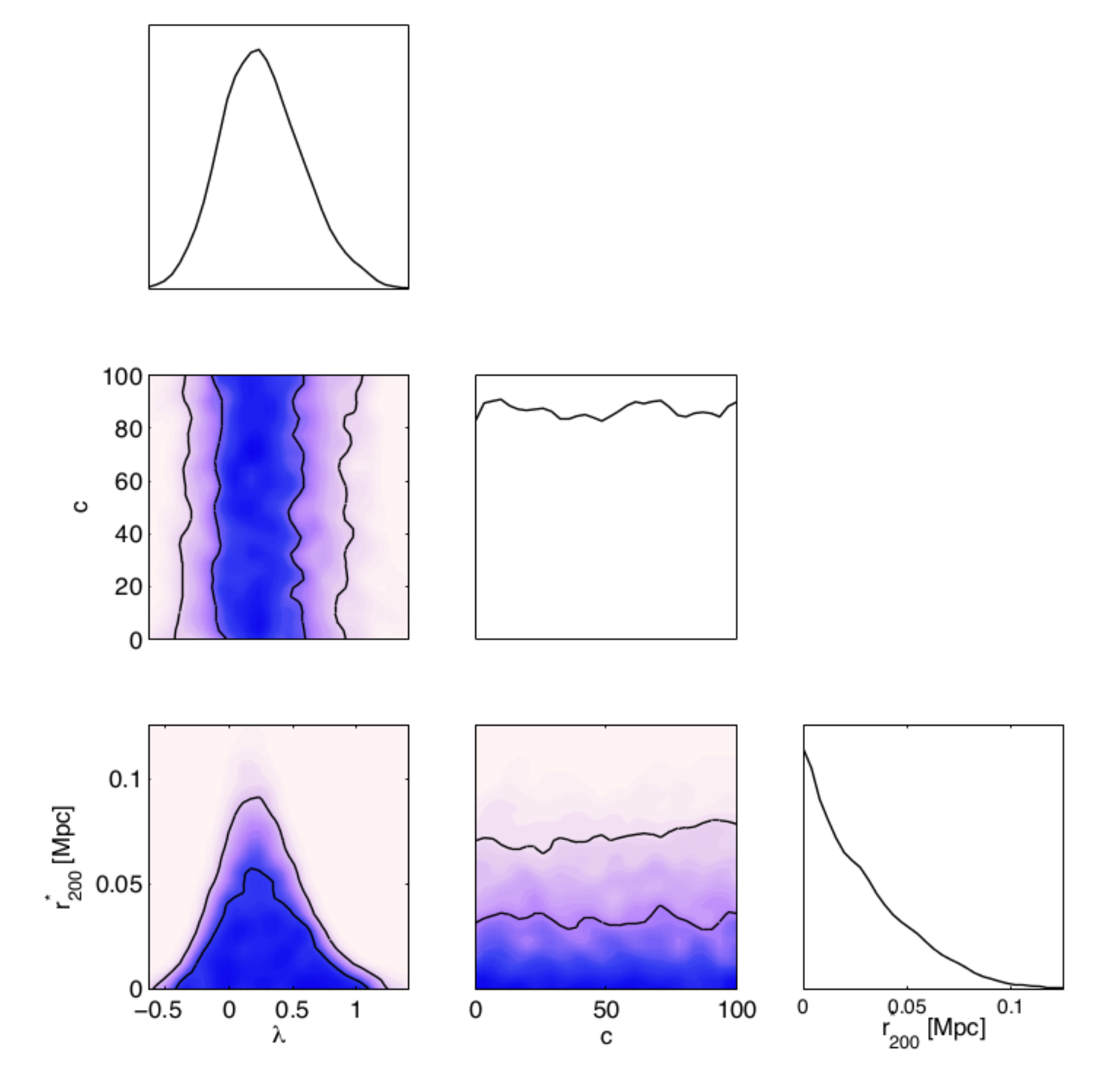}
\caption{1D and 2D marginalised posteriors distributions for the
  parameters $\mathbf{h}=\{\lambda,c,r^\ast_{200}\}$ of the NFW halo
  model, derived from the analysis of 162 simulated SNIa data
  generated assuming no lensing (right) and a NFW halo model (left). In
  the left-hand panel, true parameters are indicated by vertical
  lines and crosses in 1D and 2D plots,
  respectively. \label{fig:sim-nfw-1gal-162}}
\end{center}
\end{figure*}
\begin{figure*}
\begin{center}
\includegraphics[width=6cm,height=7cm]{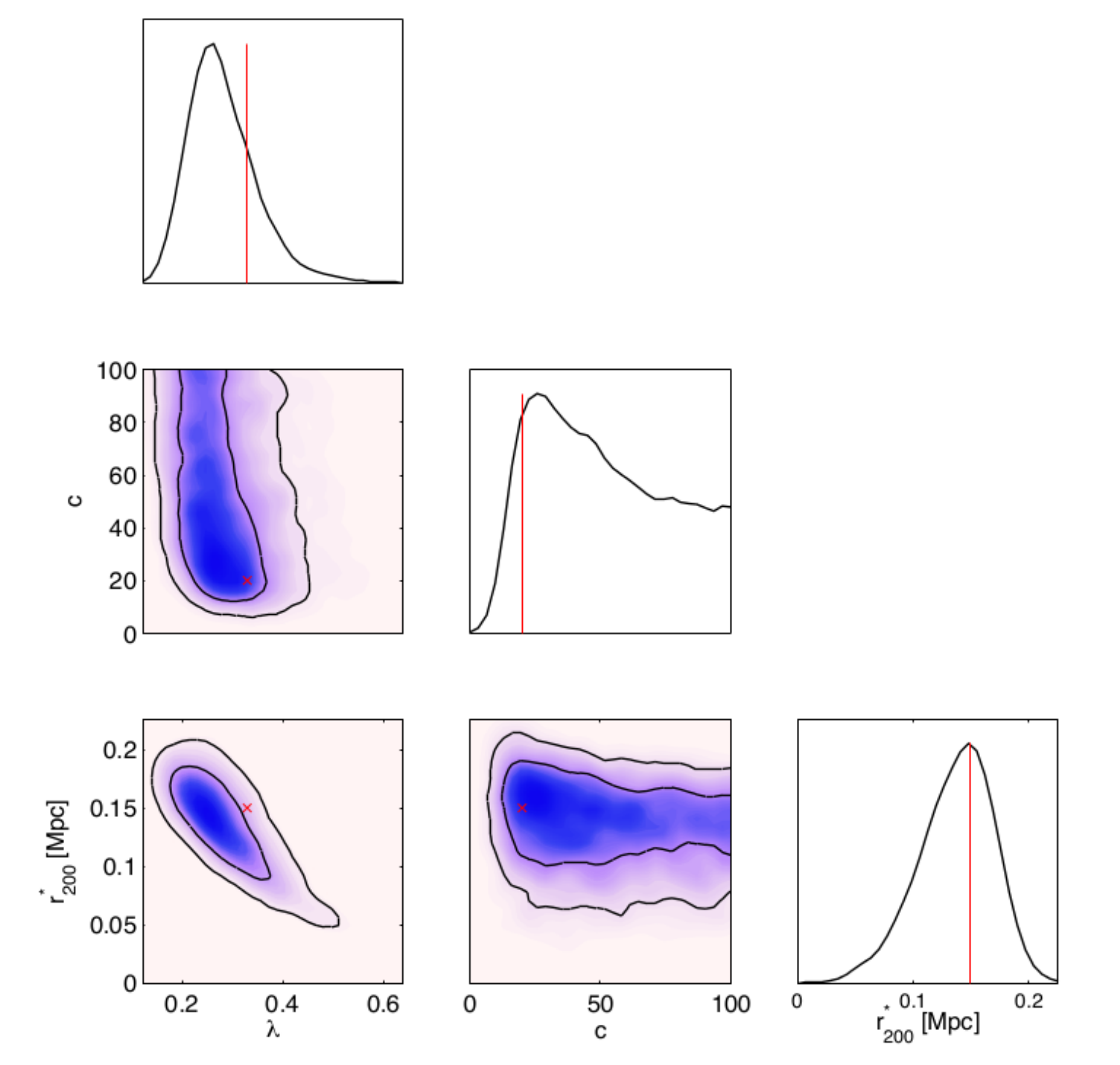}
\includegraphics[width=6cm,height=7cm]{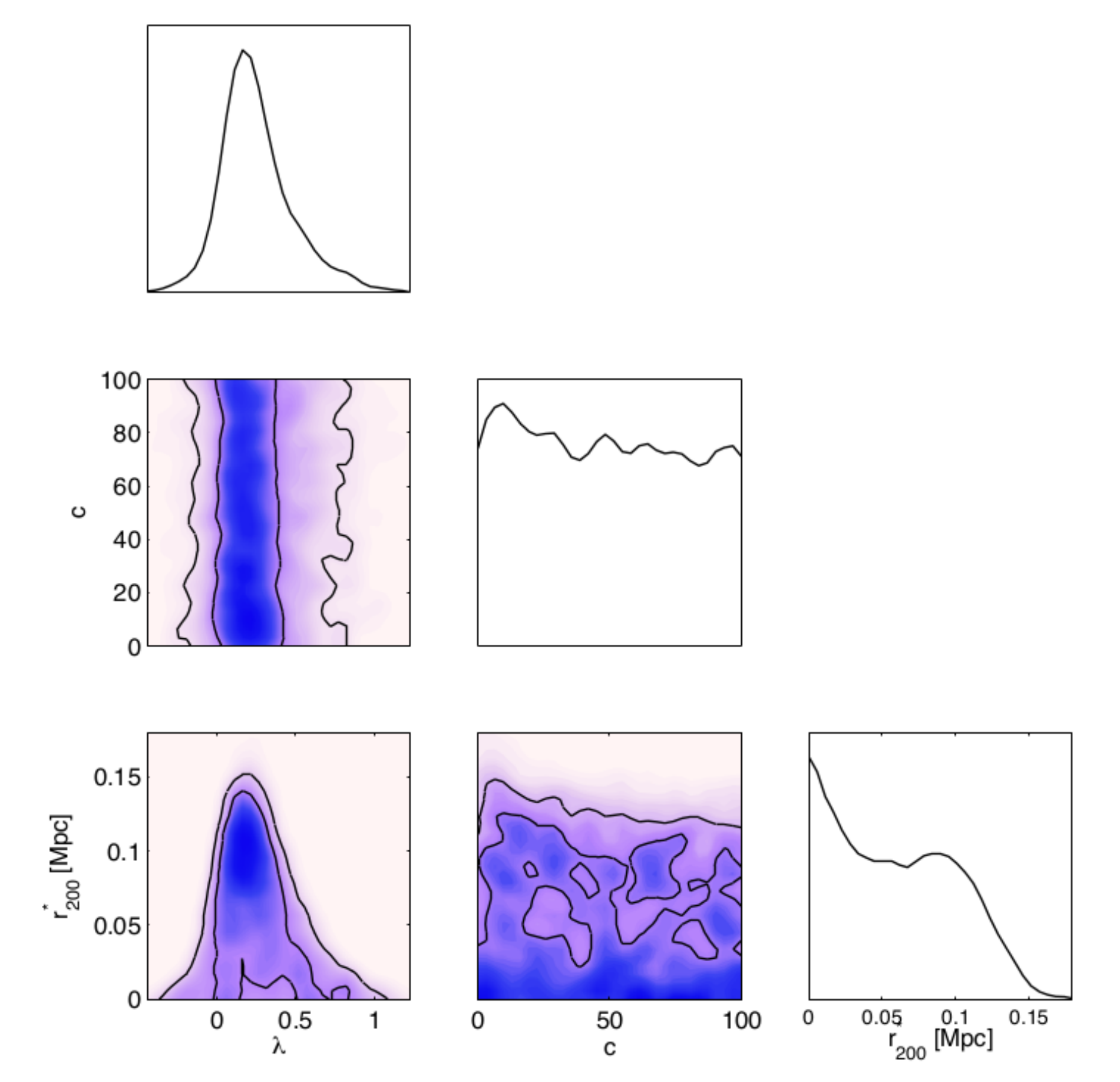}
\caption{As in Figure \ref{fig:sim-tsis-1gal-500}, but for $\mathbf{h}=\{\lambda,c,r^\ast_{200}\}$ of the NFW model. \label{fig:sim-nfw-1gal-500}}
\end{center}
\end{figure*}

To evaluate the gravitational lensing effect of an NFW halo, one must
calculate its surface density, which is given by \citep{1996A&A...313..697B}
\begin{equation}
\Sigma(\xi) = \left\{ \begin{array}{ll}
\frac{2r_{\rm{s}}\delta_{c}\rho_{\rm c}}{\left(x^{2}-1\right)}
\left[1-\frac{2}{\sqrt{1-x^{2}}}{\rm arctanh}\sqrt{\frac{1-x}{1+x}}
\hspace{0.15cm} \right] & \mbox{$\left(x < 1\right)$} \\ [5mm]
\frac{2r_{\rm{s}}\delta_{c}\rho_{\rm c}}{3} & \mbox{$\left(x = 1\right)$} \\ [2mm]
\frac{2r_{\rm{s}}\delta_{c}\rho_{\rm c}}{\left(x^{2}-1\right)}
\left[1-\frac{2}{\sqrt{x^{2}-1}}\arctan\sqrt{\frac{x-1}{x+1}}\hspace{0.15cm}
 \right] 
& \mbox{$\left(x > 1\right)$} 
\end{array}
\right.
\end{equation}
where $x = \xi/r_{\rm s}$ is a dimensionless projected radial distance
in the lens plane. The resulting convergence $\kappa_{\rm gal}(\xi)$ due to
a NFW halo is then obtained by substituting $\Sigma(\xi)$ into
Eq.~\ref{eq:conver}.

A complication arises similar to that encountered for the tSIS profile,
namely that there is potentially a relationship between the virial radius
of a NFW halo and the luminosity of the galaxy it surrounds. To allow
for and investigate this possibility, I therefore adopt the scaling
law
\begin{equation}
r_{200}=r_{200}^{*}\left( \frac{L}{L_*} \right)^{\lambda}.
\label{eq:nfwlaw}
\end{equation}
In this case, one thus seeks to constrain the three parameters
$\mathbf{h}=\{\lambda,c,r^\ast_{200}\}$.

I first apply my method to simulated data-sets of the same size (162
SNIa) as the real SNLS data, as was done for tSIS
model. Figure \ref{fig:sim-nfw-1gal-162} shows that one cannot obtain
any real constraints on the NFW halo parameters, since the
marginalised posteriors look very similar to those obtained from
simulations containing no lensing signal. Increasing the number of SNe
in the simulations to 500, I find that my method does produce
constraints on the halo parameters that are consistent with the input
values used in the simulations; this is illustrated in the left
panel of Figure \ref{fig:sim-nfw-1gal-500}. As shown in the right
panel, however, the analysis of the real SNLS data yields no useful
constraints on the halo parameters.

\subsection[Model selection between different dark matter halo\\ models]{Model selection between different dark matter halo models}
\begin{table}[tbf]
\begin{center}
\begin{tabular}{l|ccc}
\hline\hline & \multicolumn{3}{c}{Simulation model} \\ Analysis model & No
lensing & tSIS & NFW \\ \hline No lensing &
$\phantom{-}0.0$ & $\phantom{-}0.0$ & $\phantom{-}0.0$
\\ tSIS & $-1.8 $ & $-1.1 $ & $-1.3$ \\ NFW & $-3.1 $ & $-2.6 $ & $-2.3$ \\ \hline\hline
\end{tabular}
\caption{$\Delta\log\mathcal{Z}$ ($\log$-evidence value relative to
  the null evidence) for the analysis of simulated data with 162
  SNIa, with errors of 0.2.
\label{tab:simevids162}}
\end{center}
\end{table}
\begin{table}[tbf]
\begin{center}
\begin{tabular}{l|ccc}
\hline\hline
& \multicolumn{3}{c}{Simulation model} \\
Analysis model & No lensing & tSIS & NFW \\
\hline
No lensing & $\phantom{-}0.0$ 
& $\phantom{-}0.0$ 
& $\phantom{-}0.0$ \\
tSIS & $-1.7 $ & $\phantom{-}4.5 ^\ast$ &  $\phantom{-}6.9$ \\
NFW & $-3.1$ & $\phantom{-}3.6 $ &  $\phantom{-}7.2^\ast$ \\
\hline\hline
\end{tabular}
\caption{$\Delta\log\mathcal{Z}$ ($\log$-evidence value relative to
  the null evidence) for the analysis of simulated data with 500
  SNIa, with errors of 0.2. Asterisks denote the cases for which the corresponding halo
  parameter constraints are plotted in the left panels of Figs.~\ref{fig:sim-tsis-1gal-500} and
  \ref{fig:sim-nfw-1gal-500}, respectively.
\label{tab:simevids500}}
\end{center}
\end{table}
\begin{table}[tbf]
\begin{center}
\begin{tabular}{l|c}
\hline\hline
Model & $\Delta\log\mathcal{Z}$ \\
\hline
tSIS & $\phantom{-}0.2 ^\ast$\\
NFW & $-2.5 ^\ast$  \\
\hline\hline
\end{tabular}
\caption{$\Delta\log\mathcal{Z}$ ($\log$-evidence value relative to
  the null evidence for no lensing signal) for the analysis of the
  real SNIa data, with errors of 0.2. Asterisks denote the cases for which the corresponding
  halo parameters constraints are plotted in the right panels of Figs~\ref{fig:sim-tsis-1gal-500} and
  \ref{fig:sim-nfw-1gal-500}, respectively.\label{tab:realevids}}
\end{center}
\end{table}

To understand better the marginal detection (at best) of any lensing
signal in the real SNLS data, I now perform a systematic Bayesian
model comparison using my simulated data-sets. I begin by considering
simulations of the same size (each containing 162 SNIa) as the real
SNLS data. Table~\ref{tab:simevids162} lists the Bayesian log-evidence
for each analysis model, relative in each case to the null
(no-lensing) model.  In each case, the no lensing model is
preferred. This concurs with my findings for the real SNLS data, and
suggests that one cannot detect a lensing signal with data of this
quantity and quality, let alone distinguish between different halo
models.

To determine the nature of the data required to obtain a robust
detection of lensing, I analyse simulations each containing 500
SNIa. The results are given in Table \ref{tab:simevids500}. For
simulations containing no lensing signal, the method correctly
identifies this as the preferred model.  More importantly, however, my
method also prefers the models with lensing (at high significance) for
simulations that do contain a lensing signal. This demonstrates that
the method performs correctly. Moreover, the correct halo model is
also picked out, although the selection between halo models is not
robust as the log-evidence differences are quite small.

Finally, Table \ref{tab:realevids} shows the results using real data. As
mentioned above, for the tSIS halo model, there is a very
slight preference for a lensing signal, but only by 0.2 log-evidence
units, which is the level of the uncertainty in the calculation of the
evidence. For the NFW halo model, however, the presence of a lensing
signal is strongly disfavoured by $-2.5$ log-evidence units relative
to the no lensing model.

\subsection{Foreground galaxies catalogue}
\label{ch:paper2:gal}

In all the analyses described in this section, I have used the {\em
  true} galaxies from the SNLS catalogue. To understand my results further, it is of interest to examine the magnification along a
large number of lines-of-sight through these galaxies.

\begin{figure}
\begin{center}
\includegraphics[width=6cm]{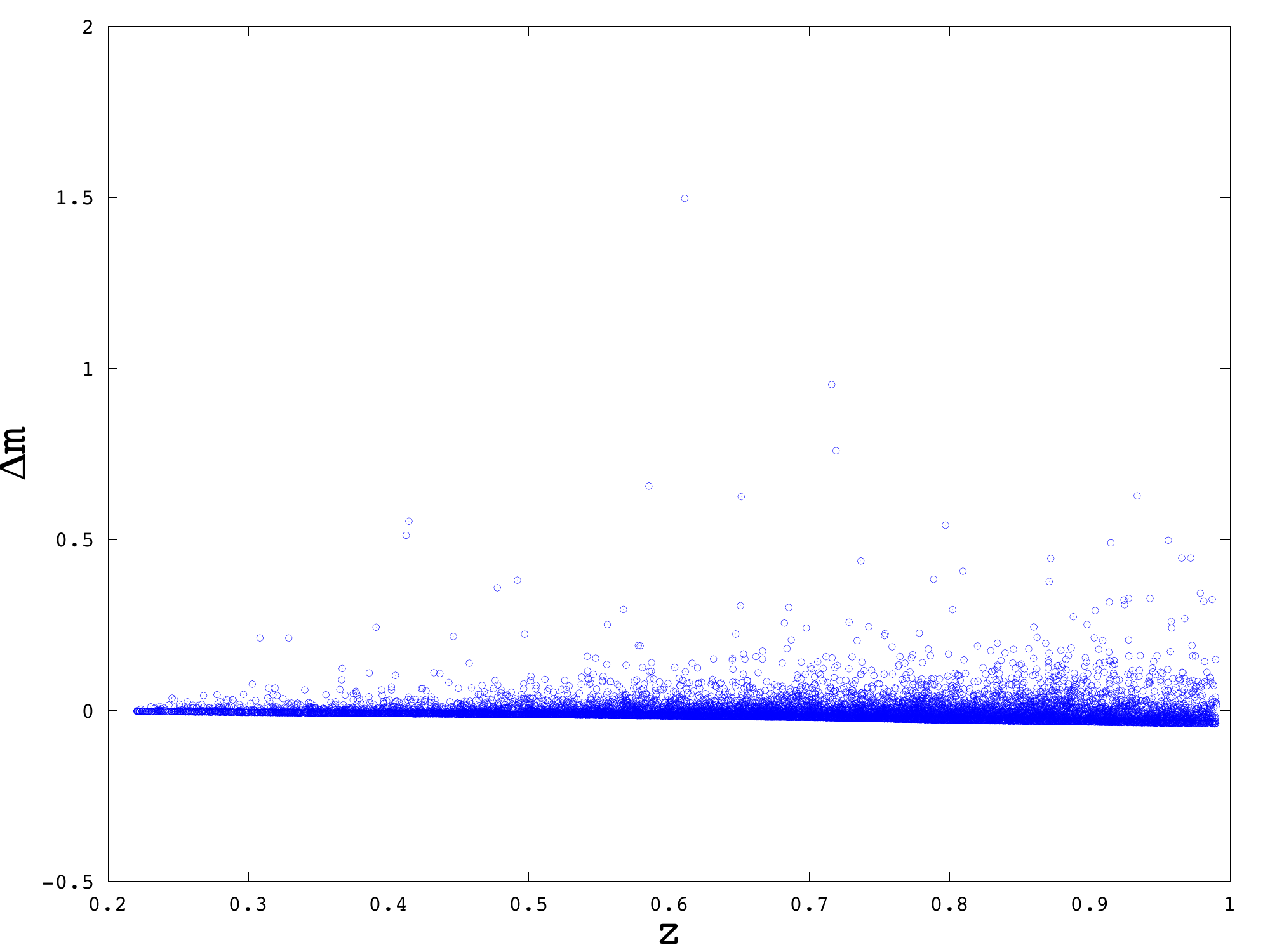}
\includegraphics[width=6cm]{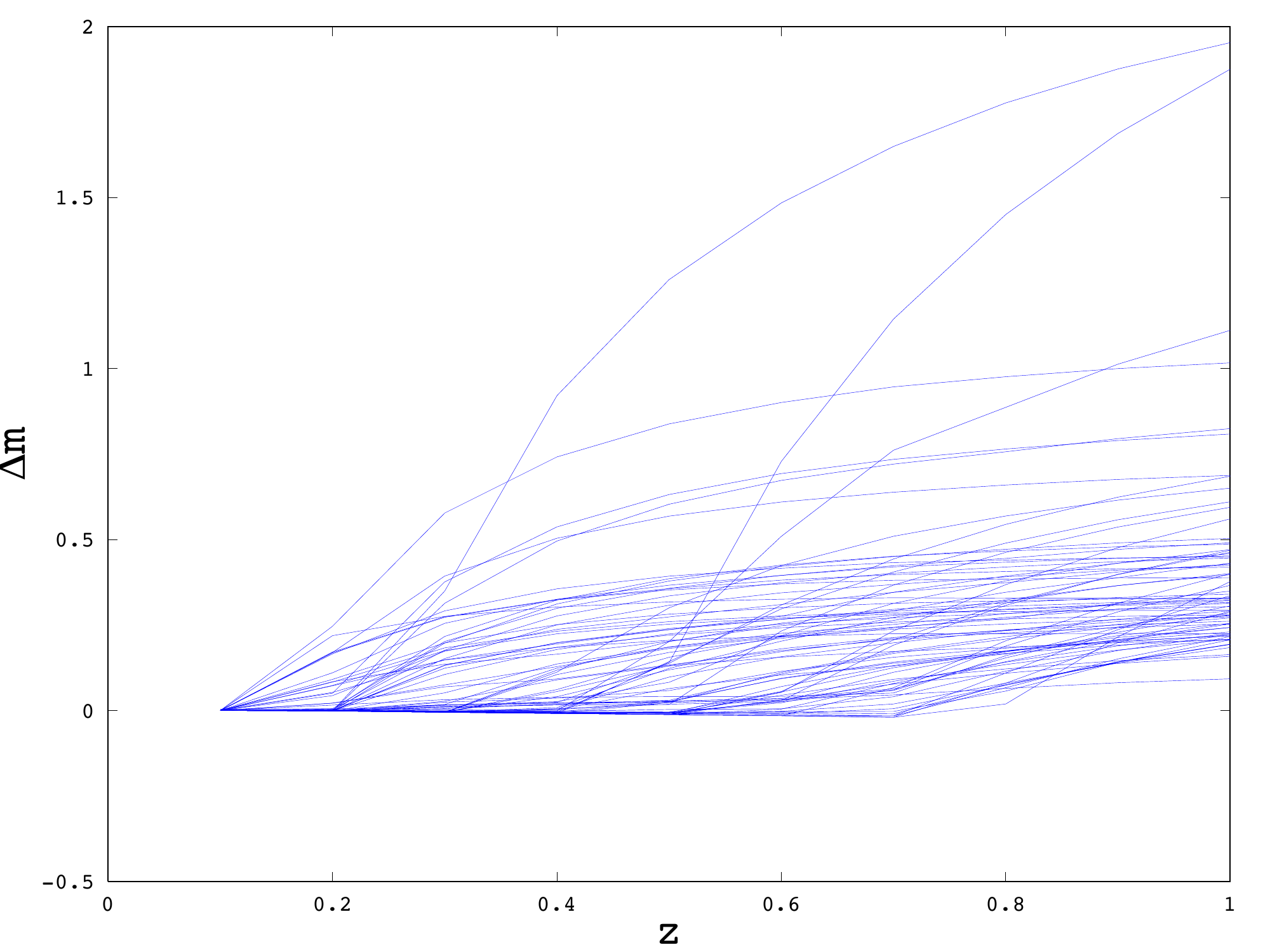}
\caption{Left: magnification of $10^4$ simulated SNIa. Right:
magnification for the 75 lines-of-sight that
  exhibit the strongest lensing effect. From Paper II.
  \label{fig:simmags}}
\end{center}
\end{figure}

In Figure~\ref{fig:simmags} (right panel), I plot the magnification
factor for each of $10^4$ simulated SNIa.  From the plot, one can see
that the background correction results in most of the SNIa being
demagnified. Nonetheless, there are a small number of SNIa that have
very large magnifications. This suggests that one should expect a
large variation in the strength of the lensing signal when analysing
SNIa catalogues containing relatively few events, such as the SNLS
data-set. The important criterion is whether the SNIa catalogue
contains one or more SNIa that are very strongly magnified. As one
observes more SNIa, one would expect the variation between randomly
constructed catalogues to diminish, and a more stable and robust
detection of lensing to be possible. This agrees with my results given
above.

Another important observation is that, for strongly magnified SNIa,
there is no clear correlation between the size of the magnification
and redshift. This suggests the counter-intuitive conclusion that
observing high-redshift SNIa may not confer any advantage in
attempting to detect a lensing signal. I examine this issue further by
plotting in Figure~\ref{fig:simmags} (right panel) the magnification as
a function of redshift along the 75 lines-of-sight that exhibit the
highest magnification. One first notices that the magnification of the
three most highly lensed lines-of-sight continues to increase markedly
up to $z=1$. Nonetheless, for the remaining lines-of-sight, the
magnification does not typically increase much beyond $z \sim 0.5$.
Again this suggests that, at least along these lines-of-sight, there
is little advantage in observing a very high-redshift SNIa in terms of
detecting a lensing signal.

One must be careful in drawing such conclusions, however, because I am
using the true galaxy catalogue from SNLS. This real catalogue will
inevitably suffer from selection effects that result in high-redshift
galaxies being under-represented. This is expected to give the above effects. One may investigate this issue further by performing the
same analysis for galaxy catalogues constructed taken from some large
numerical simulation, and this would be an interesting topic for future
research. Moreover, forthcoming surveys such as DES and Euclid will
enable us to take a significant step forward, since they will provide
not only SNIa data, but also very good measurements of foreground
galaxies.

\section{SNe photometric classification}
\label{ch:paper3}

In Section \ref{ch:SN_Clas:class}, I started a discussion about
methods that can perform photometric classification of SNe into
Ia and non-Ia types. Since such methods are becoming increasingly
necessary for the analysis of large SNIa data-sets, I also worked on
developing fully automated methods that can perform this task in a
quick, automated and robust manner. In Paper III, I present a new
method for performing automated photometric classifications of SNe
into Ia and non-Ia types. This method adopts an extremely naive
approach to the question and I do not use any prior information about
SNe physics. Thus in Paper V, I use a HNN to include
information about SN models.  Both of these methods take a two-stage
approach. First, the SN lightcurves are fitted to an analytic
parameterised function in order to standardise the number of variables
associated with each SN. The resulting fitted parameters, together
with a few further quantities associated with the fit, are then used
as the input feature vector to a classification NN and a hierarchy of
classification NNs whose output is the probability that the SN is of a
particular type.

\subsection{First step: Lightcurve fitting}
\label{ch:paper3:lc_fitting}
\begin{figure}
\begin{center}
\includegraphics[width=12.5cm]{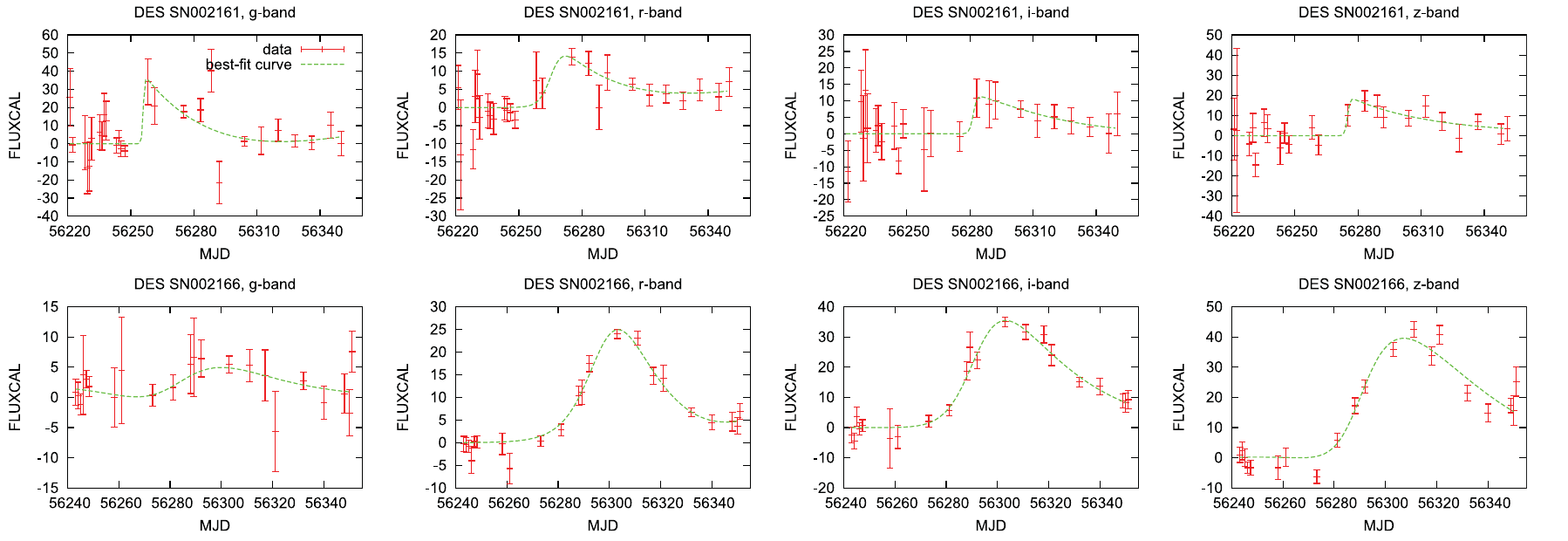}
\caption{Simulated lightcurve measurements and associated
  uncertainties (red points) in the $g$, $r$, $i$ and $z$ filters for
  a Type-Ia (top row) and non-Ia (bottom row) SN, together with the
  best-fit function (green line). From Paper III.}\label{fig:PAPER3_1}
\end{center}
\end{figure}

The form of the fitted function is given by
\begin{equation}
f(t) = A\left[1+B(t-t_1)^2\right] \frac{e^{-(t-t_0)/T_{\rm fall}}}{1+e^{-(t-t_0)/T_{\rm rise}}},
\label{eq:fitformula}
\end{equation}
where, for each SN, $t=0$ corresponds to the time of the earliest
measurement in the $r$-band lightcurve. Figure \ref{fig:PAPER3_1} shows
the best-fitting functional form in four wavebands for a typical Ia and
non-Ia SN.

For each fitted lightcurve, I construct a feature vector that contains
the mean values of the one-dimensional marginalised posteriors of each
parameter in the fitting formula of Eq.~\ref{eq:fitformula},
$\widehat{\mathbf{\Theta}} = \{\hat{A}, \hat{B}, \hat{t}_1, \hat{t}_0,
\hat{T}_{{\rm rise},}, \hat{T}_{{\rm fall},} \}$ and their standard
deviations ${\mathbf{\sigma}} = \{ \sigma_{A}, \sigma_{B},
\sigma_{t_{\rm 1}}, \sigma_{t_{\rm 0}}, \sigma_{T_{{\rm rise}}},
\sigma_{T_{{\rm fall}}} \}$. I also append to the feature vector the
number of flux measurements $n$ in the lightcurve, the
maximum-likelihood value of the fit and the Bayesian evidence of the
model. This feature vector then provides a standardised input for the
training of the NN.

\subsection{Second step: NN classification}
\label{ch:paper3:NN}

In Section \ref{NN}, I described the 3-layer feed-forward NNs that I
use in my work. In particular, I use the SkyNet package.

In the application to SN classification, it
is important to assess the quality of the network output classes by
constructing some statistical measures. The most appropriate
quantities are the completeness $\epsilon_{\rm Ia}$ (fraction of all
SNIa that have been correctly classified; also often called the
efficiency), purity $\tau_{\rm Ia} $ (fraction of all Type Ia
candidates that have been classified correctly) and figure of merit
$\mathcal{F}_{\rm Ia}$ for SNIa. These are defined as follows:
\begin{eqnarray}
\epsilon_{\rm Ia} & = & \frac{N_{\rm Ia}^{\rm true}}{N_{\rm Ia}^{\rm total}},\\
\label{eq:completeness}
\tau_{\rm Ia} & = & \frac{N_{\rm Ia}^{\rm true}}{N_{\rm Ia}^{\rm true} + N_{\rm Ia}^{\rm false}},\\
\label{eq:purity}
\mathcal{F}_{\rm Ia} & = & \frac{1}{N_{\rm Ia}^{\rm total}} \frac{(N_{\rm Ia}^{\rm true})^2}{N_{\rm Ia}^{\rm true}+WN_{\rm Ia}^{\rm false}},
\label{eq:FoM}
\end{eqnarray}
where $N_{\rm Ia}^{\rm total}$ is the total number of SNIa in
the sample, $N_{\rm Ia}^{\rm true}$ is the number of SNe correctly
predicted to be of Type Ia, $N_{\rm Ia}^{\rm false}$ is the number of
SNe incorrectly predicted to be of Type Ia and $W$ is a penalty factor
which controls the relative penalty for false positives over false
negatives.

\begin{figure}
\begin{center}
\includegraphics[width=12.5cm]{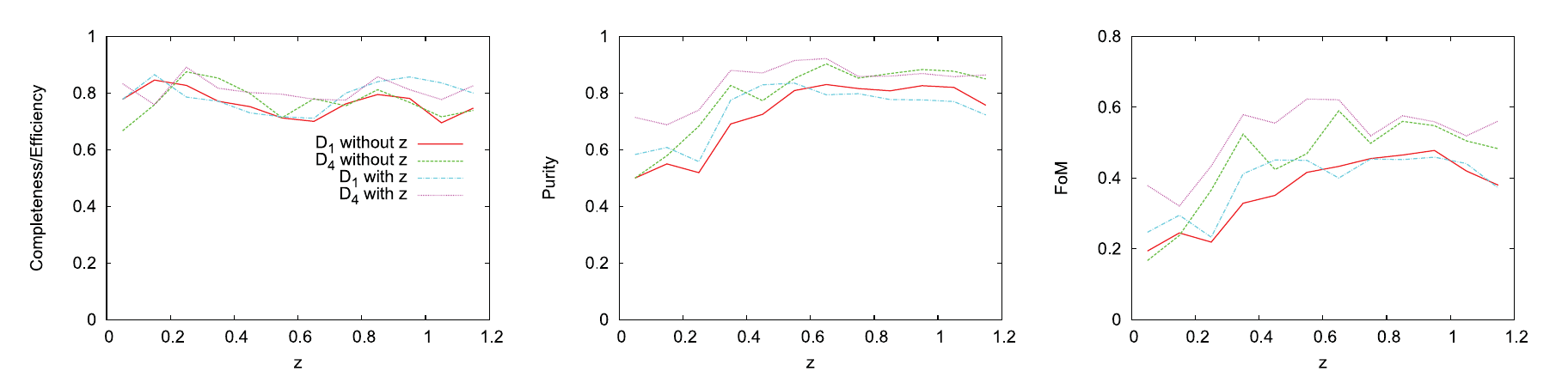}
\caption{Completeness, purity and figure of merit for SNIa
  classification as a function of redshift ($z$) from applying trained
  networks with a threshold probability ($p_{\rm th}$) of 0.5; ``with
  $z$'' and ``without $z$'' indicates that redshift information was/was
  not used in network training. Samples $\mathcal{D}_1$ and
  $\mathcal{D}_4$ use 10 and 40 per cent of the data, respectively,
  for training. From Paper III.}\label{fig:PAPER3_2}
\end{center}
\end{figure}

\subsection[SuperNova Photometric Classification Challenge \\(SNPCC)]{SuperNova Photometric Classification Challenge (SNPCC) }
\label{ch:paper3:NN:SNPCC}

I started my work on developing methods for photometric classification
of SNe by applying it to the updated simulated data-set released
following the SNPCC \citep{2010PASP..122.1415K,2010arXiv1001.5210K}. I did not apply any cuts to the
original data, so my data-set contained low signal-to-noise SNe and
very poorly-sampled lightcurves, sometimes containing very few
measured fluxes that are not necessarily measured on both sides of
peak brightness.

\subsubsection{NN}
In Paper III, I found that applying a regular classification network,
I obtain very robust classification results, namely a completeness of
0.78 (0.82), purity of 0.77 (0.82), and SNPCC figure-of-merit of 0.41
(0.50) when I use 10 (40) per cent of the data for training and assume
a canonical threshold output probability $p_{\rm th}=0.5$. I pick my
training set randomly and do not use any redshift information.  A
modest 5--10 per cent improvement in these results is achieved by also
including the SN host-galaxy redshift and its uncertainty as inputs to
the classification network, see Figure \ref{fig:PAPER3_2}.  The
quality of the classification does not depend strongly on the SN
redshift.

\subsubsection{HNN}

\begin{figure}[t]
\begin{center}
\includegraphics[width=0.229\columnwidth, angle =-90]{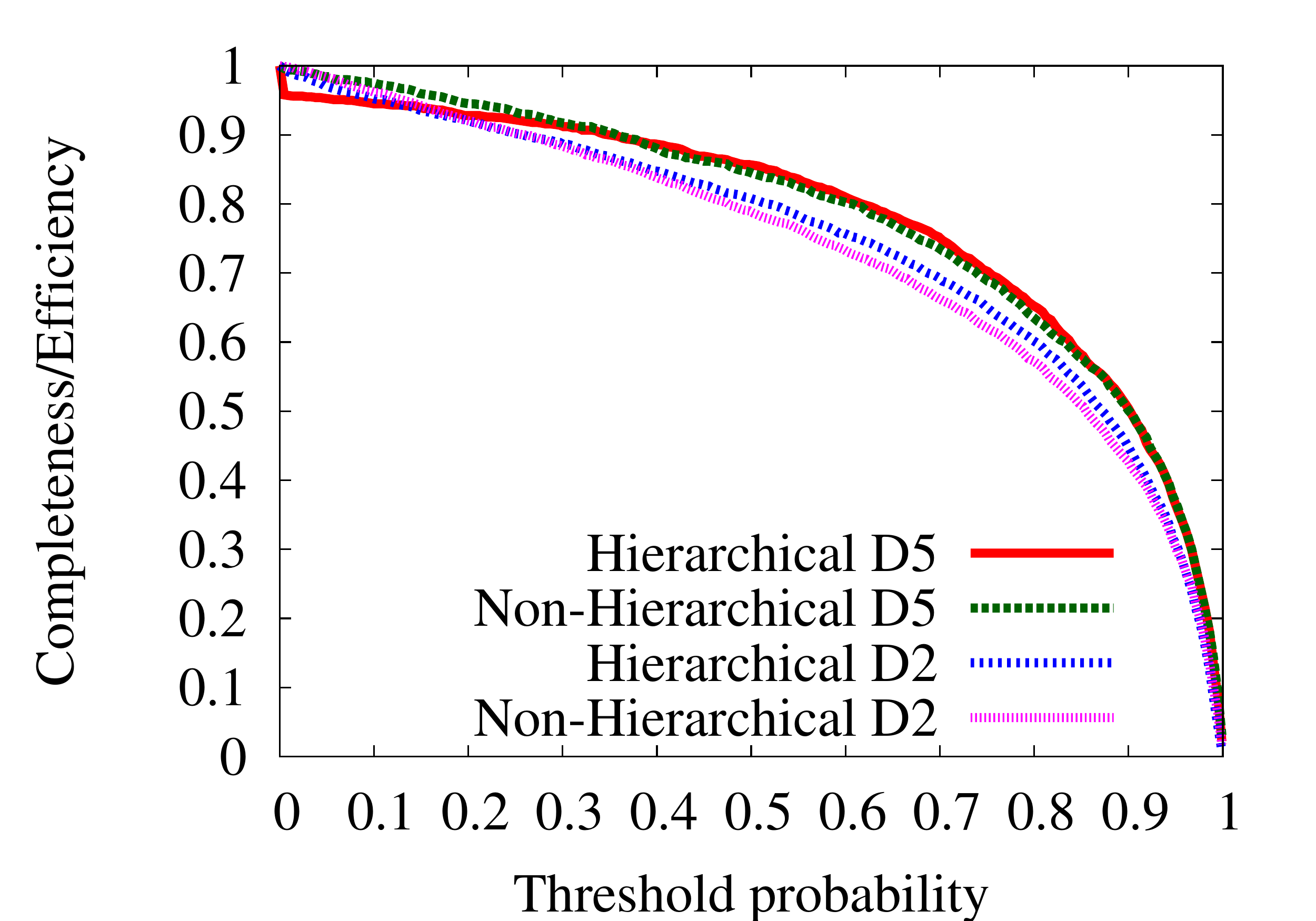}
  \includegraphics[width=0.229\columnwidth, angle =-90]{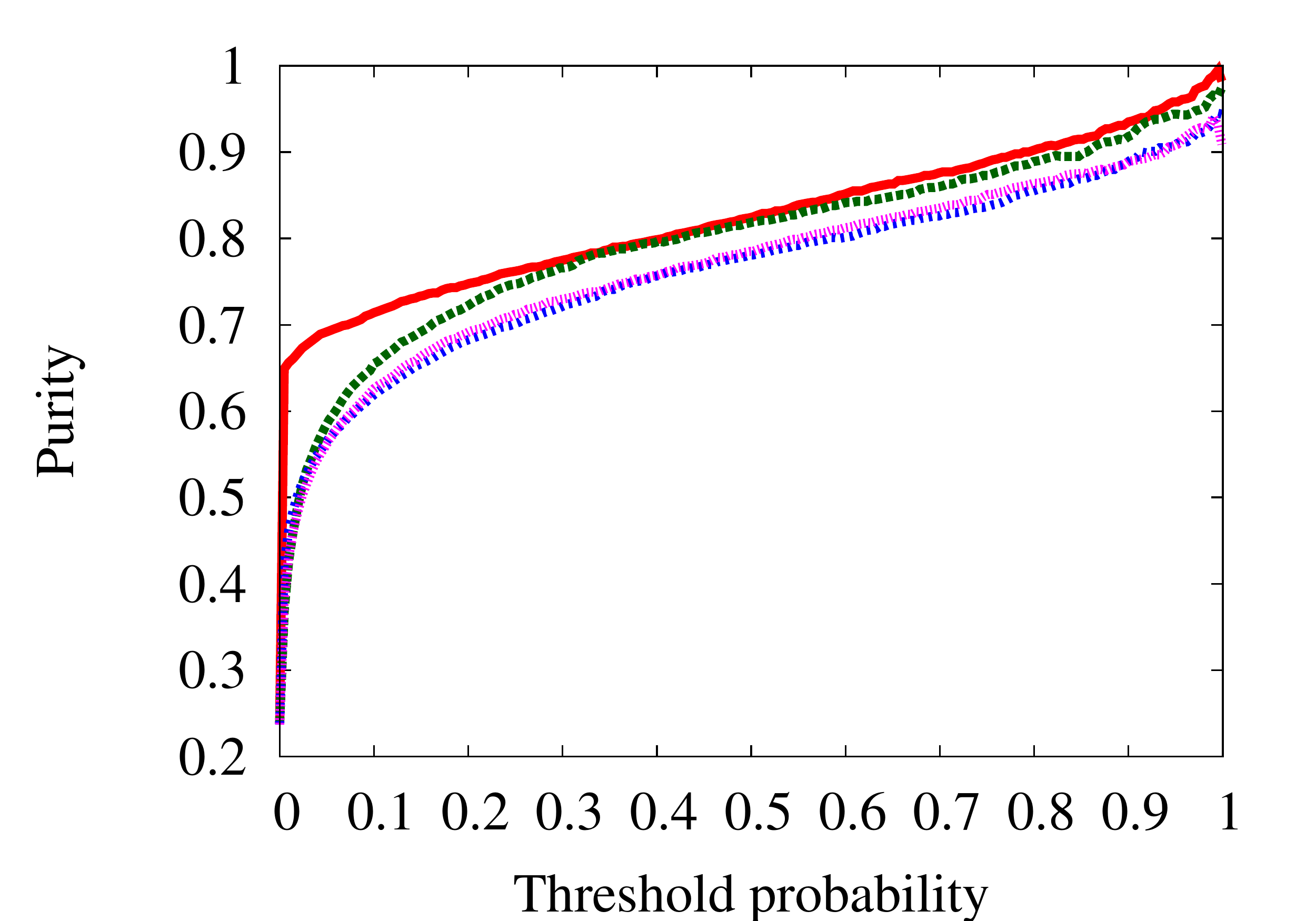}
  \includegraphics[width=0.229\columnwidth, angle =-90]{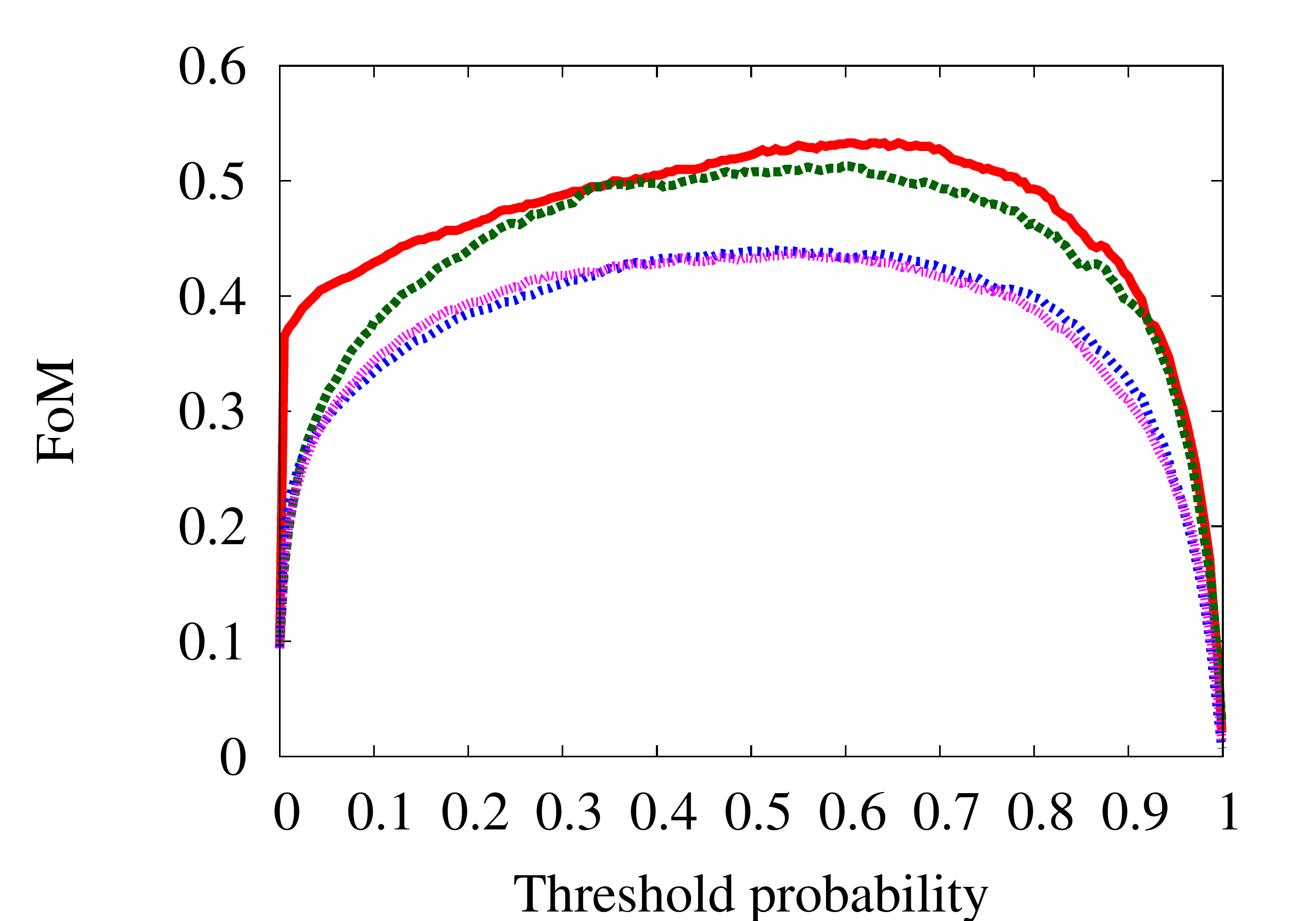}
\caption{Completeness, purity and FoM as a function of threshold
  probability $p_{\rm th}$ from applying trained networks to get
  Ia/non-Ia classification probabilities on the testing data-set from
  NN (Paper III) and HNN. No redshift information was used in network
  training. Samples $\mathcal{D}_2$ and $\mathcal{D}_5$ use 20 and 50
  per cent of the data, respectively, for training. From Paper V}
\label{fig:Results-SNPCC}
\end{center}
\end{figure}

In Paper V, I further develop the method presented in Paper III by
introducing a HNN, which accommodates the structure
shown on Figure \ref{fig:SNclass}. From Figure \ref{fig:Results-SNPCC}
we see that HNN performs better than the method in Paper III as more
training data becomes available. Even with a small amount of data,
however, HNN still performs well. These positive preliminary results
motivates further study, in particular the investigation of the
importance of the training sample: is it percentage or total number of
SNe used for training that makes the largest difference? Also I want
to test this method on a more realistic sample, for example when
measurements for some filters are missing completely.

\section{Testing consistency}

In Section \ref{JA}, I described the benefits of using a joint
analysis of SNe from different telescopes/surveys, having SNe both at
low and high redshift. A large collection of SNe over a wide range of
redshifts results in tighter constraints on cosmological parameters.
This comes at a price, however, since one must first check that the
individual SN surveys produce results that are mutually consistent.
If this is not the case, any results derived from their
combination may be misleading. Such checks for mutual consistency are
rarely performed.  In Sections \ref{sec:method:Rtest} and
\ref{sec:method:hypPar} I discussed Bayesian methods to perform such a test. 
In this Section, I apply this method to SN data.

\subsection{Consistency test based on $\chi^2$-method}
\label{ch:con:chi2}

In Paper IV, I test the mutual consistency of different SN
surveys within JLA \citep{2014arXiv1401.4064B} and Union2 compilations. The same way as in Section
\ref{ch:paper1:2bins} I separate each compilation into subsets
according to the telescope/survey with which they have been observed.
Since the $\chi^2$-method uses a non-normalised likelihood one replaces
$\Pr(\mathbf{D}|\mathbf{\Theta}, H)$ by the ``likelihood''
$\mathcal{L}(\mathbf{\Theta},\sigma_\text{int})$ discussed in Section \ref{ch:Cosmology:chi2}. In this
case, however, one can no longer interpret the terms in
Eq.~\ref{eq:R-test} directly as probabilities. Consequently, the value
of the $\mathcal{R}$ cannot be compared with the normal Jeffreys'
scale. One still expects, however, that for data-sets that are
mutually consistent the $\mathcal{R}$-value will be higher than for
inconsistent ones \cite{2011MNRAS.415..143M}. Consequently, one may still use
the $\mathcal{R}$-value, but as a one-sided test statistic in the
frequentist sense, which must be calibrated using simulations.

The distribution of $\mathcal{R}$ under the null hypothesis $H_0$ is
constructed from simulations in which the individual surveys are
mutually consistent. The $\mathcal{R}$ value obtained by analysing the
real data can then be compared with this distribution in the standard
manner. In particular, we calculate the $p$-value as:
\begin{equation}
 p = \frac{N(\mathcal{R}_{\rm s} < \mathcal{R}_{\rm r})}{N_{\rm tot}},
\label{eq:p-value}
\end{equation}
where $\mathcal{R}_{\rm s}$ and $\mathcal{R}_{\rm r}$ are the
$\mathcal{R}$ values obtained by analysing simulated and real
data-sets respectively, $N(\mathcal{R}_{\rm s} < \mathcal{R}_{\rm r})$
is the number of simulations with $\mathcal{R}$ values less than that
obtained by analysing the real data and $N_{\rm tot}$ is the total
number of simulations.

\begin{figure}[tb]
\centering
\includegraphics[width=3in]{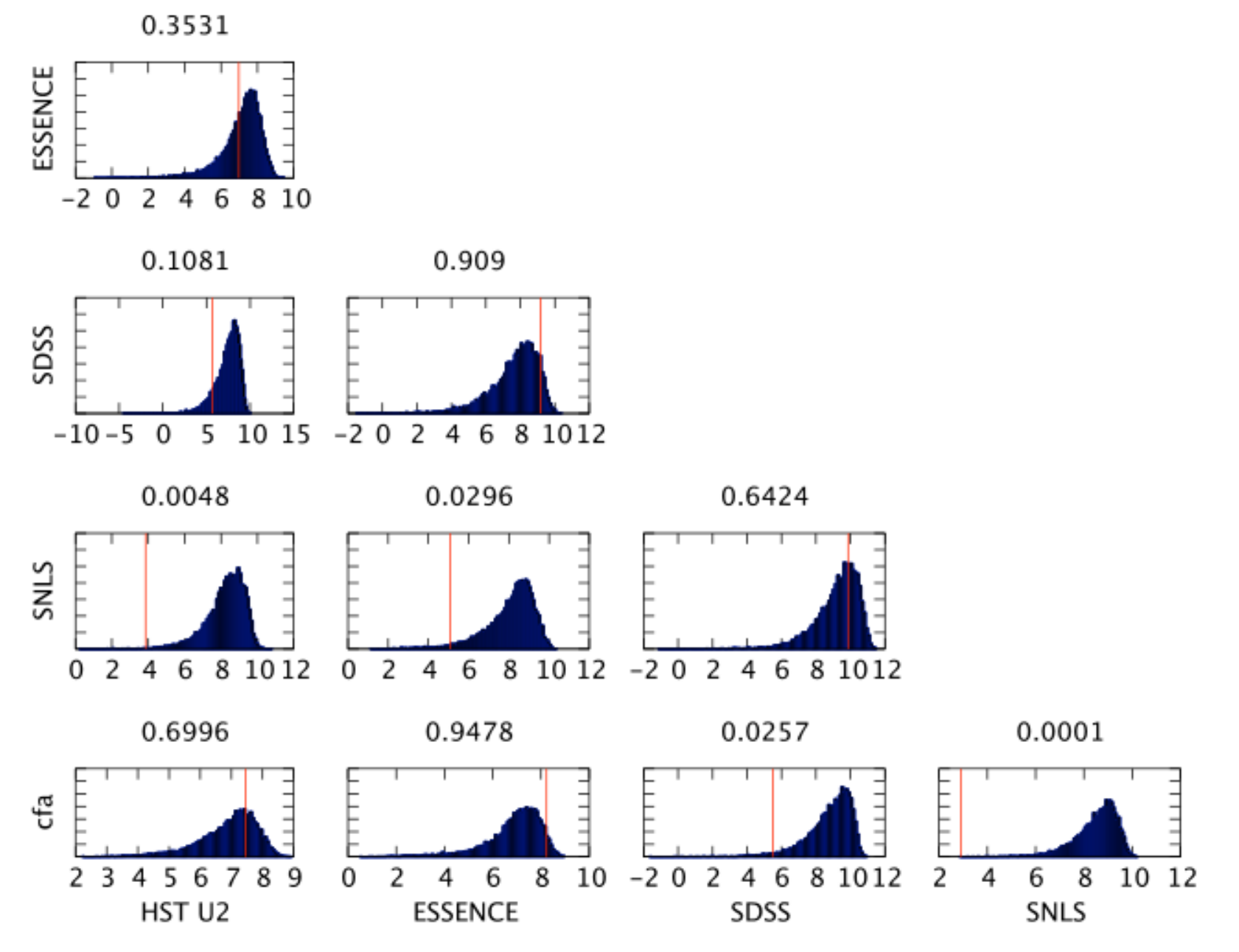} 
\includegraphics[width=0.35\linewidth]{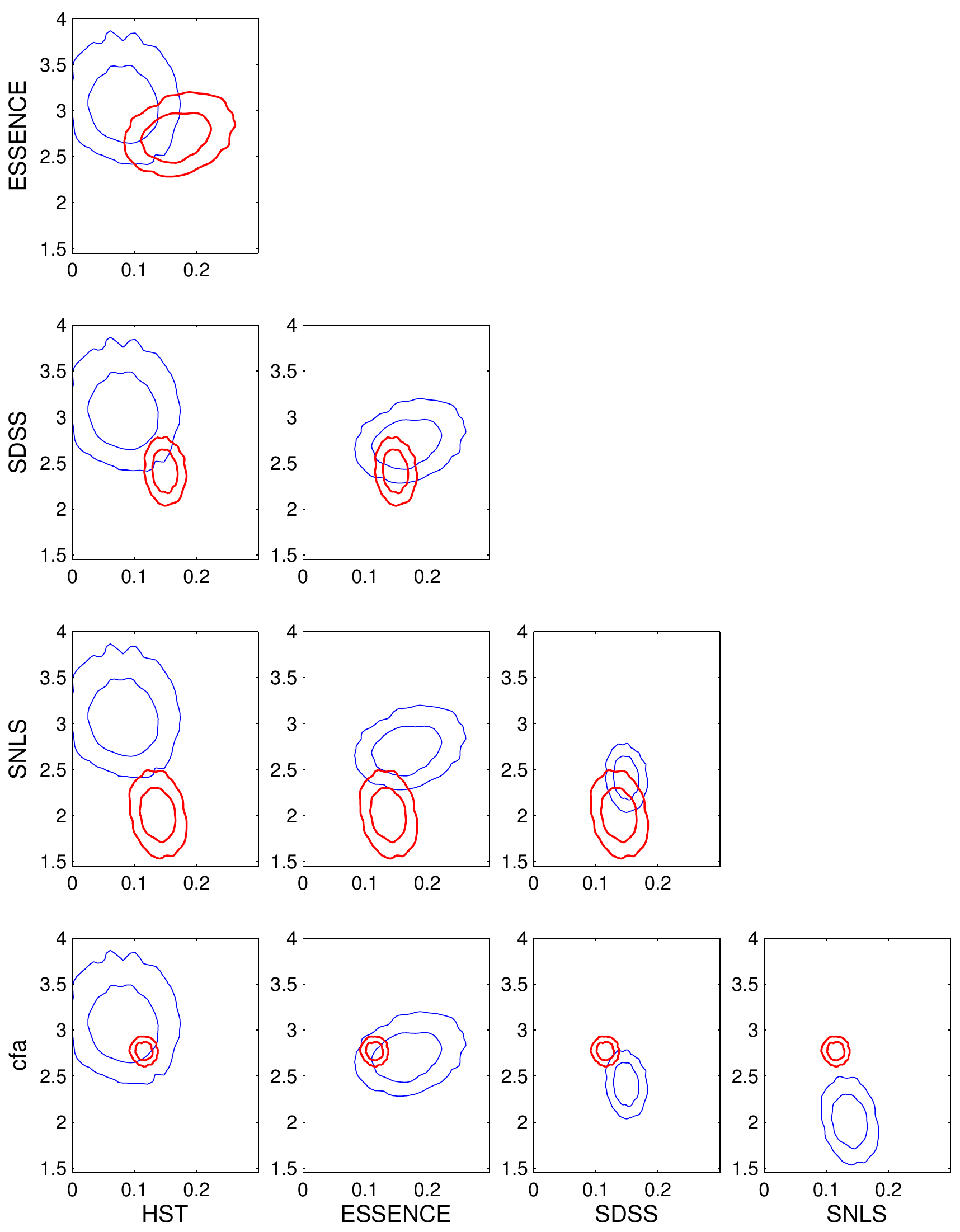} 
\caption{Left panel:  Results of the consistency test applied to survey pairs in
 the Union2 compilation. The blue histograms show the distribution of
 $\mathcal{R}$-values obtained from $10^4$ consistent simulations of each pair,
  and the red vertical line indicates the $\mathcal{R}$-value obtained from the
 real data. The corresponding one-sided $p$-value is given above each
  panel. Right panel: Two-dimensional marginalised constraints on the parameters
  $(\alpha,\beta)$ obtained from the individual constituent
 surveys contained in the Union2 catalogue. The red (blue) contours
  denote the 68 and 95 per cent confidence regions for the survey in
  that row (column). From Paper IV.\label{fig:U2}}
\end{figure}

My key finding is that the multipliers $\alpha$ and $\beta$ of the
stretch and colour corrections, respectively, are significantly
different for the Union2 catalogue (see the left panel of
Figure \ref{fig:U2}). By contrast, the JLA catalogue shows no such inconsistency. Interestingly, both catalogues show no inconsistency in
the constraints derived on cosmological parameters.  Nonetheless, the
inconsistency discovered for the Union2 catalogue means that one must
be careful interpreting the results obtained from a joint analysis of
it. The results of the consistency test for the Union2 compilation are
shown in the left panel of Figure \ref{fig:U2}, together with the
corresponding $p$-values.

\subsection{Consistency test based on BHM}
\label{ch:con:BHM}

\begin{table}
\begin{center}
\begin{tabular}{l|rrrr}
\hline\hline
 & HST & SDSS & SNLS & CfA\\
\hline
SDSS & 3.97 & --- & --- & --- \\
SNLS & 1.50 & 8.33 & --- & --- \\
Low-$z$ & 3.24 & 8.28 & -3.50 & --- \\
CfA & 3.46 & 8.09 & 1.54 & 6.13 \\[1mm]
\hline\hline
\end{tabular}
\caption{$\log(\mathcal{R})$-values for pairwise consistency tests of data-sets in the S11 catalogue. Errors on these $\log(\mathcal{R})$-values are all around $0.18$.}
\label{tab:S11_2w_2b}
\end{center}
\end{table}
\begin{table}
\begin{center}
\begin{tabular}{l|rrrr}
\hline\hline
 & ESSENCE & HST & SDSS & CfA \\
\hline
HST & 10.62 & --- & --- & --- \\
SDSS & -16.42 & 6.19 & --- & --- \\
SNLS &  4.84  &  8.01 & 13.47 & --- \\
CfA & 7.47 & 6.21 & -12.17 & -0.42 \\[1mm]
\hline\hline
\end{tabular}
\caption{$\log(\mathcal{R})$-values for pairwise consistency checks of data-sets in the Union2 catalogue. Errors on these $\log(\mathcal{R})$-values are all around $0.18$.}
\label{tab:U_2w_2b}
\end{center}
\end{table}

In contrast to the $\chi^2$-method, using BHM allows one to use the
$\mathcal{R}$-test directly, as described in Section
\ref{sec:method:Rtest}. Since I use here the same data as in Section
\ref{ch:paper1:2bins}, I will use BHM with two bins for colour
parameters as the default hereafter.

In Tables \ref{tab:S11_2w_2b} and \ref{tab:U_2w_2b}, I summarise the
results for pairwise consistency checks of the S11 and Union2 data-sets
respectively.  I should recall here that positive (negative) values of
$\log(\mathcal{R})$ give evidence in favour of consistency
(inconsistency) between data-sets with the level of consistency
(inconsistency) interpreted according to Jeffreys' scale given in Table
\ref{tab:Jeffreys}. From Tables \ref{tab:S11_2w_2b} and
\ref{tab:U_2w_2b}, we see that, as for the consistency check based on
the $\chi^2$-method, some data-sets are inconsistent within the Union2
catalogue.

The level of inconsistency between different SN data-sets in the Union2
catalogue again means that one should be careful interpreting the
results from joint analysis performed using this compilation. One should
note, however, that the inconsistent pairings derived in the BHM
analyses are different to those found using the $\chi^2$-method. This
is not surprising, as the two methods are very differently affected by
redshift dependence of SN properties, as discussed in
Chapter~\ref{ch:Cosmology}, and the BHM method used here is explicitly
allowing for such an effect. Nonetheless, in both the BHM and
$\chi^2$-method, the inconsistencies found are related to the $\beta$
parameter associated with the colour correction.

\subsection{Hyper-parameters}
\label{ch:con:hyper}

\begin{table}
\begin{center}
\begin{tabular}{l|c}
\hline\hline
Data-set & Union2  \\
\hline
ESSENCE  & $0.49 \pm 0.11$  \\
HST  & $0.65 \pm 0.18$ \\
SDSS & $0.78 \pm 0.22$ \\
SNLS  & $0.46 \pm 0.09$ \\
CfA & $0.17 \pm 0.06$ \\[1mm]
\hline\hline
\end{tabular}
\caption{Estimated values of the hyper-parameters $\gamma_{i}$ for the Union2 catalogue.\label{tab:hyp}}
\end{center}
\end{table}

Here I present results from analysing the SN data-sets from Section
\ref{ch:paper1:2bins} using the hyper-parameter approach described in
Section~\ref{sec:method:hypPar}.

In my work I will introduce the hyper-parameters $\gamma_i$, one for
each data-set, by modifying the covariance matrix $\widehat{C}_i$ of
data-set $D_i$ to become $\widehat{C}_i/\sqrt{\gamma_i}$. This
explicitly allows for the possibility that the quoted measurement
uncertainties are over or under-estimated, but may be considered more
generally as a weighting of each data-set. As discussed in
\citet{Hobson02}, the prior distribution on the hyper-parameters
$\gamma_i$ is exponential with expectation value unity. This follows
because one expects {\em a priori} the quoted measurement uncertainties
from each data-set to be neither over- nor under-estimated. With this
constraint, and the requirement that the weights are non-negative, the
correct prior distribution according to the maximum-entropy principle
is the exponential prior (see e.g. \citealt{Hobson02}).

First I perform model selection between the two hypotheses.
\begin{itemize}
\item[$H_{0}:$] The combined data-set can be described without hyper-parameters $\gamma_{i}$, i.e. the data-sets are all consistent with each other and measurement uncertainties are neither over- nor under-estimated.
\item[$H_{1}:$] The combined data-set requires the hyper-parameters $\gamma_{i}$, in order to deal with inconsistencies between data-sets and/or inaccuracies in the measurement uncertainties.
\end{itemize}

For the S11 and Union2 catalogues,
$\log(\mathcal{Z}_{H_{0}}/\mathcal{Z}_{H_{1}})$ is found to be $2.30
\pm 0.17$ and $-8.54 \pm 0.18$, respectively. These values point
towards inconsistency between different data-sets in the Union2
catalogue which is in perfect agreement with our findings in the last
section using the $\mathcal{R}$-test. The S11 catalogues did not show
any inconsistency between different data-sets according to the
$\mathcal{R}$-test, which is reinforced by their preference for a
model without hyper-parameters $\gamma_{i}$.

\begin{figure}[t]
\begin{center}
\includegraphics[width=9cm]{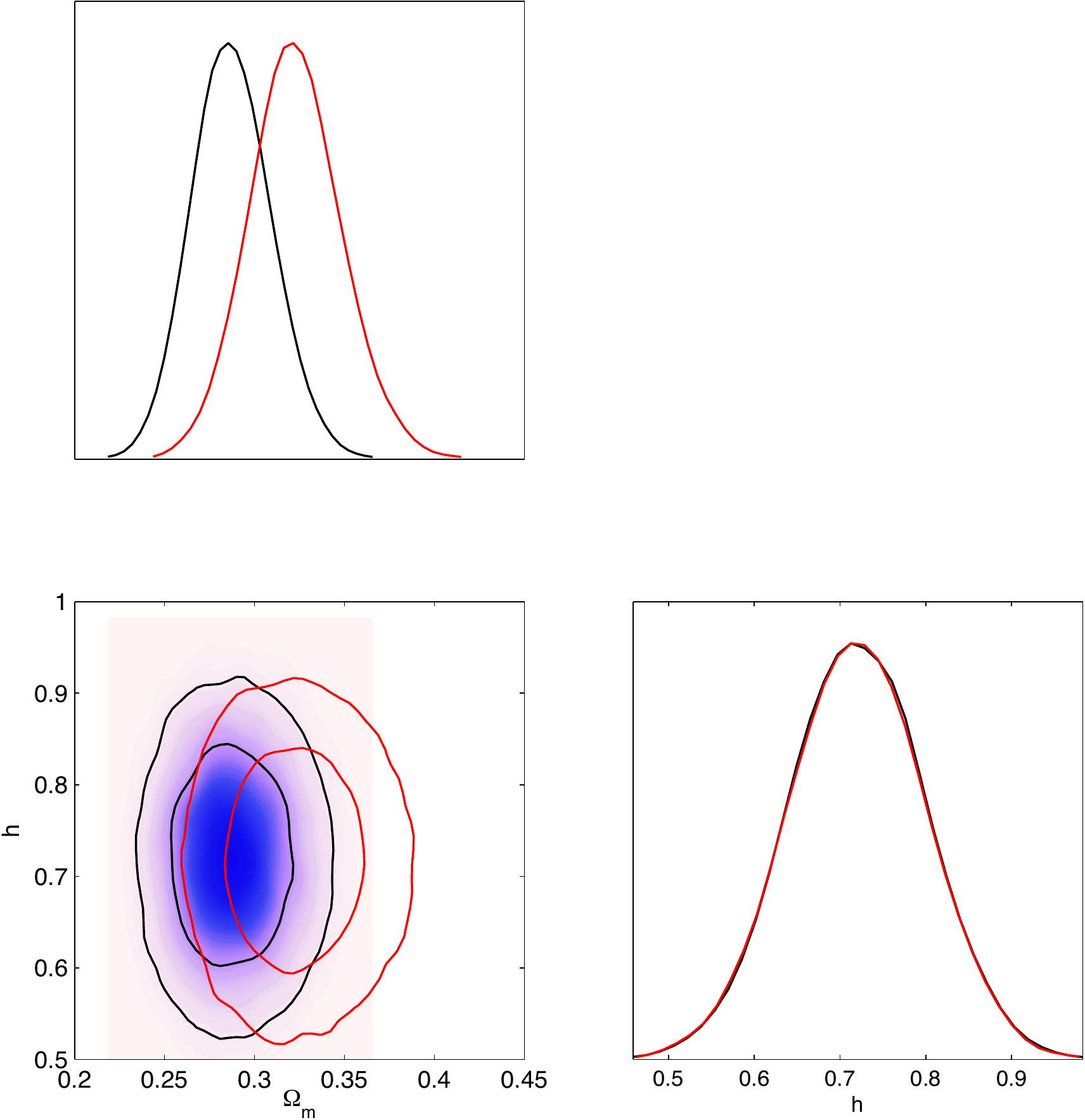}
\caption{1D and 2D marginalised posterior distributions for the matter
  density $\Omega_{\rm m,0}$ and Hubble parameter $h$ when the Union2
  catalogue is analysed with (red) and without (black)
  hyper-parameters $\gamma_{i}$.}
\label{fig:hyp}
\end{center}
\end{figure}

Estimated values of hyper-parameters $\gamma_{i}$ are given in Table
\ref{tab:hyp}. In the absence of any inconsistencies or measurement
inaccuracies, one should expect $\gamma_{i} \sim 1$, therefore any
deviation from unity provides evidence in favour of some unaccounted
systematics in the data-set. It can be seen from Table \ref{tab:hyp}
that in the Union2 catalogue, ESSENCE, SNLS and CfA, all have
$\gamma_{i}$ more than $3\sigma$ away from unity. In order to show the
effect of these systematics on cosmological parameter inferences, we
plot the marginalised posterior distributions for the matter density
$\Omega_{\rm m,0}$ and Hubble parameter $h$ when the Union2 catalogue
is analysed with and without hyper-parameters
$\gamma_{i}$. Introduction of hyper-parameters results in a 15\%
increase in the estimated value of $\Omega_{\rm m,0}$, as shown in
Figure~\ref{fig:hyp}. Hence it is important to include these effects.

%


\chapter{Outlook}
\label{ch:Outlook}

\epigraph{`Let the past look after itself, and let the present move
  forwards into the future'}{\textit{Douglas Adams}}

\section{Future problems for SN astronomy}

SN data helped observational cosmology to make a great step forward
over the last two decades. The discovery of the accelerated expansion
of the Universe undoubtedly is one of the most important breakthroughs
over the past 20 years. Unfortunately, the nature of it still remains
unknown. Nevertheless, future SN surveys certainly will help our
understanding of this great mystery of 21st century physics. Already
today, SN data are good enough to perform model selection between
different cosmological scenarios. Future SN surveys will address all
these problems, not only by increasing the number of observed SNe,
thereby improving the statistical errors, but also by focussing on
systematics. The problems discussed in Chapter \ref{ch:Cosmology}
regarding rigorous inference methods will be more important than ever,
which will inevitably lead to the development of more robust methods
for cosmological parameter inference.

 Another big question in SN studies is the nature of the SN explosion
 itself. Standard models for SNIa explosions, such as a carbon-oxygen
 white dwarf burning and the merger of two white dwarfs, are
 increasingly questioned for their universality. On the other hand,
 the nature of core-collapse SNe is an even bigger puzzle, resulting in
 our inability even to write a parametric function for the non-Ia
 lightcurves. The solution to this problem does not relate to cosmology
 {\em per se}, but will inevitably have a huge influence on cosmological
 parameter inference. Knowing an explosion mechanisms will play a
 crucial role in a better standardisation of SNe.
 
 Observing SNe in multiple filters proved to be key to their
 standardisation. Future surveys will increase the number of filters,
 which will not only help to better standardise events, but will open a
 window for spectroscopy-free SNe cosmology. Having measurements of
 lightcurves in multiple filters in the future will allow us to make
 classification based only on photometry. Already, some photometric
 classification methods seems to be very competitive.  Improving these
 methods will be one of the main objectives for future photometric
 surveys. For example, as was mentioned before, knowing the exact
 explosion mechanism for different types of SNe will be a great
 advantage, since having a parametric function for all SN types will
 allow us to make model selection between different SN types, as
 opposed to performing the comparison using template methods.
 
In discussing spectroscopy-free cosmology, however, it is important
not to belittle the importance of high-quality spectroscopic data.
Such data is a cornerstone of SN cosmology. New experiments, such as
iPTF and ZTF, will have on-line spectroscopic follow-up for all detected
SNe. Having high-resolution spectroscopic data at the epochs as close
as possible to the explosion will give us invaluable information to
study the explosion mechanisms. Having a large data-set of
spectroscopically-classified SNe is also of a great importance for 
developing the photometric classification algorithms mentioned above.
 
 Another interesting path for SN cosmology is to try to standardise
 non-Ia SNe. As I discussed in Section \ref{ch:IIp_cosmology}, SNII-P
 have proved to be viable candidates for making cosmological
 inferences. With a better understanding of ``standard'' SN types and
 with ``modern'' SNe, such as SLSNe, or types yet to be discovered, we can
 try in the future to make a cosmological inference based on more than
 one type, or even using all of them.

One might summarise the current status of SN cosmology by saying
that, although it blossomed in 1998, the fruits are only now becoming
ripe.

\section{My interest for the coming years}

The Bayesian methods I developed during my PhD years are ideally
suited for investigating the problems discussed in the previous
section.  Since the statistical methods I am working with are very
general in nature, I would like to broaden my research interests into
other neighbouring topics, such as the CMB and large-scale structures,
as well as more areas in gravitational lensing. In particular, I wish
to apply my experience with Bayesian parameter estimation and model
selection to these areas. Building on my work with SNIa, one area of
considerable interest would be to apply my Bayesian methods for
testing the consistency of data-sets to a wider range of cosmological
probes. There are already hints from Planck and SNIa data that the two
data-sets disagree on the value of the Hubble parameter. This could be
a real cosmological effect, or the result of undiagnosed
systematics. Also, as cosmological data-sets improve, it will become
possible to distinguish the standard concordance cosmology from
alternative models at greater significance. As the signal-to-noise
ratio of data improves, one (somewhat ironically) has to be {\em more}
careful in performing statistical analyses, as systematic effects
begin to emerge from the noise. My Bayesian methods are again ideally
suited to investigate such problems.

\textit{As the final words of my ``The supernova cosmology cookbook'', I would
  like to quote my favorite physicist, Lev Davidovich Landau:
  ``Cosmologists are often in error, but never in doubt''. }

\textit{\crussian{Финальными словами моей ``Поваренной книги о сверхновых в кос\-мологии: Байесовские численные рецепты'' я хочу процицировать моего любимого физика Льва Давидовича Ландау ``Астрономы часто ошибаются, но никогда не сомневаются''.}}

\appendix
\chapter{Princes cake}
\label{ch:appA}

\section{Ingredients}
Cake:
\begin{itemize}
\item 4 eggs
\item 2 dl sugar
\item     1 dl flour
\item     1 dl potato flour
\item     2 teaspoons baking powder
\end{itemize}
  
Filling:
\begin{itemize}
\item 4 sheets of gelatine
\item  2 dl vanilla sauce
\item     2 teaspoons vanilla sugar
\item     3 dl whipping cream
\end{itemize}

Garnish:
\begin{itemize}
\item 300 g marzipan
\item green and red food dye
\item    icing sugar
\end{itemize}

\section{What to do}
Cake:
\begin{enumerate}
\item  Preheat the oven to 175 C.
\item Grease and dust a round shape, about 2.5 liters.
\item Beat eggs and sugar until fluffy. Mix the two flours and baking
  powder and fold into the batter and mix well. Pour into mold. Bake
  in the lower part of the oven for about 40 minutes. Turn the cake and
  let it cool.
\end{enumerate}
Filling: 
\begin{enumerate}
\item Add the gelatine leaves to cold water. Boil the mix of the vanilla sauce
  according to package directions. Remove the gelatine leaves and
  squeeze them well. Place them in the hot vanilla sauce.
\item Whip the cream until thick. Stir in the vanilla sauce when the cream starts to
  thicken. Let the cream become almost solid.
\end{enumerate}

Cut the cake into 3 layers (for 1 cake). The top should be slightly
less than 1 inch thick. Flatten the layers with filling
between. Save a bit of it. Let the filling be slightly higher in
the middle so the cake will be puffy when the uppermost thin layer
is put on. Spread the rest of the cream on top and around the edge.

Garnish:
\begin{enumerate}
\item Colour about $70\%$ of the marzipan green and the remaining marzipan red (it will become more pink than red). 
\item Roll the green marzipan into a round, thin, smooth sheet.
It should be enough to cover the entire cake with an even thickness. 
\item Cut out a circle sector with an area of about $30\%$ of the area of the full sheet.
\item Roll the pink marzipan to a sheet big enough to cover the piece which was cut out from the green.
\item Cut out a circle segment with the same size as that which was removed before. 
\item Cover the cake with the the green and pink marzipan pieces.
\item   Sift icing sugar on top.
\end{enumerate}

As done in many Swedish bakeries, put on top a pretty pink marzipan rose. The best is to find a rose with the mass of approximately $5\%$ of the total marzipan mass composing your cosmology cake.

\section{When to eat}
Princess cake is perfect for any occasion, but best (in the author's
opinion) with a cup of Swedish filtered coffee on a cold winter day which lasts for four months.


\setlength{\bibsep}{0pt}            
\renewcommand{\bibname}{References} 

\def\reff@jnl#1{{\rm#1\/}}
\def\aj{\reff@jnl{AJ}}                 
\def\araa{\reff@jnl{ARA\&A}}           
\def\apj{\reff@jnl{ApJ}}               
\def\apjl{\reff@jnl{ApJ}}              
\def\apjs{\reff@jnl{ApJS}}             
\def\ao{\reff@jnl{Appl.Optics}}        
\def\apss{\reff@jnl{Ap\&SS}}           
\def\aap{\reff@jnl{A\&A}}              
\def\aapr{\reff@jnl{A\&A~Rev.}}        
\def\aaps{\reff@jnl{A\&AS}}            
\def\azh{\reff@jnl{AZh}}               
\def\baas{\reff@jnl{BAAS}}             
\def\jcap{\reff@jnl{JCAP}}           
\def\jrasc{\reff@jnl{JRASC}}           
\def\memras{\reff@jnl{MmRAS}}          
\def\mnras{\reff@jnl{MNRAS}}           
\def\pra{\reff@jnl{Phys.Rev.A}}        
\def\prb{\reff@jnl{Phys.Rev.B}}        
\def\prc{\reff@jnl{Phys.Rev.C}}        
\def\prd{\reff@jnl{Phys.Rev.D}}        
\def\prl{\reff@jnl{Phys.Rev.Lett}}     
\def\pasp{\reff@jnl{PASP}}             
\def\pasj{\reff@jnl{PASJ}}             
\def\qjras{\reff@jnl{QJRAS}}           
\def\skytel{\reff@jnl{S\&T}}           
\def\solphys{\reff@jnl{Solar~Phys.}}   
\def\sovast{\reff@jnl{Soviet~Ast.}}    
\def\ssr{\reff@jnl{Space~Sci.Rev.}}    
\def\zap{\reff@jnl{ZAp}}               
\def\nat{\reff@jnl{Nature}}            

\end{document}